\documentclass[twoside,11pt]{article}

\usepackage{blindtext}

\usepackage{amsmath,amssymb}

\usepackage{natbib}
\usepackage{mathrsfs}
\usepackage{bm, xcolor}
\usepackage{mathrsfs}
\usepackage{verbatim}
\usepackage{longtable}
\usepackage{threeparttable}
\usepackage{color}
\usepackage{amsmath}

\usepackage{float} \usepackage{algorithm} \usepackage{algpseudocode} 

\usepackage{enumerate}
\usepackage{amssymb}
\usepackage{amsbsy}
\usepackage{amsmath}
\usepackage{amsfonts}
\usepackage{latexsym}
\usepackage{graphicx}
\usepackage{xifthen} \usepackage{color}
\usepackage{manfnt}
\usepackage{ifthen}
\usepackage{bm}
\usepackage{caption}
\usepackage{subcaption}
\usepackage{chngcntr}
\usepackage{apptools}

\newcommand{\mc}[1]{\mathcal{#1}}

\def\mbi#1{\boldsymbol{#1}}

\newcommand{\norm}[1]{\left\|{#1}\right\|} \newcommand{\lone}[1]{\norm{#1}_1}

\newcommand{\absbigg}[1]{\left \lvert{#1}\right\rvert}
\newcommand{\firstbigg}[1]{\left ({#1}\right)}
\newcommand{\secondbigg}[1]{\left \{{#1}\right\}}
\newcommand{\thirdbigg}[1]{\left [{#1}\right]}

\newcommand{\defeq}{:=}

 \newcommand{\wt}[1]{\widetilde{#1}}

\newcommand{\reals}{\mathbb{R}} 
   
\newcommand{\naturals}{\mathbb{N}}

\renewcommand{\iff}{\Leftrightarrow}

\newcommand{\cP}{\mathcal{P}}

\newcommand{\E}{\mathbb{E}} \renewcommand{\P}{\mathbb{P}} \newcommand{\var}{{\rm Var}} \newcommand{\cov}{\mathop{\rm Cov}}  \newcommand{\simiid}{\stackrel{\rm iid}{\sim}}

\newcommand{\indic}[1]{\mbf{1}\left\{#1\right\}}

\newcommand{\cd}{\stackrel{d}{\rightarrow}}

\newcommand{\cp}{\stackrel{p}{\rightarrow}}
\newcommand{\ucd}{\stackrel{d}{\rightsquigarrow}} 

    \newcommand{\floor}[1]{\left\lfloor{#1} \right\rfloor}
\newcommand{\ceil}[1]{\left\lceil{#1} \right\rceil}

\providecommand{\argmax}{\mathop{\rm argmax}}

\makeatletter
\def\singlespace{\def\baselinestretch{1}\@normalsize}

\newtheorem{assumption}{Assumption}\@addtoreset{equation}{section}

\renewcommand{\hat}{\widehat}

\def\singlespace{\def\baselinestretch{1}\@normalsize}

\makeatother

\def\newpage{\vfill\eject}

\def\leqsim{\stackrel{<}{\sim}}
\def\geqsim{\stackrel{>}{\sim}}
\def\wt{\widetilde}

\newdimen\biblioindent    \biblioindent=30pt

 at 7truept

\def\beqr{\begin{eqnarray}}
	\def\eeqr{\end{eqnarray}}
\def\beqrs{\begin{eqnarray*}}
	\def\eeqrs{\end{eqnarray*}}

\def\beq{\begin{equation}}
\def\eeq{\end{equation}}
\def\beqn{\begin{eqnarray}}
\def\eeqn{\end{eqnarray}}
\def\beqnn{\begin{eqnarray*}}
\def\eeqnn{\end{eqnarray*}}

\def\wt{\widetilde}

\def\A{{\bf A}}

\def\Z{{\bf Z}}
\def\z{{\bf z}}

\def\calI{\mathcal{I}}

\def\calP{\mathcal{P}}

\def\calN{\mathcal{N}}

\def\indic{\mathbb I} 
\def\defeq{\stackrel{\mathrm{def}}{=}}  

\makeatother
 \usepackage{jmlr2e}
\usepackage{multirow}
\allowdisplaybreaks[4]

\usepackage{diagbox}

\usepackage[symbol]{footmisc}

\usepackage{lastpage}
\jmlrheading{26}{2025}{1-\pageref{LastPage}}{3/24; Revised 6/25}{7/25}{21-0000}{Rohit Kanrar, Feiyu Jiang and Zhanrui Cai}

\ShortHeadings{changeAUC: change-point detection using auc}{Kanrar, Jiang and Cai}
\firstpageno{1}

\begin{document}
\setlength{\abovedisplayskip}{3pt}
\setlength{\belowdisplayskip}{3pt}
\title{Model-free Change-Point Detection Using \\AUC of a Classifier}

\author{\name Rohit Kanrar\email rohitk@iastate.edu \\
       \addr Department of Statistics\\
        Iowa State University\\
       Ames, IA 50011-1090, USA
       \AND
       \name Feiyu Jiang$^{*}$\email jiangfy@fudan.edu.cn \\
       \addr School of Management\\ Fudan University\\
       Shanghai, 200433, China
       \AND
       \name Zhanrui Cai$^{*}$\email zhanruic@hku.hk \\
       \addr Faculty of Business and Economics\\  University of Hong Kong\\
      Hong Kong, China}

\footnote[0]{$^*$: Corresponding authors.}

\editor{Mladen Kolar}

\maketitle

\begin{abstract}In contemporary data analysis, it is increasingly common to work with non-stationary complex data sets. These data sets typically extend beyond the classical low-dimensional Euclidean space, making it challenging to detect shifts in their distribution without relying on strong structural assumptions. This paper proposes a novel offline change-point detection method that leverages classifiers developed in the statistics and machine learning community.  With suitable data splitting, the test statistic is constructed through sequential computation of the Area Under the Curve (AUC) of a classifier, which is trained on data segments on both ends of the sequence. It is shown that the resulting AUC process attains its maxima at the true change-point location, which facilitates the change-point estimation. The proposed method is characterized by its complete nonparametric nature, high versatility, considerable flexibility, and absence of stringent assumptions on the underlying data or any distributional shifts. Theoretically, we derive the limiting pivotal distribution of the proposed test statistic under null, as well as the asymptotic behaviors under both local and fixed alternatives. The localization rate of the change-point estimator is also provided. Extensive simulation studies and the analysis of two real-world data sets illustrate the superior performance of our approach compared to existing model-free change-point detection methods.
\end{abstract}


\newpage

\section{Introduction} \label{sec:intro}
In this paper, we focus on detecting distributional changes in a sequence of independent elements  $\{\Z_t\}_{t=1}^T$ taking value in a measurable space $(E,\mathcal{E})$, which can be formulated  as, 
\begin{align} \label{eq:hyp_single_cp}
    \begin{split}
        H_0: & \Z_t \sim \calP_X \ \text{for all} \ 1 \leq t \leq T\\
        H_1: & \Z_t \sim \begin{cases}
        \calP_X \ \text{for} \ 1 \leq t \leq t_0 \\
        \calP_Y \ \text{for} \ t_0+1 \leq t \leq T,\quad 1\leq t_0<T \text{  is unknown,}
    \end{cases}
    \end{split}
\end{align}  
where $\calP_X\neq \calP_Y$ are two unknown distributions.  
The hypothesis testing problem \eqref{eq:hyp_single_cp} is typically known as the change-point detection problem, with a vast literature devoted to this area and broad applications in bioinformatics, climate science, finance, geographic studies, signal processing and robotics, among many other areas. 

Existing change-point detection methods often impose strong structural assumptions regarding the distributions $\cP_X$ and $\cP_Y$, and are typically designed for specific scenarios. Most approaches concentrate on either the first- or second-order moments within the classical Euclidean space or rely on parametric assumptions about the data generation process; see, e.g., \cite{aue_horvath_jtsa_2013_reviewpaper}, \cite{chen_gupta_2012_book}, and \cite{truong2020selective} for reviews. In recent times, it has become increasingly common to collect data in complex forms, often characterized by high dimensionality and even deviations from the conventional Euclidean space.   For example, financial analysts often encounter high-dimensional stock data to detect distributional shifts during significant economic events like the ``Great Recession" \citep{chakraborty_zhang_2021}. Researchers frequently use satellite image data to investigate the impact of human activities on various environments \citep{labuzzetta_zhu_2024_sdsi, moore_chu_zhu_2025+}. In urban studies, researchers analyze transportation data for disruptions or changes in traffic patterns \citep{chu_chen_2019_annals}. These examples highlight the formidable challenge of devising a universally applicable change-point detection framework that accommodates diverse data dimensions and types. Given these observations, we propose a novel change-point detection framework named \texttt{changeAUC} for detecting general distributional changes, regardless of the dimensions and types of data.

\subsection{Our Contributions} \label{subsec:our_contributions}
Our proposed method \texttt{changeAUC} is motivated by the recent success of classification algorithms in high-dimensional two-sample testing problems \citep{hu_lei_2023_jasa} and independence testing problems \citep{cai2022model,cai_et_al_2022}. Specifically, \texttt{changeAUC}  involves a four-step procedure: 1) data splitting by trimming at the beginning and end of the sequence; 2) training the classifier using splits from beginning and end; 3) recursive splitting of data based on the candidate change-point location; and 4) testing the accuracy of observations in the middle based on AUC, the Area Under the ROC Curve. The key idea behind the classification accuracy test is the following: if the two groups of samples are indeed heterogeneous, a complex classifier should be able to distinguish the underlying distributions, yielding higher accuracy on the test set. We leverage this intuition by training a classifier once, followed by recursively conducting two-sample testing across time points by measuring classification accuracy, thus enabling our objective of change-point detection. 
The proposed method enjoys several appealing advantages. 
    \begin{itemize}
        \item {\sc completely nonparametric}. It does not impose any parametric assumption on the data distributions.  Moreover, it does not impose any structure on the distributional shift, e.g., change occurs in mean or covariance,  change is sparse or dense.  
        \item {\sc highly versatile}. Our approach applies to different data types, encompassing classical low-dimensional Euclidean, high-dimensional, and non-Euclidean data such as images and texts. The sole prerequisite is the availability of a suitable classifier. 
        \item {\sc considerably flexible}. It can be fine-tuned for any particular application by selecting an appropriate classifier. Such flexibility also enables effective incorporation of prior knowledge into the analysis. For instance, a Convolutional Neural Network (CNN) can identify change-points in images or videos. Our method does not necessitate a consistently accurate classifier.  As long as the classifier captures some signals between two distributions, \texttt{changeAUC} can effectively detect change-points.
        \item {\sc distribution-free}. Our test statistic has a pivotal limiting distribution under the null. In fact, it is shown to be a functional of Brownian motion, which implies that the critical values can be numerically tabulated. 
    \end{itemize}

Rigorous theoretical justifications have been provided to support the good performance of our proposed method.  We use empirical process theory \citep{vaart2023empirical} to obtain the pivotal limiting distribution of the test statistic under null.   The asymptotic behaviors of the test statistic are also derived under local and fixed alternatives. Furthermore, we establish the consistency and localization rate of the maximizer of the AUC process for estimation purposes. Extensive numerical studies on both simulated data and two real applications further corroborate the theoretical findings. The empirical results underscore the effectiveness of our testing framework, demonstrating its ability to control size performance and showcasing competitive power performance compared to existing methods.

\subsection{Related Work} \label{sec:review}
The origin of offline change-point problems can be dated back to the seminal work of \cite{page_1954_biometrika}. The classical approaches to tackling change-point problems are typically limited to low dimensional quantities, such as mean and variance, or specific parametric models. We refer to \cite{aue_horvath_jtsa_2013_reviewpaper}, \cite{chen_gupta_2012_book}, and \cite{truong2020selective}  for comprehensive reviews. With modern data collection techniques, high-dimensional and non-Euclidean data have become ubiquitous in many scientific areas. In particular, for high dimensional data,  extensive works have devoted to mean changes \citep{jirak_annals_2015, cho_fryzlewicz_2015_jrssb,wang_samsworth_jrssb_2018,enikeeva_harchaoui_annals_2019,yu_chen_2021_jrssb, wang2022inference};  covariance changes \citep{avanesov_bazun_electronicJ_2018,steland_2019_jma,dette_et_al_2020_jasa,li_li_fryzlewicz_2022_jbes}; and other model parameters \citep{liu2020unified, xu2024change}. For non-Euclidean data, much effort has been put into Hilbert space, and we mention \cite{berkes2009detecting, aue2018detecting} for functional mean changes and \cite{jiao2023break} for functional covariance changes.   We also note \cite{dubey2020frechet} and \cite{jiang2023two} for recent developments in change-point detection of first- and second-order moments for non-Euclidean data in metric space, where classical algebraic operators, such as addition and multiplication, do not apply. 

Previous works mainly focus on detecting only one or two specific types of changes, which may lack power when changes occur in other aspects of the data.  Nonparametric methods have emerged in recent years as a powerful counterpart to parametric methods for detecting general types of change. Most methods are appropriate for the low dimensional Euclidean space, see, e.g., \cite{kawahara_sugiyama_2012_sadm}, \cite{liu_et_al_2013_neural_networks}, \cite{matteson_james_2014_jasa}, \cite{zou_et_al_annals_2014}, \cite{ arlot_et_al_jmlr_2019}, \cite{padilla_et_al_2021_ieee}. We are only aware of a few recent developments in high-dimensional data; e.g., \cite{chakraborty_zhang_2021} uses generalized energy distance.  However, their method can only capture the pairwise homogeneity  (heterogeneity under $H_1$)  of the marginal distributions in high dimensional $(\calP_X,\calP_Y)$, which may lack power when  $(\calP_X,\calP_Y)$ correspond to the same marginal distributions but differ in other aspects. Some graph-based tests have been proposed recently by \cite{chen_zhang_2015_annals} and \cite{chu_chen_2019_annals}. Yet, as pointed out by  \cite{chakraborty_zhang_2021}, they may lack the power to detect changes in higher-order moments or when the data dimension is high with sparse distributional shifts. In contrast, our proposed approach takes advantage of contemporary classifiers within the statistics and machine learning community. This adaptability allows us to detect changes beyond marginal distributions and moments, setting our method apart in terms of its capability to uncover variations in high-dimensional data and even non-Euclidean data.  These findings are further corroborated by simulation studies in Section \ref{sec:simulation}.

We also mention a few recent contributions that borrow the strength of classifiers in the context of two-sample testing and change-point problems. Since \cite{kim_et_al_2020_annals}, there has been research effort that employ classifiers to conduct two-sample tests, see, e.g. \cite{kim_et_al_2018_ejs} and references therein. However, the extension to the change-point context seems largely unexplored, except for \cite{lee2023training}, \cite{wang_et_al_2023_ieee}, \cite{londschien_buhlman_kovacs_2022}, and \cite{li_paul_fryzlewicz_wang_2022}.

In particular, for online data, both \cite{lee2023training} and \cite{wang_et_al_2023_ieee} leverage neural networks to estimate likelihood ratio functions for change-point detection.  For offline data, \cite{londschien_buhlman_kovacs_2022} employs the random forest classifier for detecting change-points by introducing a classifier-based nonparametric likelihood ratio, while  
\cite{li_paul_fryzlewicz_wang_2022} proposes to use neural networks for detecting change-points in time series under supervised setting.  Compared with offline methods, our proposed framework differs in several key aspects. 1) Both of the aforementioned methods implicitly assume Euclidean data type, whereas our approach additionally accommodates non-Euclidean data with appropriate classifiers. 2) Neither of the aforementioned methods offers asymptotic results, leading to increased computational demands through permutation or conservative type-I error controls. In contrast,  our method offers an asymptotic approach by providing a pivotal limiting null distribution, facilitating practical statistical inference. Therefore, our results complement their works, particularly in the theoretical aspect.
3) Unlike the nonparametric likelihood used by \cite{londschien_buhlman_kovacs_2022} and the empirical risk minimizer-based classification accuracy used by \cite{li_paul_fryzlewicz_wang_2022}, we measure the classification accuracy using the AUC. Compared with likelihood-ratio-based methods, which can be difficult to compute in high-dimensional settings or when specifying the likelihood is challenging, the use of AUC from a classifier offers greater robustness by relieving reliance on specific distributional assumptions. In addition, the rank-sum comparison in AUC provides robustness against estimation error of likelihood ratio  and classification probability, meaning that the classifier does not need to be strictly consistent. It allows for weakly trained classifiers without extensive hyperparameter tuning. More discussions are deferred to Section \ref{sec:methods}. Lastly, we mention a similar data-splitting approach implemented by \cite{gao_wang_shao_2023_arxiv}, where the authors propose a dimension-agnostic change-point test statistic by projecting middle observations to the direction determined by observations at both ends of the sequence. However, it is designed only for detecting a change in the mean, while our method is more general and uses rank-sum comparison through AUC.  

The rest of this article is organized as follows.  
Section \ref{sec:methods} introduces our proposed method with comprehensive justifications for each step. Theoretical analysis is conducted in Section \ref{sec:theory}. Section \ref{sec:simulation} presents the finite sample performance of our framework. Section \ref{sec:real_data} demonstrates the application of our method to two real data sets. Section \ref{sec:discussion} concludes with discussions and future directions. All technical proofs are deferred to the Appendix.

\section{Methods} \label{sec:methods}
Section \ref{subsec:proposed_method} outlines \texttt{changeAUC} for detecting at most one change-point in a sequence of independent observations. Section \ref{subsec:clf_accuracy} provides a detailed justification of the steps of our proposed framework. Section \ref{subsec:extn_multiple_cp} extends the method to estimate multiple change-points.

\subsection{Proposed Method} \label{subsec:proposed_method}
Consider a sequence of independent random elements $\{\Z_t\}_{t=1}^T$ defined on a common probability space $(\Omega, \mc{F}, \P)$ and taking value in a measurable space $(E, \mc{E})$.  Let  $\cP_X$ and $\cP_Y$ be two different probability distributions on the same probability space such that $\cP_X, \cP_Y$ are absolutely continuous with respect to each other, i.e., $\cP_X \ll \cP_Y$ and $\cP_Y \ll \cP_X$.   To test \eqref{eq:hyp_single_cp}, our proposed methodology consists of the following four steps:

\begin{itemize}
    \item[1.] \textbf{Sample Splitting:} For a small trimming parameter $\epsilon \in (0, 1/2)$ with $m = \floor{T\epsilon}$, data are split into three parts: $D_0 \defeq \{\Z_t\}_{t=1}^{m}$, $D^v \defeq \{\Z_{t}\}_{t=m+1}^{T-m}$ and $D_1 \defeq \{\Z_{t}\}_{t=T-m+1}^{T}$.
    
    \item[2.] \textbf{Training Classifier:} Labels $v_t = 0$ for $\Z_t \in D_0$ and $v_t = 1$ for $\Z_t \in D_1$ are assigned and a classifier is trained based on observations from beginning and end of the sequence, i.e., $\secondbigg{(v_t, \Z_t): \Z_t \in D_0 \cup D_1}$. 
    
     For the steps below, let $\hat{\theta}(\cdot)$ be the estimated conditional probability distribution associated with the classifier. Given a new observation $\Z=\z\in E$, $\hat{\theta}(\z) = \hat{\P}\thirdbigg{v=1 \vert \Z=\z}$ is the estimated probability of $\Z=\z$ being generated by the same distribution as $D_1.$
    \item [3.] \textbf{Recursive Splitting:} For another small trimming parameter $\eta \in (0, 1/2-\epsilon)$ and the set of candidate change-points $\calI_{cp} \defeq \{\floor{Tr}: r \in [\epsilon+\eta, 1 - \epsilon - \eta]\}$, validation data $D^v$ are split into two parts for each $k \in \calI_{cp}$: $D_0^v(k) \defeq \{\Z_{t}\}_{t=m+1}^{k}$ and $D_1^v(k) \defeq \{\Z_{t}\}_{t=k+1}^{T-m}$.
    
    \item[4.] \textbf{Measuring Accuracy:} For each $k \in \calI_{cp}$, labels $v_t = 0$ for all $\Z_t \in D_0^v(k)$ and $v_t = 1$ for all $\Z_t \in D_1^v(k)$ are assigned to calculate the Area Under the Receiver Operating Characteristic (ROC) curve (AUC) using $\{(v_t, \Z_t): \Z_t \in D_0^v(k) \cup D_1^v(k)\}$: \begin{equation}\label{Psik}
        \hat{\Psi}(k) \defeq \frac{1}{(k-m)(T-m-k)} \sum_{i=m+1}^{k} \sum_{j=k+1}^{T-m} \indic\secondbigg{\hat{\theta}(\Z_i) < \hat{\theta}(\Z_j)}.
    \end{equation}
\end{itemize} Steps 3-4 are repeated for all candidate change-points to obtain AUCs, $\{\hat{\Psi}(k)\}_{k \in \mathcal{I}_{cp}}$.  

\begin{remark} \label{rmk:tuning_param}
The choices of $\epsilon$ and $\eta$ are subjective. A moderate value of $m=\floor{T\epsilon}$ is essential for appropriate classifier training. A smaller choice of $\epsilon$ may result in power loss, whereas a larger $\epsilon$ requires the change-point to exist in the middle of the sequence. In practice, we propose to fix $\epsilon=0.15$. However, choosing a smaller $\epsilon$ in our framework is feasible, particularly when the sample size is substantial or if the classification algorithm performs well with a smaller training sample. Notably, under the supervised setting or when knowledge transfer is plausible through a pre-trained classifier,  a much smaller value of $\epsilon$ is allowed. The role of $\eta$ works as the trimming parameter commonly adopted in change-point literature, see \cite{andrews1993tests}. We propose to fix $\eta=0.05$ for practical implementations. 
\end{remark}

Under $H_0$ with no change-point, $\hat{\Psi}(k)$ is expected to be close to 0.5 for all values of $k,$ resembling a random guess. In contrast, under $H_1$, $\hat{\Psi}(k)$ is anticipated to exceed 0.5 for $k$ near the true change-point $t_0$, allowing for differentiation between the two distributions. Therefore, the test statistic is formulated by choosing the maximal AUC, $$\hat{Q}_T\defeq\max_{k\in \calI_{cp}} \hat{\Psi}(k).$$ The resulting maximizer, $\hat{R}_T\defeq\argmax_{k\in \calI_{cp}}\hat{\Psi}(k)$, is then used for change-point estimation. We expect to find more evidence against $H_0$ in \eqref{eq:hyp_single_cp} for a large value of $\hat{Q}_T$. Section \ref{sec:theory} rigorously investigates the asymptotic behavior of $\hat{Q}_T$ under the null hypothesis $H_0$ and establishes that, after appropriate scaling, it converges weakly to a pivotal process:  \begin{equation} \label{eq:pivotal_trailer}
    \sup_{k \in \calI_{cp}} \secondbigg{\sqrt{T}(\hat{\Psi}(k) - 1/2)} \cd \sup_{r \in [\gamma, 1-\gamma]} G_0(r) \quad \text{as} \quad T \to \infty,
\end{equation} where $\gamma \defeq \epsilon + \eta$, and $G_0(\cdot)$ is a pivotal functional of standard Brownian motions. Therefore, we can simulate the theoretical quantiles and devise a test for $H_0$. Specifically, we generate  $10^5$  i.i.d. standard normal random variables  to approximate one realization
of the standard Brownian motion path. Let $Q(1-\alpha)$ be the $100(1-\alpha)\%$ quantile of $\sup_{r \in [\gamma, 1-\gamma]} G_0(r)$.   Based on $10^5$ Monte Carlo simulations, Table \ref{tab:theo_quantiles} outline the value of $Q(1-\alpha)$ for $\alpha=20\%, 10\%, 5\%, 1\%$  and 0.5\% respectively. Consequently, a level-$\alpha$ test for detecting a single change-point rejects $H_0$ if $ \sqrt{T} \firstbigg{\hat{Q}_T - 1/2} \geq Q(1-\alpha)$. A comprehensive view of steps involved in \texttt{changeAUC} is presented in Algorithm \ref{alg:detect_single_cp}. 

\begin{table}[ht]
\centering
\begin{tabular}{rrrrrr}
  \hline
$\alpha$ & 20\% & 10\% & 5\% & 1\% & 0.5\% \\ 
  \hline
$Q(1-\alpha)$ & 2.231 & 2.664 & 3.040 & 3.784 & 4.051 \\ 
   \hline
\end{tabular}
\caption{Simulated quantiles of $\sup_{r\in [\gamma, 1-\gamma]} G_0(r)$ for $\gamma=\epsilon+\eta, \epsilon=0.15, \eta=0.05$.}
\label{tab:theo_quantiles}
\end{table}

\begin{algorithm}
\caption{\texttt{changeAUC}: single change-point detection.}\label{alg:detect_single_cp}
    \begin{algorithmic}[1]
        \Require Data $\{\Z_t\}_{t=1}^{T}$; choice of classifier; choices of tuning parameters, $\epsilon \in (0, 1/2)$ and $\eta \in (0, 1/2-\epsilon)$; significance level $\alpha$.
        \State Split the data into $D_0, D^v, D_1$ using $m=\floor{T\epsilon}$ and obtain $\calI_{cp}$ from $D^v$ using $\eta$.
        \State Train the classifier based on $D_0$, $D_1$ to obtain $\{\hat{\theta}(\Z_t): \Z_t \in D^v\}$.
        \For{$k \in \calI_{cp}$} 
            \State Split $D^v$ into $D_0^v(k)$, $D_1^v(k)$.
            \State Calculate $\hat{\Psi}(k)$ in \eqref{Psik}, based on $\{\hat{\theta}(\Z_t): \Z_t \in D_0^v(k)\}$ and $\{\hat{\theta}(\Z_t): \Z_t \in D_1^v(k)\}$. 
        \EndFor
        \State Compute $\hat{Q}_T \defeq \max_{k \in \calI_{cp}} \hat{\Psi}(k)$ and $\hat{R}_T \defeq \argmax_{k \in \calI_{cp}} \hat{\Psi}(k)$.
        \If{$\sqrt{T}(\hat{Q}_T-1/2) > Q(1-\alpha)$}
        \State{\Return $\hat{R}_{T}$ as a significant change-point}
        \EndIf
    \end{algorithmic}
\end{algorithm}

\subsection{Classification Accuracy for Detecting a Single Change-point} \label{subsec:clf_accuracy}
In practice, both $\cP_X, \cP_Y$ and $t_0$ are unknown. To test \eqref{eq:hyp_single_cp}, it is important to distinguish $\cP_Y$ from $\cP_X$. The fundamental feature of \texttt{changeAUC} lies in training an appropriate classifier to discern the heterogeneity between the two unknown distributions. Let $\theta(\cdot)$ be the true conditional probability corresponding to $\hat{\theta}(\cdot)$. 
When infinite samples are drawn from two distributions, $\mathcal{P}_X$ and $\mathcal{P}_Y$, and each sample is associated with a label $v=0$ or $v=1$ depending on its origin, we obtain the true conditional probability, $\theta(\z) \defeq \P \thirdbigg{v=1|\Z=\z}$ where $\Z=\z \in E$ is a new observation. Moreover, $\theta(\cdot)$ is related to the true likelihood ratio, \begin{equation} \label{eq:true_lik_ratio_candidate_cp}
    L(\z) \defeq \frac{\theta(\z)}{1-\theta(\z)} = \frac{d\cP_Y}{d\cP_X}(\z),
\end{equation} and $L(\z) = 1$ almost surely if $\cP_X = \cP_Y$. Otherwise, $L(\z)$ is expected to be larger (resp. smaller) than one if $\Z \sim \cP_Y$ (resp. $\Z \sim \cP_X$). Therefore, $L(\z)$, or equivalently $\theta(\z)$, serves as a one-dimensional projection of the original observation $\z$. Proposition \ref{prop:unif_oracle_auc_bound} below states the close relationship between such projection and the total variation between two distributions. It provides an initial motivation for employing a rank-sum comparison of $L(\cdot)$.

\begin{proposition} \label{prop:unif_oracle_auc_bound}
    For two probability measures $\cP_X$ and $\cP_Y$ on a common probability space $(\Omega, \mc{F}, \P)$ such that $\cP_Y \ll \cP_X$, with the Radon-Nikodyn derivative $d\cP_Y/d\cP_X(\cdot)$ having a continuous distribution under $\cP_X$, let $\Z_X \sim \cP_X$ and $\Z_Y \sim \cP_Y$ be independent. Then, \begin{equation} \label{eq:bound_pop_auc_by_tv}
        \frac{1}{2} d_{tv}(\cP_X, \cP_Y) \leq \P \secondbigg{ \frac{d\cP_Y}{d\cP_X}(\Z_X) < \frac{d\cP_Y}{d\cP_X}(\Z_Y) } - \frac{1}{2} \leq d_{tv}(\cP_X, \cP_Y) ,
    \end{equation} where, $d_{tv}(\cP_X, \cP_Y)$ is 
    the total variation distance between $\cP_X$ and $\cP_Y$.
\end{proposition} 

Given a large total variation distance between the two probability distributions, we expect that an empirical estimate of the probability in \eqref{eq:bound_pop_auc_by_tv} will capture the heterogeneity between $\cP_X$ and $\cP_Y$. Note that the probability in \eqref{eq:bound_pop_auc_by_tv} involves a rank comparison of the likelihood ratio, $L(\cdot) \defeq (d\cP_Y/d\cP_X)(\cdot)$.
For low-dimensional Euclidean data, it is common to estimate  $\cP_X$ and $\cP_Y$ directly, see e.g. \cite{padilla_et_al_2021_ieee}. However, for high-dimensional Euclidean data, direct estimation of the individual likelihood functions is challenging and usually inconsistent due to the curse of dimensionality. Instead, it is possible to approximate the likelihood ratio $L(\cdot)$ by plugging a surrogate estimator $\hat{\theta}(\cdot)$ in \eqref{eq:true_lik_ratio_candidate_cp}: \begin{equation} \label{eq:estimate_lik_ratio_classifier}
    \hat{L}(\z) \defeq \hat{\frac{d\cP_Y}{d\cP_X} }(\z) = \frac{\hat{\theta}(\z)}{1-\hat{\theta}(\z)} .
\end{equation}
The two-sample rank-sum comparison of $\hat{L}(\mbi{z})$ is equivalent to the AUC \citep{fawcett_2006_pattrec_letter, chakravarti_et_al_2022}. Specifically, for each change-point candidate $k$, based on the validation data  $D_0^v(k)$ and $D_1^v(k)$, provided $\hat{\theta}(\Z_t) \in (0, 1) \ \forall t=m+1, \ldots, T-m$, we have \begin{align} \label{eq:prelim_auc_rank_sum}
    \begin{split}
        \hat{\Psi}(k) &\defeq \frac{1}{(k-m)(T-m-k)} \sum_{i=m+1}^{k} \sum_{j=k+1}^{T-m} \indic\secondbigg{\hat{\theta}(\Z_i) < \hat{\theta}(\Z_j)} \\
        &= \frac{1}{(k-m)(T-m-k)} \sum_{i=m+1}^{k} \sum_{j=k+1}^{T-m} \indic\secondbigg{\frac{\hat{\theta}(\Z_i)}{1-\hat{\theta}(\Z_i)} < \frac{\hat{\theta}(\Z_j)}{1-\hat{\theta}(\Z_j)}} \\
        &= \frac{1}{(k-m)(T-m-k)} \sum_{i=m+1}^{k} \sum_{j=k+1}^{T-m} \indic\secondbigg{\hat{L}(\Z_i) < \hat{L}(\Z_j)},
    \end{split}
\end{align} where, the second equality holds since, $x \to x / (1-x)$ is a monotone transformation for $x \in (0, 1)$. 
Intuitively, if $k$ is indeed a change-point, $\Psi(k)$ can be viewed as an empirical estimator of $\P \secondbigg{ {d\cP_Y}/{d\cP_X}(\Z_X) < {d\cP_Y}/{d\cP_X}(\Z_Y) }$ based on a second-order U-statistic. Therefore, the AUC-based test can be regarded as an alternative approximation to the likelihood-ratio-based test. Indeed, there is another line of research that directly targets the likelihood ratio, such as in \cite{kawahara_sugiyama_2012_sadm}, \cite{liu_et_al_2013_neural_networks}, and \cite{wang_et_al_2023_ieee}. However, these methods still rely on specifying a model structure for the likelihood ratios, which may be susceptible to systematic bias. On the contrary, our methods rely on ranks instead of explicit likelihood values or ratios, offering greater robustness to distributional assumptions. For instance, our approach remains effective even when $\widehat{L}(\cdot)$ is estimated with bias or is proportional to some unknown constant. This makes it more advantageous when the true data distribution is complex or when specifying the likelihood  (ratio) is challenging. Furthermore, the rank-sum comparison also allows the incorporation of weakly trained classifiers. We include more discussion on this direction in Remark \ref{rmk:auc_premature_training}. 

\begin{remark} \label{rmk:auc_premature_training}
  Compared with methods using exact values of estimated (conditional) probabilities  $\widehat{\theta}(\Z_t)$'s, the AUC-based test evaluates the (conditional) rank order of the data rather than relying on these exact values, making it less sensitive to estimation errors due to sample splitting. This is corroborated in Section \ref{subsec:sim_non_Euclidean} where directly using outputs of the classifier within the CUSUM statistic can lead to distorted type-I error control. In the case of a single change-point at $t_0$,  estimation is achieved when the sets of conditional probabilities, $\secondbigg{\hat{\theta}(\Z_t): \Z_t \in D_0^v}$ and $\secondbigg{\hat{\theta}(\Z_t): \Z_t \in D_1^v}$ are well separated. Unlike approaches requiring these sets to be close to 0 and 1, respectively, as typically obtained through training a consistent classifier, our framework exhibits robust performance even when such conditions are not strictly met. In particular, Section \ref{sec:theory} below highlights that the use of rank-sum comparison alleviates the necessity for a consistently estimated $\hat{\theta}(\cdot)$.
\end{remark}

\subsection{Extension to Multiple Change-points} \label{subsec:extn_multiple_cp}
It is feasible to integrate \texttt{changeAUC} into existing algorithms designed for estimating multiple change-points, such as wild binary segmentation \citep{fryzlewicz2014wild}, seeded binary segmentation (SBS) \citep{kovacs_et_al_2023_biometrika}, isolating single change-points \citep{anastasiou2022detecting}, among others. In this paper, we employ the SBS to estimate multiple change-points in two real data sets in Section \ref{sec:real_data}. The SBS procedure works as follows. For a sequence of length $T$, the SBS recursively applies a single change-point test statistic such as ours to a set of predetermined intervals $\calI_{\lambda}$ defined as: \begin{equation}
    \calI_{\lambda} = \bigcup_{k=1}^{\ceil{\log_{1/\gamma}(T)}} \calI_{k},\quad \calI_{k} = \cup_{i=1}^{T_k} \{ (\floor{(i-1)s_k}, \floor{(i-1)s_k + l_k}) \},
\end{equation} where $\lambda \in [1/2, 1)$ is known as a decay parameter, and each $\calI_{k}$ is a collection of sub-intervals of length $l_k=T \gamma^{k-1}$ evenly shifted by $s_k=(T-l_k)/(T_k-1)$.  The SBS finds the largest value of test statistics evaluated at all sub-intervals and compares it with a predetermined threshold level $\Delta$. Once the maximum test statistic exceeds $\Delta$, a change-point is claimed.  The data are further divided into two sub-samples accordingly, and the same procedure is applied to each sub-samples. \cite{kovacs_et_al_2023_biometrika} specifically addresses multiple change-point detection in the mean only. Their threshold $\Delta$ may not be directly applicable when integrating \texttt{changeAUC} into their algorithm, necessitating new theoretical developments for overall control of type-I error in the presence of multiple change-points beyond the mean of distributions. We defer this aspect to future work. Instead, we follow the approach proposed by \cite{dubey2023change} and perform a permutation test to obtain the threshold $\Delta$.  Algorithm \ref{alg:detect_multiple_cp} outlines the procedure for this approach. In practice,  $\Delta$ is set to be the $90\%$ quantile of the permutation null distribution of the maximal test statistics, and   $\gamma=1/\sqrt{2}$. 

\begin{algorithm}[!htp]
\caption{\texttt{changeAUC-SBS}: multiple change-points detection.}\label{alg:detect_multiple_cp}
    \begin{algorithmic}[1]
        \Require $l$ and $u$: lower and upper boundaries of time index; all parameters of \texttt{changeAUC} in Algorithm \ref{alg:detect_single_cp}, minLen: minimum length of intervals, $\gamma$: decay parameter, $\hat{\tau}$: initial set of change-points, $B$: number of permutations.
        \Procedure{\texttt{changeAUC-SBS}}{l, u, $\gamma$, $\hat{\tau}$, minLen}
        \If{$u-l < \text{minLen}$} 
        \State STOP
        \Else
        \State calculate seeded intervals $\calI_{\lambda} = \{[l_i, u_i]: i \in |\calI_{\lambda}|\}$.
        \For{$i = 1, \ldots, |\calI_{\lambda}|$}
            \State Apply \texttt{changeAUC} on $\{\z_t: t \in [l_i, u_i] \cap \naturals\}$.
            \State Denote the maximum AUC as $\hat{Q}_i$ and maximizer as $\hat{R}_i$.
        \EndFor
        \For{$b=1, \ldots, B$}
            \State Permute $\{\z_t: t \in [l, u] \cap \naturals\}$. Apply \texttt{changeAUC} on the permuted sample.
            \State Denote $\hat{Q}^{(b)}$ as the maximum AUC. 
        \EndFor
        \State Calculate $\Delta$: $90\%$ quantile of $\{\hat{Q}^{(b)}\}_{b=1}^{B}$.
        \If{$\max_{i \in \{1, \ldots, |\calI_{\lambda}|\}} \hat{Q}_i \geq \Delta$}
        \State{Fix $m_Q = \argmax_{i \in \{1, \ldots, |\calI_{\lambda}|\}} \hat{Q}_i$ and $\hat{\tau} = \hat{\tau} \cup \hat{R}_{m_Q}$}
        \State{Implement \texttt{changeAUC-SBS}(l, $\hat{R}_{m_Q}$, $\gamma$, $\hat{\tau}$, minLen).}
        \State{Implement \texttt{changeAUC-SBS}($\hat{R}_{m_Q}+1$, u, $\gamma$, $\hat{\tau}$, minLen).}
        \Else
        \State break
        \EndIf
        \EndIf
        \Return $\hat{\tau}$: final set of change-points.
        \EndProcedure
    \end{algorithmic}
\end{algorithm}

\section{Theoretical Justification} \label{sec:theory} 
In this section, we investigate the theoretical properties of \texttt{changeAUC}. Section \ref{subsec:asymp_dist} derives the limiting null distribution of the test statistic, which justifies our claims in \eqref{eq:pivotal_trailer}.  Section \ref{subsec:uniform_weak_conv_alternative} investigates the power behavior under fixed and local alternatives. Section \ref{subsec:weak_consistency} presents the consistency property of the change-point estimator $\hat{R}_T$. In what follows, $\P_*, \ \E_*, \ \var_{*} \ \text{and}, \ \cov_{*}$ respectively denote the conditional probability, expectation, variance, and covariance given the trained classifier on the training data $D_0\cup D_1$. 

\subsection{Limiting Null Distribution} \label{subsec:asymp_dist}
Recall that $\hat{Q}_T \defeq \max_{k \in \calI_{cp}} \hat{\Psi}(k)$ is the proposed statistic for testing $H_0$ vs $H_1$. With suitable scaling, it is customary to study the limiting behavior of the AUC process: \\$\{\hat{\Psi}(\floor{Tr})\}_{r \in [\gamma, 1-\gamma]}$, where $\gamma \defeq \epsilon + \eta$. In Theorem \ref{thm:unif_weak_conv}, we show that $\{\sqrt{T}(\hat{\Psi}(\floor{Tr})-1/2)\}_{r \in [\gamma, 1-\gamma]}$ converges uniformly to a pivotal process with respect to $\P_*$.

\begin{theorem} \label{thm:unif_weak_conv}
 Suppose $\widehat{\theta}(z)$ is continuous with respect to  $\mathcal{P}_X$. Then, under $H_0$,
    \begin{equation*}
        \left\{ \sqrt{T}\left(\hat{\Psi}(\floor{Tr}) - 1/2\right)\right\}_{r\in [\gamma,1-\gamma]} \ucd \{G_0(r)\}_{r\in [\gamma,1-\gamma]},\quad \text{in  } L^{\infty}([0,1]), 
    \end{equation*} under $\mathbb{P}_*$, where for a  standard Brownian Motion (BM) $B(\cdot)$, $G_0(\cdot)$ is defined as \begin{equation} \label{eq:defn_g_0_null}
    G_0(r) \defeq \frac{1}{\sqrt{12}} \thirdbigg{ \frac{B(1-\epsilon) - B(r)}{1-\epsilon-r} - \frac{B(r)-B(\epsilon)}{r-\epsilon} }, \quad \ \text{for} \ r \in [\gamma, 1-\gamma].
    \end{equation}
\end{theorem} Here, $G_0(r)$ is a functional of standard BM that does not depend on the classifier, data dimension, or data distribution. Theorem \ref{thm:unif_weak_conv} leads us to propose a computationally feasible yet statistically sound testing framework using simulated theoretical quantiles in Table \ref{tab:theo_quantiles}.

In Theorem \ref{thm:unif_weak_conv}, we do not require that $\widehat{\theta}(\cdot)$ is consistently estimated. In fact, under $H_0$, a random guess is also applicable. The key assumption lies in the continuity of $\widehat{\theta}(\cdot)$ with respect to  $\mathcal{P}_X$, which is typically satisfied for commonly used classifiers such as logistic regression, random forest, and deep neural networks.  When there exists a point mass, we can inject a random tie-breaking ranking scheme, and all the theory still goes through. 
\begin{remark}
    Throughout the paper, we assume temporal independence among data. While this assumption simplifies the analysis, we briefly discuss how our methodology and theory could be extended to accommodate the time series setting. First, the independence assumption ensures a unified constant variance (1/12 appeared in $G_0(r)$) for the rank-sum statistics under $H_0$. For time series data, however, a (long-run) variance estimator is required and typically involves additional tuning parameters, such as the bandwidth of a kernel function. In such cases, a self-normalization \citep{zhao_jiang_shao_2022_jrssb} technique can be employed to mitigate these challenges. Second,  we may require additional trimming parameters to control for the serial dependence of $D^v$ on $D_0\cup D_1$; see similar treatments in \cite{gao_wang_shao_2023_arxiv}. Otherwise, the estimated classifier $\widehat{\theta}(\cdot)$ is correlated with  $\Z_t\in D^v$, which greatly complicates the theoretical analysis. Note in the independent setting, this issue is automatically resolved. Third, the concentration inequalities and empirical process theory developed in this paper must be modified accordingly to account for temporal dependence. This is left for future research.
\end{remark}

\subsection{Uniform Weak Convergence Under Local and Fixed Alternative} \label{subsec:uniform_weak_conv_alternative}
Next, we investigate the limiting behaviors of the test under the alternative hypothesis of at most one change-point. Let the true change-point be $t_0 = \floor{\tau T}$ for some $\tau \in [\gamma, 1-\gamma]$. The observations are generated from the following setup: \begin{equation*}
    \Z_t \simiid \begin{cases}
        \cP_X \ \text{for} \ 1 \leq t \leq t_0 \\
        \cP_Y \ \text{for} \ t_0+1 \leq t \leq T.
    \end{cases}
\end{equation*} Let us recall the likelihood ratio, $L(\cdot) = ({d\cP_Y}/{d\cP_X})(\cdot)$ from \eqref{eq:true_lik_ratio_candidate_cp}. The following proposition quantifies  how it is connected to the total variation distance between $\cP_X$ and $\cP_Y$ and thus  the departure of $H_1$ from $H_0$. 

\begin{proposition} \label{prop:likelihood_bound}
    Consider two probability measures $\cP_X$ and $\cP_Y$ that satisfy the conditions in Proposition \ref{prop:unif_oracle_auc_bound}.
    Let $\Z_X, \Z_X' \simiid \cP_X$ be independent random elements, and define \begin{equation} \label{eq:defn_delta}
    \delta \defeq \E \absbigg{ L(\Z_X) - L(\Z'_X) }, \quad \text{for} \ \Z_X, \Z'_X \simiid \cP_X. 
\end{equation} Then, for the  total variation distance $d_{tv}(\cP_X, \cP_Y)$ between $\cP_X$ and $\cP_Y$, we have  $$
 2 d_{tv}(\cP_X, \cP_Y) \leq  \delta \leq 4 d_{tv}(\cP_X, \cP_Y).
$$
\end{proposition} 
Note that $\delta = 0$ if and only if $L(\cdot)$ is a constant measurable function, which is equivalent to $\cP_X \equiv \cP_Y$ almost surely. 
 Therefore, $\delta$ can measure how much the alternative hypothesis $H_1$ deviates from the null hypothesis $H_0$. 
 
 To facilitate the analysis of power behavior, we let  $\delta=\delta_T\geq 0$ that allows the signal to be changing with the sample size.  In Section \ref{subsec:fix_alt}, we present the limiting behavior of $\hat{Q}_T$ under the fixed alternative framework $\sqrt{T}\delta_T\to\infty$.  Section \ref{subsec:local_alt} investigates the limiting behavior of $\hat{Q}_T$, under a local alternative framework $\sqrt{T}\delta_T\to C$ for a fixed constant $C>0$.

\subsubsection{Fixed Alternative} 
\label{subsec:fix_alt}
\begin{assumption}\label{assump:fixed_alt}
    For independent random elements $\Z_X \sim \cP_X$, $\Z_Y \sim \cP_Y$, let us define $\mu_{*} \defeq \P_{*} \firstbigg{ \hat{\theta}(\Z_X) < \hat{\theta}(\Z_Y) }$ and $\mu \defeq \P\firstbigg{ \theta(\Z_X) < \theta(\Z_Y) }$. Let us recall $\delta=\delta_T$ from \eqref{eq:defn_delta}. As $T \to \infty$, there exists some $\sqrt{1/T}\ll c_T<\delta_T/4$, such that  $\P \thirdbigg{\absbigg{ \mu_{*} - \mu } \leq \frac{\delta_T}{4} - c_T } \to 1$.
\end{assumption}
Assumption \ref{assump:fixed_alt} requires that, with high probability, the true signal order $\delta_T$ (hence  $d_{tv}(\mathcal{P}_X,\mathcal{P}_Y)$ due to Proposition \ref{prop:likelihood_bound}) should dominate the average approximation error incurred by estimating conditional probability using a classifier, which is denoted as $\mu_{*} - \mu$. The condition $\sqrt{1/T} \ll c_T<\delta_T/4$ is mild in that $\delta_T\gg \sqrt{1/T}$ under the fixed alternative. It requires that the estimated classifier can capture some signal from data other than purely statistical estimation error of order $\sqrt{1/T}$. Note that when $\delta_T$ is a fixed constant, a constant bound for the  approximation error (of the trained classifier) is sufficient to ensure both high power in the testing and the localization accuracy in the estimation (see Theorem \ref{thm:weak_consistency+} below). This highlights the robustness of our framework to approximation errors in the classifier, as  the classifier is not required to be consistently trained. Instead, we only require the 
estimated $\hat{\theta}(\cdot)$ to preserve the monotonicity of $\theta(\cdot)$ up to some tolerable error.  This observation resonates with the assertions in Section \ref{subsec:clf_accuracy} and is further underscored in Remark \ref{rmk:auc_premature_training}. In practice, the choice of the classifier could benefit from prior knowledge of the data.

The following theorem ensures the consistency of our test under the fixed alternative.
\begin{theorem} \label{thm:fixed_alt}
    Under $H_1$, suppose Assumption \ref{assump:fixed_alt} holds and $\sqrt{T}\delta_T \to \infty$. Then, \begin{equation*}
        \sup_{r \in [\gamma, 1-\gamma]} \secondbigg{ \sqrt{T} \firstbigg{ \hat{\Psi}(\floor{Tr}) - \frac{1}{2} } } \to \infty, \quad \text{in probability} \ \P_*.
    \end{equation*}
\end{theorem}

\subsubsection{Local Alternative}  \label{subsec:local_alt}
\begin{assumption} \label{assump:local_alt}
    For random elements $\Z_1, \Z_2 \in E$ there exists $\zeta > 0$ such that the following holds true: (1) $\E_* \absbigg{ \indic\secondbigg{\hat{\theta}(\Z_1) < \hat{\theta}(\Z_2) } - \indic\secondbigg{\theta(\Z_1) < \theta(\Z_2) } } = O_{\P_*}(T^{-\zeta})$. Moreover, for $\Z_1 \sim \cP_X, \ \Z_2 \sim \cP_Y$ and  $\mu_{*} = \P_{*} \firstbigg{ \hat{\theta}(\Z_1) < \hat{\theta}(\Z_2)}$, $\mu = \P \firstbigg{ \theta(\Z_1) < \theta(\Z_2)}$, the following holds true: (2) $\mu_{*} - \mu = o_{\P_*}(T^{-1/2}).$
\end{assumption}
    Assumption \ref{assump:local_alt}(1) is slightly stronger than  Assumption \ref{assump:fixed_alt}, and requires that $\hat{\theta}(\cdot)$ can consistently preserve the monotonicity of $\theta(\cdot)$.  In the context of the local alternative hypothesis, where $\delta_T$ approaches zero as rapidly as $1/\sqrt{T}$,  a more stringent condition in   Assumption \ref{assump:local_alt}(2) is imposed on the average approximation error $(\mu_* - \mu)$. Notably, the super consistency in Assumption \ref{assump:local_alt}(2) is applied to the expected rank monotonicity of conditional probabilities rather than directly to the conditional probabilities. 
Similar assumptions have been recently used and verified in tests of independence (Theorem 7 in \cite{cai_et_al_2022}) and two-sample conditional distribution testing (Proposition 1 in \cite{hu_lei_2023_jasa}).

\begin{theorem} \label{thm:unif_weak_conv_alt}
    Suppose Assumption \ref{assump:local_alt} is true.  Under the local alternative such that $\sqrt{T}\delta_T \to C$, for some constant $C > 0$, then the following holds true: \begin{equation*}
     \secondbigg{\sqrt{T}\firstbigg{\hat{\Psi}(\floor{Tr}) - \frac{1}{2}}}_{r \in [\gamma, 1-\gamma]} \ucd \secondbigg{\Delta_G(r;\tau, \epsilon) + G_0(r)}_{r \in [\gamma, 1-\gamma]}, \quad \text{in probability} \ \P_*,
\end{equation*} where $G_0(\cdot)$ is defined in Theorem \ref{thm:unif_weak_conv}, and \begin{equation*}
    \Delta_G(r; {\tau}, \epsilon) \defeq \frac{C}{4}\begin{cases}
        (1-\epsilon-\tau)/(1-\epsilon-r) & \quad \text{for} \ \gamma < r < \tau < 1 - \gamma, \\
       (\tau-\epsilon)/(r-\epsilon) & \quad \text{for} \ \gamma < \tau< r < 1-\gamma.
    \end{cases}
\end{equation*} 
\end{theorem}

\subsection{Consistency of Change-point Estimator} \label{subsec:weak_consistency}
The primary objective of this paper is to introduce a testing framework for a single change-point. Motivated by favorable localization performance observed in simulation studies, we establish the consistency result for estimating the change-point using $\hat{R}_T = \argmax_{k \in \calI_{cp}} \hat{\Psi}(k)$. 
\begin{theorem} \label{thm:weak_consistency+}
 Let $t_0=\floor{\tau T}$ be the true change-point. Under the fixed alternative  $\delta_T\gg\sqrt{1/T}$, if Assumption \ref{assump:fixed_alt} holds with $ \sqrt{\log T/T}\ll c_T<\delta_T/4$, then \begin{equation*}
    \P_* \thirdbigg{ \absbigg{ \hat{R}_T - t_0 } < C_Lc_T^{-2}\log T } \geq 1- o(1),
\end{equation*} where  $C_{L}>0$ is a constant that only depends  on the trimming parameters $\epsilon$, $\eta$, and relative location of the true change-point $\tau.$
\end{theorem}

Theorem \ref{thm:weak_consistency+} establishes the localization rate of the change-point estimator $\hat{R}_T$. As an immediate consequence,  we obtain its weak consistency, i.e., $\hat{R}_T/ T \to_p \tau$.  In Theorem \ref{thm:weak_consistency+}, the multiplicative order $c_T^{-2}$ in the localization rate reflects the influence of classifier accuracy and the signal strength, where $c_T$ quantifies the gap between the approximation error of the trained classifier $|\mu^*-\mu|$, and the signal $\delta_T$  in Assumption \ref{assump:fixed_alt}. The additional condition on $c_T\gg \sqrt{\log T/T}$ is mild because if $c_T=O(\sqrt{\log T/T})$, the localization rate becomes trivial. To gain more insights, if $\delta_T \asymp c_T$, i.e. the approximation error of the classifier is negligible compared with the signal, the localization rate simplifies to $O_p(\delta_T^{-2}\log T)=O_p(d_{tv}^{-2}\log T)$, which according to \cite{padilla_et_al_2021_ieee} and \cite{yu2020review}, is minimax optimal up to a logarithmic factor in the univariate setting. Otherwise, if $c_T=o(\delta_T),$  the estimation effect of the classifier dominates the signal in the localization rate, which results in suboptimal performance of change-point estimation. In addition, Theorem \eqref{thm:weak_consistency+} does not necessarily contradict the localization rate in \cite{padilla_et_al_2021_ieee} for multivariate data, where the signal is characterized by the ($L_{\infty}$-type) supremum distance between densities, whereas our approach uses the ($L_{1}$-type) total variation distance.  While it would be interesting to provide  alternative lower bounds in this context, we defer this to future work.

\section{Numerical Simulations} \label{sec:simulation}
In this section, we examine the finite sample performance of our proposed method. The performance on high-dimensional Euclidean data is investigated in Section \ref{subsec:sim_Euclidean} whereas that on the non-Euclidean data using artificially constructed images from the CIFAR10 database \citep{cifar10} is presented in Section \ref{subsec:sim_non_Euclidean}. Both \texttt{R} and \texttt{Python} code developed for \texttt{changeAUC} is available at \url{https://github.com/rohitkanrar/changeAUC}.

\subsection{High Dimensional Euclidean Data} \label{subsec:sim_Euclidean}
First, we examine the size and power performance of the proposed test in high-dimensional Euclidean data. We consider three classifiers, briefly summarized below:  \begin{itemize}
    \item \textbf{Regularized Logistic:} (denoted by \texttt{Logis}) We apply the regularized logistic classifier with the LASSO penalty. \texttt{R} package \texttt{glmnet} is used \citep{glmnet}.
    \item \textbf{Fully-connected Neural Network:} (denoted by \texttt{Fnn}) We use a fully connected neural network with three hidden layers, ReLU activation and the binary cross-entropy loss. Additionally, a LASSO penalty is enforced in each layer. Python framework \texttt{Tensorflow} is used.
    \item \textbf{Random Forest:} (denoted by \texttt{Rf}) We employ a random forest classifier. \texttt{R} package \texttt{randomForest} \citep{randomForest} is used. Tuning parameters are set to default values as pre-fixed in the package. 
\end{itemize}

For size performance, we consider the following data generating processes (DGP) with  $T\in \{1000, 2000\}$ and $p\in \{10, 50, 200, 500, 1000\}$,
\begin{enumerate}[]
\item \textbf{Standard Null:} $\{\Z_t\}_{t=1}^{T} \simiid \calN_{p}(\bm{0}_p, I_p)$;
\item \textbf{Banded Null:} $\{\Z_t\}_{t=1}^{T} \simiid \calN_{p}(\bm{0}_p, \Sigma)$  with $\Sigma=(\sigma_{ij})_{i,j=1}^p$,  $\sigma_{ij}=0.8^{|i-j|}$;
\item \textbf{Exponential Null:} $\{\Z_{tj}\}_{t=1}^{T} \simiid \mathrm{Exp}(1)$ for $j=1,\cdots,p.$
\end{enumerate}

Table \ref{tab:size_Euclidean} presents the empirical rejection rates across 1000 replications at $5\%$ significance level.  The results demonstrate that all the considered classifiers successfully control the type-I error in all settings. 

\begin{table}[ht]
\centering
\scalebox{0.77}{\begin{tabular}{|c|ccc|ccc|ccc|}
  \hline
 & \multicolumn{3}{c|}{Standard Null} & \multicolumn{3}{c|}{Banded Null} & \multicolumn{3}{c|}{Exponential Null} \\
  \hline
$(T,p)\backslash$ Classifier& \texttt{Logis} & \texttt{Rf} & \texttt{Fnn} & \texttt{Logis} & \texttt{Rf} & \texttt{Fnn} & \texttt{Logis} & \texttt{Rf} & \texttt{Fnn} \\  
\hline
(1000,10) & 3.6 & 2.5 & 4.4 & 3.6 & 2.5 & 3.9 & 3.6 & 4.7 & 4.5 \\ 
  (1000,50) & 4.4 & 3.5 & 3.9 & 4.4 & 3.5 & 3.8 & 5.7 & 4.1 & 5.0 \\ 
  (1000,200) & 4.4 & 4.1 & 5.4 & 4.4 & 4.1 & 4.5 & 5.2 & 5.1 & 5.0 \\ 
  (1000,500) & 5.7 & 4.5 & 3.5 & 5.7 & 4.5 & 3.7 & 4.5 & 4.4 & 4.4 \\ 
  (1000,1000) & 3.8 & 5.1 & 3.6 & 3.8 & 5.1 & 4.4 & 4.1 & 4.7 & 3.7 \\ 
  (2000,10) & 3.8 & 3.1 & 2.8 & 3.8 & 3.1 & 4.8 & 4.4 & 3.7 & 4.6 \\ 
  (2000,50) & 4.4 & 4.6 & 4.0 & 4.4 & 4.6 & 4.5 & 4.6 & 5.1 & 5.0 \\ 
  (2000,200) & 5.2 & 4.7 & 3.5 & 5.2 & 4.7 & 5.0 & 5.2 & 4.7 & 4.5 \\ 
  (2000,500) & 4.2 & 3.9 & 4.3 & 4.2 & 3.9 & 3.3 & 4.0 & 5.6 & 3.6 \\ 
  (2000,1000) & 4.0 & 4.1 & 4.6 & 4.0 & 4.1 & 5.0 & 4.3 & 4.6 & 3.8 \\ 
   \hline
\end{tabular}}
\caption{Size performance of AUC over 1000 Monte Carlo replications, at 5\% level.}
\label{tab:size_Euclidean}
\end{table}

Next, we study the power performance in the high-dimensional setting. We fix the change-point location at $t_0=\floor{T/2}$ with data length $T=1000$, and data dimension $p\in \{500,1000\}$, such that  $\{\Z_t\}_{t=1}^{t_0} \simiid \calN_{p}(\bm{0}_p, I_p)$. For the data after the change-point, the following data-generating processes are considered. 

\begin{enumerate}[]
    \item \textbf{Dense Mean Change:}   $\{\Z_t\}_{t=t_0+1}^{T} \simiid \calN_{p}(\mbi{\mu}, I_p)$ with $\mbi{\mu} = (\mu_i)_{i=1}^p$, $\mu_{i} = 2/\sqrt{\floor{p/5}}$ for $1\leq i\leq \floor{p/5}$, and $\mu_{i} = 0$ for $1+\floor{p/5}\leq i\leq p$;

    \item \textbf{Sparse Mean Change:}   $\{\Z_t\}_{t=t_0+1}^{T} \simiid \calN_{p}(\mbi{\mu}, I_p)$ with $\mbi{\mu} = (\mu_i)_{i=1}^p$, $\mu_{i} = 2/\sqrt{\floor{p/100}}$ for $1\leq i\leq \floor{p/100}$, and $\mu_{i} = 0$ for $1+\floor{p/100}\leq i\leq p$;
    
    \item \textbf{Dense Cov Change:}  $\{\Z_t\}_{t=t_0+1}^{T} \simiid \calN_{p}(\bm{0}_p, \Sigma)$  with $\Sigma=(\sigma_{ij})_{i,j=1}^p$,  $\sigma_{ij}=0.1^{\indic(i\neq j)}$;
    \item \textbf{Banded Cov Change:}  $\{\Z_t\}_{t=t_0+1}^{T} \simiid \calN_{p}(\bm{0}_p, \Sigma)$  with $\Sigma=(\sigma_{ij})_{i,j=1}^p$,  $\sigma_{ij}=0.8^{|i-j|}$; 
    \item \textbf{Dense Diag Cov Change:}   $\{\Z_t\}_{t=t_0+1}^{T} \simiid \calN_{p}(\bm{0}_p, \Sigma)$, with $\Sigma=(\sigma_{ij})_{i,j=1}^p$,   $\sigma_{ii} = 1 + 5/\sqrt{\floor{p/5}}$ for $1 \leq i \leq \floor{p/5}$, $\sigma_{ii} = 1$ for $\floor{p/5} + 1 \leq i \leq p$, and $\sigma_{ij} = 0$ for $i \neq j$;
    \item\textbf{Sparse Diag Cov Change:} $\{\Z_t\}_{t=t_0+1}^{T} \simiid \calN_{p}(\bm{0}_p, \Sigma)$, with $\Sigma = (\sigma_{ij})$, where $\sigma_{ii} = 1 + 5/\sqrt{\floor{p/100}}$ for $1 \leq i \leq \floor{p/100}$, $\sigma_{ii} = 1$ for $\floor{p/100} +1 \leq i \leq p$, and $\sigma_{ij} = 0$ for $i \neq j$;
    \item \textbf{Dense Distribution Change:}   $\{\Z_{tj}\}_{t=t_0+1}^{T} \simiid  \text{Exp(1)}-1$ for $j=1, \ldots, \floor{p/5}$ and $\{\Z_{tj}\}_{t=t_0+1}^{T} \simiid \calN_{1}(0, 1)$ for $j=\floor{p/5}+1, \ldots, p$;
    \item \textbf{Sparse Distribution Change:}  $\{\Z_{tj}\}_{t=t_0+1}^{T} \simiid   \text{Exp(1)}-1$ for $j=1, \ldots, \floor{p/100}$ and $\{\Z_{tj}\}_{t=t_0+1}^{T} \simiid \calN_{1}(0, 1)$ for $j=\floor{p/100}+1, \ldots, p$.
    \end{enumerate}
Note that for the distributional changes, the first- and second-order moments of the underlying distributions remain the same before and after the change-point.

In addition to the proposed methods,  we consider the following four competitors among many existing tests for detecting change-points in high-dimensional data.
\begin{itemize}
    \item A graph-based method \texttt{gseg} proposed by \cite{chen_zhang_2015_annals} and \cite{chu_chen_2019_annals}, and is implemented by the \texttt{gSeg} package in \texttt{R}. Here, the graph is constructed using the minimal spanning tree, and two edge-count scan statistics are considered: weighted (\texttt{gseg\_wei}) and max-type (\texttt{gseg\_max}). 
    \item A generalized energy distance-based method proposed by \cite{chakraborty_zhang_2021} (denoted as \texttt{Hddc}). The default distance metric $\gamma(z, z') = \lone{z-z'}^{1/2}$ is used to calculate the interpoint distance. 
    \item A density-ratio based method applied on sliding windows, proposed by \cite{wang_et_al_2023_ieee} (denoted as \texttt{NODE}).  Deep neural network with default configuration is employed to estimate the density ratios. 
    \item A two-step nonparametric likelihood with the random forest-based method proposed by \cite{londschien_buhlman_kovacs_2022} (denoted as \texttt{changeforest}). All tuning parameters are fixed to the default values suggested by the authors. 
\end{itemize} 

We first assess the computational efficiency of the proposed methods in comparison to existing ones. To illustrate the idea, we use a 16-core AMD EPYC 7542 CPU with 32 GB RAM for all methods except for  \texttt{Fnn}, which is performed using a single-core Intel Xeon 6226R CPU with a single NVIDIA Tesla V100 GPU. Data generation follows the ``Dense Mean" setup, with performance under other settings being similar and thus omitted. 

Table \ref{tab:time} presents the average computation time for all five methods over 1000 replications, with standard deviations shown in parentheses. It is evident that our proposed methods \texttt{Logis}, \texttt{Fnn} and \texttt{Rf}, along with \texttt{gseg}, exhibit considerable speed, making them feasible for implementation within a reasonable time. However, owing to the recursive calculation of $U$-statistics, \texttt{Hddc} is notably sluggish, making it nearly impractical for implementation on personal laptops. Although both \texttt{changeforest}  and \texttt{Rf} apply the random forest, the former takes more time due to permutation-based testing procedure and two-step search. Note that  \texttt{NODE} also trains a deeper neural network like our method \texttt{Fnn}. The former one is a sliding window-based search procedure, which explains its high computational cost. 
\begin{table}[!htp]
\centering
\resizebox{1\textwidth}{!}{\begin{tabular}{cccccccc}
\hline
$p$                   & \texttt{gseg} &  \texttt{Hddc} & \texttt{NODE} & \texttt{changeforest} &  \texttt{Logis} & \texttt{Fnn} & \texttt{Rf} \\ \hline
\multirow{2}{*}{500}  & 1.473         & $>10^4 $          & 235.1      & 4.258        & 0.101                           & 2.539                         & 0.697                                                                             \\
                      & (0.013)       & $(>3\times 10^4)$     & (0.38)   & (0.17)      & (0.031)                         & (0.351)                       & (0.013)                                                                             \\ \hline
\multirow{2}{*}{1000} & 1.476         & $>6\times 10^4$      & 553.2     & 8.751     & 0.122                           & 2.592                         & 1.3                                                                                   \\
                      & (0.012)       & $(>8 \times 10^4)$     & (8.5)   & (0.05)     & (0.036)                         & (0.338)                       & (0.020)                                                                          \\ \hline
\end{tabular}
}
\caption{Average computation time (in seconds) across 1000 replications from the ``Dense Mean Change" setup with $T=1000$ (standard deviations inside parentheses).  }
\label{tab:time}
\end{table}

To further examine the effectiveness of these approaches, following \cite{chakraborty_zhang_2021}, we use Adjusted Rand Index (ARI) \citep{morey_agresti_1984_ari} to measure the accuracy of the estimated change-points, which is commonly used to evaluate clustering algorithms. In the context of change-point problems, an ARI close to zero indicates a larger disparity between the estimated and true change-points or detects a change-point without any actual change. Conversely, an ARI of one indicates a perfect estimation of true change-points. We fix ARI as zero if no significant change-point is detected.   
\begin{figure}[!htp]
    \centering
\includegraphics[width=\textwidth]{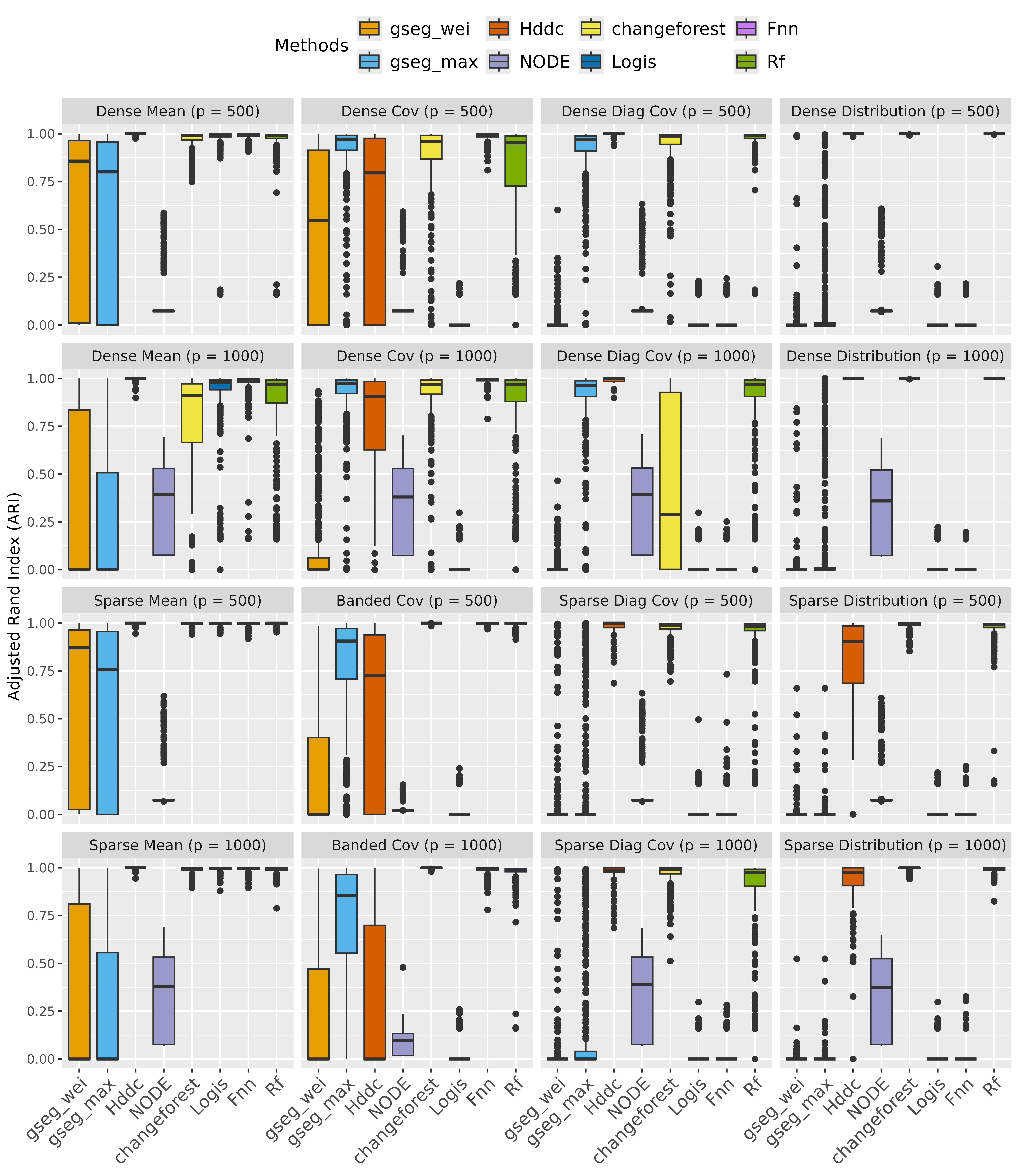}
     \caption{Boxplots of ARI values across 500 repetitions. The first and third rows correspond to dimension $p=500$, and the second and fourth rows to dimension $p=1000$, respectively. The columns represent different DGPs under consideration. Each boxplot contains eight methods considered/compared in this paper.  }
    \label{fig:boxplot_sig} 
\end{figure}

The empirical performance from 500 Monte Carlo repetitions is depicted in Figure \ref{fig:boxplot_sig} through boxplots.  We list our findings as follows. 1) Our proposed methods (including \texttt{Logis}, \texttt{Fnn} and \texttt{Rf}) together with \texttt{Hddc}  and \texttt{changeforest} are powerful in detecting both dense and sparse mean changes. 2) The  \texttt{gseg}  methods lose power in distributional changes, while \texttt{Hddc} worsens when changes occur in dense or banded covariance matrices.  
This corroborates the findings in \cite{chakraborty_zhang_2021}. 3) \texttt{NODE} does not perform well in any setup, and we conjecture this is due to the curse of high dimensionality. 4) \texttt{changeforest} and \texttt{Rf} demonstrate competitive performance across all types of changes, with \texttt{Rf} showing a slight advantage in the “Dense Diag Cov Change” setup. 5) Finally, we mention that the performance of our methods is varied; it depends on the chosen classifiers,  especially when changes occur in higher-order moments or distributions. In particular, we note that \texttt{Logis} performs well in detecting mean changes only and loses power in other setups. As for \texttt{Fnn}, it is most powerful in detecting changes in dense or banded covariances, while its performance deteriorates in other settings. We conjecture that this is due to the necessity of certain dependence among the data dimensions for a neural network to train effectively. On the other hand, the performance of \texttt{Rf} is quite robust in all settings.   In summary, if certain prior knowledge of the data is available, our proposed framework can work effectively when paired with a suitable classification algorithm. 

\subsection{Non-Euclidean CIFAR10 Data} \label{subsec:sim_non_Euclidean}
\begin{figure}[!htp]
    \centering
    \includegraphics{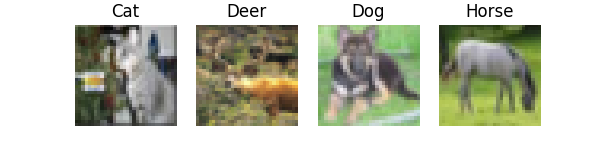}
    \caption{RGB 32-by-32 images of a Cat, Deer, Dog, and Horse from CIFAR10 Database.}
    \label{fig:mnist_example}
\end{figure}

In this section, we examine the size and power performance of the proposed method in high-dimensional non-Euclidean data. In particular, we detect change-points in artificially constructed sequences of images of various animals (dog, cat, deer, horse, etc.) from the CIFAR10 database \citep{cifar10}, see Figure \ref{fig:mnist_example} for visual illustration. Since the ground truth is known, we can accurately assess both the size and power of the method. To this end, we apply two pre-trained deep convolutional neural network (CNN) classifiers. \begin{itemize}
    \item \textbf{vgg16, vgg19:} \cite{simonyan_zisserman_vgg16} propose a deep CNN classifier with 16-19 hidden layers and a small 3-by-3 convolution filter. In the machine-learning community, it is a common practice to use pre-trained deep CNN classifiers such as \texttt{vgg16} and \texttt{vgg19} to classify images in new data sets. Here, we adopt the pre-trained weight `imagenet' \citep{deng_et_al_2009}, which is default in \texttt{Tensorflow}.
\end{itemize}

To evaluate the size performance, we randomly sample 1000 images of deer and dogs in two separate sequences of images. 
As for the alternative, we randomly choose 500 images of two different animal categories for each replication and arrange them so that the first 500 images correspond to one animal and the next 500 to the other. Hence, the true change-point is located at $t_0=500.$  We consider three choices for such pairs: ($\text{Cat} \to \text{Dog}$), $(\text{Deer} \to \text{Dog})$ and $(\text{Dog} \to \text{Horse})$. 

Out of all the methods considered, we consider two graph based methods \texttt{gseg\_wei} and \texttt{gseg\_max}, and the \texttt{changeforest} method for comparison. Here,  we perform a row-major vectorization of the three-dimensional tensor to apply \texttt{changeforest} on image data.  The proposed method based on AUC is denoted as \texttt{vgg16} and \texttt{vgg19}. The default `imagenet' embeddings are frozen to train classifiers, and only the last layer of the neural network is modified during back-propagation. A single feed-forward layer with 512 neurons is added on top of the embeddings. To further demonstrate the robustness of our AUC-based statistic,  we additionally include a baseline algorithm where the classical CUSUM statistic is computed using the one-dimensional embedding, denoted as \texttt{vgg16 cusum} and \texttt{vgg19 cusum}. Appendix \ref{subsec:vgg_cusum} includes more details on this statistic and its implementation.

\begin{table}
\centering
\scalebox{0.8}{\begin{tabular}{|l|l|p{14mm}|p{14mm}|p{23mm}|p{14mm}|p{14mm}|p{14mm}|p{14mm}|}
\hline
                      & Cases         & \texttt{gseg\_wei} & \texttt{gseg\_max} & \texttt{changeforest} & \texttt{vgg16 cusum} & \texttt{vgg19 cusum} & \texttt{vgg16} & \texttt{vgg19} \\ \hline
\multirow{2}{*}{Size} & Deer   & 4.3\%                               & 4.9\%                               & 6\%                      & 0\% & 0\% & 4\% & 3.8\%                          \\
                      & Dog     & 4.2\%                               & 4.6\%                               & 5\%          & 0\% & 0\% & 3.4\% & 4\%                          \\ \hline
\multirow{3}{*}{ARI}  & Cat $\to$ Dog    & 0.378                              & 0.289                              & 0.952              & 0.002 & 0 & 0.965 & 0.963                         \\
                      & Deer $\to$ Dog   & 0.976                              & 0.976                              & 0.992           & 1 & 1 & 0.997 & 0.996                         \\
                      & Deer $\to$ Horse & 0.970                              & 0.968                              & 0.992       & 1 & 1 & 0.994 & 0.993                         \\ \hline
\end{tabular}}
\caption{Size and Power performance of all methods: the first two rows correspond to the size, and the last three rows correspond to average ARI values.}
\label{tab:cifar10_v2}
\end{table} 

Table \ref{tab:cifar10_v2} summarizes the performance of all methods. The first two rows consist of the empirical rejection rate at the significance level $5\%$ when no change-point exists, and the last three rows enlist average ARI across 500 replications when there is a change-point. We summarize our finding as follows. 1) The testing framework of \texttt{gseg} using permutation-based p-values successfully controls the type-I error, and they can perform well when the change type is relatively easy to detect but fail in the most challenging scenario (Cat  $\to$ Dog). 2) The proposed methods \texttt{vgg16} and \texttt{vgg19}, along with \texttt{changeforest} perform favorably in all scenarios, with \texttt{vgg} methods showing a slight advantage in the case (Cat  $\to$ Dog).  3) For the CUSUM-based testing procedure using direct output probabilities of \texttt{vgg} classifiers, we observe that they fail to control the Type-I error, and has no power in the case of   (Cat  $\to$ Dog). We conjecture that this issue arises from the estimation effect of the classifier, which disrupts the use of classical CUSUM pivotal limiting distribution. This further corroborates the robustness of the AUC-based framework.

\section{Real Data Applications} \label{sec:real_data}
In this section, we present two real data applications and compare the performance of our proposed method with other methods considered in Section \ref{sec:simulation}. 

\subsection{US Stock Market During the Great Recession} \label{subsec:real_data_recession}
Detecting shifts in financial markets is crucial for policymakers and investors. In this section, we apply the proposed methods to the NYSE and NASDAQ stock data from 2005 to 2010.   Our study builds upon \cite{chakraborty_zhang_2021} where stock price changes of 72 companies under the ``Consumer Defensive" sector are analyzed.  Here, we expand our focus to a larger data set comprising daily log returns of 289 companies across the ``Healthcare", ``Consumer Defensive," and ``Utilities" sectors from January 2005 to December 2010. These sectors were identified as ``Recession-Proof Industries" during the ``Great Recession" from December 2007 to June 2009, during which the US Federal Government implemented fiscal stimulus packages to stabilize the economy. Table \ref{tab:recession_table} outlines the estimated change-points using \texttt{gseg} type methods, \texttt{Hddc}, \texttt{NODE}, \texttt{changeforest} and \texttt{Rf}.  Each method is embedded into the  Seeded Binary Segmentation \citep{kovacs_et_al_2023_biometrika} algorithm to estimate multiple change-points.   From the table, we find that estimated change-points using our method \texttt{Rf} are aligned with important historical dates during the recession. For example, in early August 2007\footnote{\url{http://news.bbc.co.uk/1/hi/business/7521250.stm}} (identified by 07-25-2007), the market experienced a liquidity crisis, and in September 2008 (identified by 09-09-2008) the international banking crisis happened caused by the bankruptcy of ``Lehman Brothers" \citep{wiggins2014lehman}. However, the change-points reported by the \texttt{gseg} methods differ significantly. Given that for stock price data, changes in daily log returns typically manifest in higher-order moments beyond the mean, we conjecture that this discrepancy is due to their impaired performance in estimating distributional changes, as indicated in Section \ref{subsec:sim_Euclidean}.  The change-points detected by \texttt{Hddc}  are close to ours, albeit a miss in 2008; while \texttt{changeforest} reports no change in 2007.   Finally, we note \texttt{NODE} fails to detect any changes. 
  
\begin{table}
        \centering
    \begin{tabular}{|l|p{25mm}|p{25mm}|p{25mm}|p{25mm}|p{25mm}}
    \hline
         Years & 2006 & 2007 & 2008 & 2009  \\
\hline
         \texttt{gseg\_wei} & 11-29 & 02-12 & 05-30 & 12-02 \\
         \hline
         \texttt{gseg\_max} & 12-18 & 02-12 & 05-30 & 11-19 \\
         \hline
\texttt{Hddc} & 08-09 & 07-25 & - & 07-13 \\
         \hline
         \texttt{NODE} & - & - & - & - \\
         \hline
         \texttt{changeforest} & 08-10 & - & 01-02 & 06-04 \\
         \hline
         \texttt{Rf} & 08-09 & 07-25 & 09-09 & 08-07  \\
         \hline
    \end{tabular}
    \caption{Estimated change-points (in MM-DD format) of US Stock Data }
    \label{tab:recession_table}
\end{table}

\subsection{New York Taxi Trips During the COVID-19 Pandemic}
\begin{figure}
     \centering
     \begin{subfigure}[b]{0.16\textwidth}
         \centering
         \includegraphics[width=\textwidth]{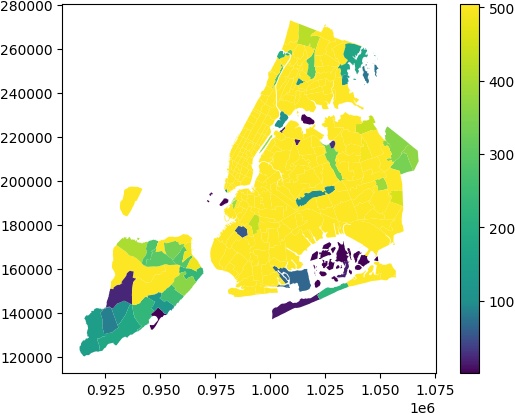}
         \caption{2019-01-15}
         \label{fig:2019pre}
     \end{subfigure}
     \hfill
     \begin{subfigure}[b]{0.16\textwidth}
         \centering
         \includegraphics[width=\textwidth]{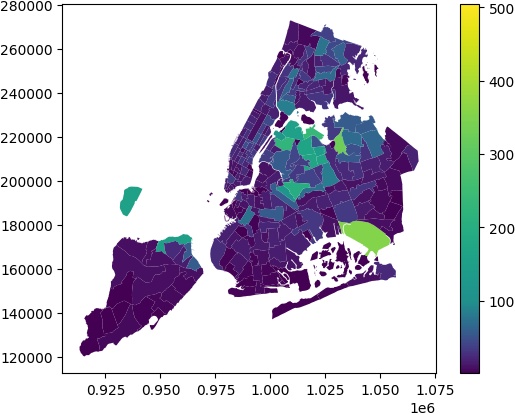}
         \caption{2019-02-15}
         \label{fig:2019post}
     \end{subfigure}
     \hfill
     \begin{subfigure}[b]{0.16\textwidth}
         \centering-
         \includegraphics[width=\textwidth]{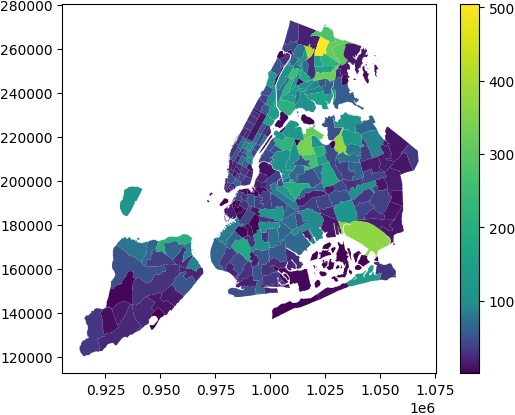}
         \caption{2019-08-05}
         \label{fig:2019pre2}
     \end{subfigure}
     \hfill
     \begin{subfigure}[b]{0.16\textwidth}
         \centering
         \includegraphics[width=\textwidth]{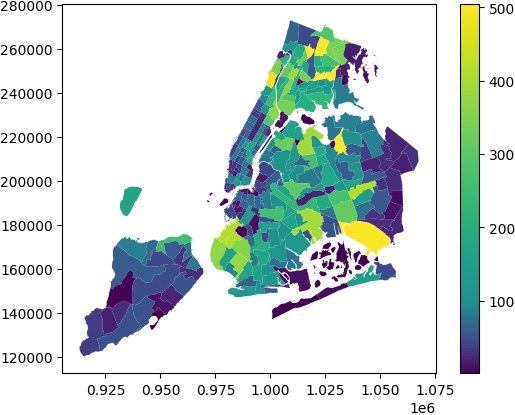}
         \caption{2019-09-05}
         \label{fig:2019post2}
     \end{subfigure}
     \hfill
     \begin{subfigure}[b]{0.16\textwidth}
         \centering
         \includegraphics[width=\textwidth]{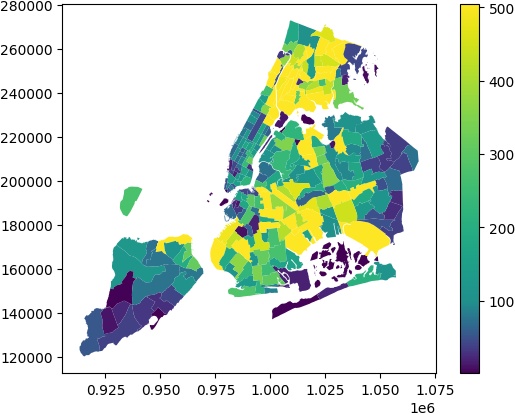}
         \caption{2020-03-05}
         \label{fig:2020pre}
     \end{subfigure}
     \hfill
     \begin{subfigure}[b]{0.16\textwidth}
         \centering
         \includegraphics[width=\textwidth]{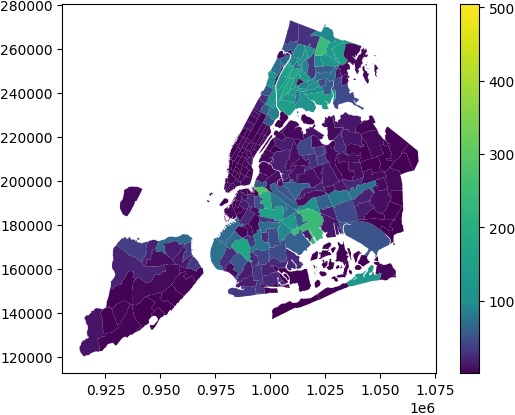}
         \caption{2020-04-05}
         \label{fig:2020post}
     \end{subfigure}
         \caption{Areal heatmaps of daily taxi trip counts in New York City on six different days before and during the COVID-19 pandemic.}
             \label{fig:heatmaps}
\end{figure}

The COVID-19 Pandemic significantly impacted New York State, particularly in early 2020. The state, especially New York City (NYC), became an epicenter for the virus, with strict lockdown measures implemented to curb the spread. This study investigates the impact of several lockdowns enforced in NYC during the COVID-19 pandemic. We collect For-Hire Vehicle (FHV) Trip records during 2018 and 2022 from the NYC Taxi and Limousine Commission (TLC) Trip Record Database.\footnote{\url{https://www.nyc.gov/site/tlc/about/tlc-trip-record-data.page}} These records contain crucial details such as pickup and drop-off times, drop-off location, trip distances, and other information. To harness the geographical insights within the data set, we employ the name of the borough associated with each drop-off point to generate daily areal heat maps. These heat maps visually depict the estimated spatial density of drop-off counts; see Figure \ref{fig:heatmaps} for six such heat-maps.  In \cite{chu_chen_2019_annals}, the same data source (during different time horizons and different types of taxi) is analyzed by summarizing the information on the number of taxi drop-offs into a 30 by 30 matrix. 

Table \ref{tab:taxi_table} showcases the estimated change-points obtained through our method with the \texttt{vgg16} classifier and \texttt{gseg} type methods. Notably, the change-points identified by both methods closely align with each other except for 2021.  These significant shifts correspond to events such as the 2019 North American cold wave (2019-01-31), the potential impact of summer vacation (2019-08-31), and the implementation of NYC's first lockdown (2020-03-17).
For instance, Figure \ref{fig:2019pre} and \ref{fig:2019post} depict the areal heat maps on January 15, 2019 (before the heatwave) and February 15, 2019 (after the heatwave), respectively, revealing a noteworthy decrease in taxi usage. Subsequently, areal heat maps in Figure \ref{fig:2019pre2} and  \ref{fig:2019post2} illustrate an increasing effect on taxi usage following the end of summer vacation. Additionally, Figure \ref{fig:2020pre} and \ref{fig:2020post} present the areal heat maps on March 5, 2020 (before the lockdown) and April 5, 2020 (after the lockdown), respectively, indicating a significant decline in taxi usage.  However, we find the estimated change-points by \texttt{changeforest} differ significantly from the other two methods.

\begin{table}
    \centering
    \begin{tabular}{|l|p{15mm}|p{25mm}|p{25mm}|p{23mm}|p{15mm}|p{15mm}}
    \hline
         Years & 2018 & 2019 & 2020 & 2021 & 2022  \\
         \hline
\texttt{gseg\_wei} & - & 01-31, 10-01 & 03-16, 08-01 & 07-01 & 04-01 \\
         \hline
         \texttt{gseg\_max} & - & 01-31, 10-01 & 03-16, 08-01 & 07-01 & 04-01 \\
\hline
         \texttt{changeforest} & - & 04-01 & 06-25 & 09-01 & - \\
         \hline
         \texttt{vgg16} & - & 01-31, 08-31 & 03-17, 11-29 & 10-31 & 04-03 \\
         \hline
    \end{tabular}
    \caption{Estimated change-points (in MM-DD format) of New York taxi data}
    \label{tab:taxi_table}
\end{table}

\section{Discussion} \label{sec:discussion}
In this paper, we propose a novel framework for offline change-point detection based on the AUC of classifiers. Despite the vast literature on change-point detection, our approach stands out in four key aspects.   First, our method is entirely nonparametric, as it does not impose any parametric assumptions on the data distribution or the structure of distributional changes. Second, by leveraging the success of classifiers, the proposed framework is widely applicable to diverse data dimensions and types. Third, by selecting an appropriate classifier, prior information can be integrated to enhance the test power. Last but not least, with suitable sample splitting and trimming, the test statistic converges to the supremum of a pivotal Gaussian process, which is independent of the choice of classifier and thus can be easily implemented. Another key advantage of our proposed method is that it can incorporate weakly trained and fine-tuned pre-trained classifiers. Rank-sum comparison embedded in the AUC makes this integration robust to estimation error in classification probabilities.

To conclude, we mention a few directions that warrant future investigation. First, it should be noted that the real data analysis in Section \ref{sec:real_data} may exhibit temporal dependence. Extending the proposed method to accommodate weakly dependent time series would be an intriguing endeavor. Second, it may be feasible to reduce the computational burden by drawing inspiration from \cite{londschien_buhlman_kovacs_2022}. Instead of computing the AUC for all potential change-points, it could be computed for initial points, such as the quartiles of the sequence, followed by an extensive search in proximity to the initial candidate change-point that yields the highest AUC. However, exploring the theoretical properties of such a two-stage approach may be challenging.  Third, the current setting is unsupervised and, therefore, requires certain sample splitting and trimming; it would be interesting to investigate the asymptotic properties of the test statistic under the supervised setting, as in \cite{li_paul_fryzlewicz_wang_2022}. Fourth, it will be useful to extend our framework to the online setting \citep{cai2020online}. We leave these directions to future research.  

\acks{The authors thank the action editor and two anonymous referee for their helpful comments that greatly improved the quality of this work. Kanrar thanks the Information Technology Services at Iowa State University and the University of Hong Kong for providing access and support to use the High-performance computing clusters. Jiang’s work is supported in part by  National Key R\&D Program of China (No. 2024YFA1015700), and  NSFC (Nos.  12201124, 12331009, 72271060, and 72432002). Cai's work is supported in part by the University of Hong Kong Seed Fund for Basic Research for New Staff.}

\newpage

\appendix
\textbf{Overall Organzation of Appendix:} The Appendix contains all technical proofs for the paper. Appendix \ref{sec:prop_proofs} provides with proofs of Proposition \ref{prop:unif_oracle_auc_bound} and \ref{prop:likelihood_bound}. Appendix \ref{sec:thm_proofs} contains all the proofs of main theorems. Appendix \ref{sec:lemmas} lists some useful lemmas, and  Appendix \ref{sec:lemma_proofs} gives their proofs. Appendix \ref{sec:add_simu} outlines additional simulation results and more details on CUSUM statistic used in Section \ref{sec:simulation}.

\section{Proof of Propositions} \label{sec:prop_proofs} \subsection{Proof of Proposition \ref{prop:unif_oracle_auc_bound}} 
\begin{proof}
   Let $\Z_X, \Z'_X \simiid \cP_X$ and $\Z_Y \sim \cP_Y$ be independent random elements.   Note that, \begin{flalign*}
        \begin{split}
            d_{tv}(\cP_X, \cP_Y) &= \sup_{A} \absbigg{\cP_Y(A) - \cP_X(A)} = \frac{1}{2} \int_{x \in E} \absbigg{ d\cP_Y(x) - d\cP_X(x) }dx  \\ 
            &= \frac{1}{2} \int_{x \in  E} \absbigg{ \frac{d\cP_Y}{d\cP_X}(x) - 1} d\cP_X(x) = \frac{1}{2} \E \absbigg{\frac{d\cP_Y}{d\cP_X}(\Z_X) - 1}.
        \end{split}
    \end{flalign*}   Then, by triangle inequality, $$\E\absbigg{\frac{d\cP_Y}{d\cP_X}(\Z_X) - \frac{d\cP_Y}{d\cP_X}(\Z_X')} = \E\absbigg{\frac{d\cP_Y}{d\cP_X}(\Z_X) - 1 + 1 - \frac{d\cP_Y}{d\cP_X}(\Z_X')} \leq  2 \E \absbigg{\frac{d\cP_Y}{d\cP_X}(\Z_X) - 1}.$$  Moreover, note that, $\E [ \frac{d\cP_Y}{d\cP_X}(\Z_X)] = 1$, hence \begin{flalign*}
        \E\absbigg{\frac{d\cP_Y}{d\cP_X}(\Z_X) - 1} = \E\absbigg{\frac{d\cP_Y}{d\cP_X}(\Z_X) - \E(\frac{d\cP_Y}{d\cP_X}(\Z_X'))} \leq \E\absbigg{\frac{d\cP_Y}{d\cP_X}(\Z_X) - \frac{d\cP_Y}{d\cP_X}(\Z_X')},
    \end{flalign*} where the last inequality holds due to the conditional Jensen's inequality. Combining the last two inequalities, we have, \begin{equation} \label{eq:big_eq1_proposition}
        2 d_{tv}(\cP_X, \cP_Y)  \leq \E\absbigg{\frac{d\cP_Y}{d\cP_X}(\Z_X) - \frac{d\cP_Y}{d\cP_X}(\Z_X')} \leq  4 d_{tv}(\cP_X, \cP_Y).
    \end{equation} Therefore, the claimed result holds if we can show \begin{flalign}\label{eq:big_eq2_proposition}
        \begin{split}
                &\frac{1}{2} - \P\secondbigg{ \frac{d\cP_Y}{d\cP_X}(\Z_Y) < \frac{d\cP_Y}{d\cP_X}(\Z_X) } \\=& \frac{1}{2} - \E \frac{d\cP_Y}{d\cP_X}(\Z_X) \indic\secondbigg{\frac{d\cP_Y}{d\cP_X}(\Z_X) < \frac{d\cP_Y}{d\cP_X}(\Z_X')} = \frac{1}{4} \E \absbigg{\frac{d\cP_Y}{d\cP_X}(\Z_X) - \frac{d\cP_Y}{d\cP_X}(\Z_X')}.  
        \end{split}
    \end{flalign} 
To prove \eqref{eq:big_eq2_proposition}, we first note \begin{flalign*}
        \begin{split}
            &\E \frac{d\cP_Y}{d\cP_X}(\Z_X) \indic\secondbigg{\frac{d\cP_Y}{d\cP_X}(\Z_X) < \frac{d\cP_Y}{d\cP_X}(\Z_X')}\\ =& \int_{x, x' \in  E} \frac{d\cP_Y}{d\cP_X}(x) \indic\secondbigg{\frac{d\cP_Y}{d\cP_X}(x) < \frac{d\cP_Y}{d\cP_X}(x')} d\cP_X(x) d\cP_X(x') \\
            =& \int_{x, x' \in  E} \indic\secondbigg{\frac{d\cP_Y}{d\cP_X}(x) < \frac{d\cP_Y}{d\cP_X}(x')} d\cP_Y(x) d\cP_X(x') 
           =  \P\secondbigg{ \frac{d\cP_Y}{d\cP_X}(\Z_Y) < \frac{d\cP_Y}{d\cP_X}(\Z_X) },
        \end{split}
    \end{flalign*} 
which implies the first equality in \eqref{eq:big_eq2_proposition}.   
    For the second equality, we have, 
    \begin{flalign*}
        \begin{split}
            &\E \absbigg{\frac{d\cP_Y}{d\cP_X}(\Z_X) - \frac{d\cP_Y}{d\cP_X}(\Z_X')}\\ = & \E(\frac{d\cP_Y}{d\cP_X}(\Z_X) - \frac{d\cP_Y}{d\cP_X}(\Z_X'))\indic\secondbigg{\frac{d\cP_Y}{d\cP_X}(\Z_X) > \frac{d\cP_Y}{d\cP_X}(\Z_X')} \\&+ \E(\frac{d\cP_Y}{d\cP_X}(\Z_X') - \frac{d\cP_Y}{d\cP_X}(\Z_X))\indic\secondbigg{\frac{d\cP_Y}{d\cP_X}(\Z_X) < \frac{d\cP_Y}{d\cP_X}(\Z_X')} \\
            =& 2 \E(\frac{d\cP_Y}{d\cP_X}(\Z_X) - \frac{d\cP_Y}{d\cP_X}(\Z_X'))\indic\secondbigg{\frac{d\cP_Y}{d\cP_X}(\Z_X) > \frac{d\cP_Y}{d\cP_X}(\Z_X')} \\
            =& 2 \thirdbigg{\E \frac{d\cP_Y}{d\cP_X}(\Z_X)\indic\secondbigg{\frac{d\cP_Y}{d\cP_X}(\Z_X) > \frac{d\cP_Y}{d\cP_X}(\Z_X')} - \E \frac{d\cP_Y}{d\cP_X}(\Z_X')\indic\secondbigg{\frac{d\cP_Y}{d\cP_X}(\Z_X) > \frac{d\cP_Y}{d\cP_X}(\Z_X')} } \\
            =& 2 \thirdbigg{\E \frac{d\cP_Y}{d\cP_X}(\Z_X')\indic\secondbigg{\frac{d\cP_Y}{d\cP_X}(\Z_X') > \frac{d\cP_Y}{d\cP_X}(\Z_X)} - \E \frac{d\cP_Y}{d\cP_X}(\Z_X')\indic\secondbigg{\frac{d\cP_Y}{d\cP_X}(\Z_X) > \frac{d\cP_Y}{d\cP_X}(\Z_X')} } \\
            =& 2 \thirdbigg{1 - 2 \E \frac{d\cP_Y}{d\cP_X}(\Z_X')\indic\secondbigg{\frac{d\cP_Y}{d\cP_X}(\Z_X) > \frac{d\cP_Y}{d\cP_X}(\Z_X')} },
        \end{split}
    \end{flalign*} where the second and the fourth equality hold because  $\Z_X$ and $\Z_X'$ are iid.
\end{proof} 

\subsection{Proof of Proposition \ref{prop:likelihood_bound}}
\begin{proof}This is a direct consequence of Proposition \ref{prop:unif_oracle_auc_bound} and Lemma \ref{lemma:reln_mu_delta}.\end{proof}

\section{Proof of Theorems} \label{sec:thm_proofs}
\textbf{Notations:} We denote $a(T)\lesssim b(T)$ if there exists $T_0\in\mathbb{N}_+$ and $C>0$ such that for any $T\geq T_0$, $a(T)\leq C b(T)$. 
 We denote $a(T)\asymp b(T)$ if $a(T)\lesssim b(T)$ and $b(T)\lesssim a(T)$. 
 From Section \ref{sec:theory}, let us recall the additional notations defined earlier. Let $\P_*, \ \E_*, \ \var_{*} \ \text{and}, \ \cov_{*}$ be the conditional probability, expectation, variance, and covariance given the trained classifier and the training data $D_0\cup D_1$. For an arbitrary random element $\Z \in E$, let $F_{X, *}(\cdot)$ and $F_{Y,*}(\cdot)$ be the conditional cumulative distribution functions (CDF) of $\hat{\theta}(\Z)$ when $\Z \sim \cP_X$ and $\Z \sim \cP_Y$, respectively. For independent random elements $\{\Z_t\}_{t=1}^{T}$, we define  $\hat{W}^X_t \defeq F_{X,*}(\Z_t)$ and $\hat{W}^Y_t \defeq F_{Y,*}(\Z_t)$ for all $t=1, \ldots, T$. Also define, $\gamma \defeq \epsilon + \eta$. In all of the proofs, we first establish the asymptotic results with respect to the conditional probability measure, $\P_{*}$ ( w.r.t. the training set). To emphasize the distinction between the $\P$ and $\P_*$, we use separate notations, $o_{\P}(\cdot), O_{\P}(\cdot)$ and $o_{\P_*}(\cdot), O_{\P_*}(\cdot)$. In most of our technical proofs, we first establish results under $\P_*$. The result can then be extended to $\P$. Note that the set of conditional probabilities obtained by the classifier, $\secondbigg{\hat{\theta}(\Z_t): t=\floor{Tr}}_{r \in [\epsilon, 1-\epsilon]}$, is a set of constant measurable transformation of observations in $D^v$, with respect to $\P_{*}$.

\subsection{Proof of Theorem \ref{thm:unif_weak_conv}} \label{subsec:unif_weak_conv_proof} 
\begin{proof}
  Under $H_0$, it follows that, $\{\Z_t\}_{t=1}^{T} \simiid \cP_X$.  Given $\hat{\theta}(\cdot)$,  $F_{X,*}(\cdot)$ is the conditional CDF of $\widehat{\theta}(\mathbf{Z})$. Since $\hat{\theta}(\cdot)$ is continuous w.r.t. $\mathcal{P}_X,$ we apply the probability integral transformation to have, $\{F_{X,*}(\hat{\theta}(\Z_t))\}_{t=m+1}^{T-m}\defeq\{\hat{W}^X_t\}_{t=m+1}^{T-m} \simiid \text{Uniform}(0,1)$ under $\P_*.$ Then, we can write, \begin{align} \label{eq:null_remaind_term}
    \begin{split}
       &\indic\secondbigg{\hat{\theta}(\Z_i) < \hat{\theta}(\Z_j)} - 1/2  \\
        =& (\hat{W}^X_j - 1/2) + (1/2 - \hat{W}^X_i) + \underbrace{\firstbigg{\indic\secondbigg{\hat{\theta}(\Z_i) < \hat{\theta}(\Z_j)} - 1/2 - \hat{W}^X_j + \hat{W}^X_i}}_\text{$\defeq \hat{h}^0(\Z_i, \Z_j)$}.
    \end{split}  
\end{align} For $k \in \calI_{cp}$, we have, \begin{flalign*}
    \begin{split}
        \hat{\Psi}(k) - \frac{1}{2} &= \frac{1}{(k-m)(T-m-k)} \sum_{i=m+1}^{k} \sum_{j=k+1}^{T-m}\left[\indic\secondbigg{\hat{\theta}(\Z_i) < \hat{\theta}(\Z_j)} - 1/2\right] \\ 
        &= P_{T}^{0}(k) + R_{T}^{0}(k),
    \end{split}
\end{flalign*}
where the projection term and the remainder term are,
\begin{align} \label{eq:decompose_auc_projection_remainder}
    \begin{split}
        P_T^0(k) &= \frac{1}{T-m-k} \sum_{j=k+1}^{T-m} \firstbigg{ \hat{W}^X_j - 1/2 } + \frac{1}{k-m} \sum_{i=m+1}^{k} \firstbigg{ 1/2 - \hat{W}^X_i}, \\ 
        R_{T}^0(k) &= \frac{1}{(k-m)(T-m-k)} \sum_{i=m+1}^{k} \sum_{j=k+1}^{T-m} \hat{h}^{0}(\Z_i, \Z_j).
    \end{split}
\end{align} For  $\epsilon \in (0, 1/2), \ \eta \in (0, 1/2-\epsilon)$, and $\gamma=\epsilon+\eta$, define a piece-wise constant stochastic processes on $[\gamma, 1-\gamma]$ such that  $\wt{P}_T^0(r) \defeq P_T^0(\floor{Tr})$ and $\wt{R}_T^0(r) \defeq R_T^0(\floor{Tr})$, $\forall \ r \in [\gamma, 1-\gamma]$. We consider two processes, $\{\wt{P}_T^0(r)\}_{r \in [\gamma, 1-\gamma]}$ and $\{\wt{R}_T^0(r)\}_{r \in [\gamma, 1-\gamma]}$, separately and study the uniform weak convergence of the projection and reminder term. In particular, we show: \begin{enumerate}
\item \textbf{Weak Convergence of Projection Term:} \begin{equation} \label{eq:proj_null_proof_statement}
        \secondbigg{ \sqrt{T} \wt{P}_T^0(r) }_{r \in [\gamma, 1-\gamma]} \ucd \secondbigg{G_0(r)}_{r \in [\gamma, 1-\gamma]}, \quad \text{where,}
    \end{equation} \begin{equation*}
        G_0(r) \defeq \frac{1}{\sqrt{12}} \thirdbigg{ \frac{B(1-\epsilon) - B(r)}{1-\epsilon-r} - \frac{B(r)-B(\epsilon)}{r-\epsilon} }, \quad \ \text{for} \ r \in [\gamma, 1-\gamma].
     \end{equation*} 
    \item \textbf{Uniform Convergence of Remainder Term to Zero:} \begin{equation} \label{eq:remaind_null_proof_statement}
        \sup_{r \in [\gamma, 1-\gamma]} \secondbigg{\sqrt{T} \wt{R}_T^0(r)}\cp 0.
    \end{equation} It is sufficient to show the following two intermediate results: \begin{enumerate}
        \item \textbf{Finite-dimensional Weak Convergence to Zero:} $$\sqrt{T} \wt{R}_T^0(r) \cp 0, \ \forall \ r \in [\gamma, 1-\gamma].$$
        \item \textbf{Stochastic Equicontinuity:} For any $x>0$, $$
        \lim_{\delta_T\to 0}\limsup_{T\to\infty}  \P_* \left(\sup_{|r-r'|\leq \delta_T, \gamma\leq r,r'\leq 1-\gamma} \sqrt{T}| \wt{R}_T^0(r) - \wt{R}_T^0(r') |>x \right)=0.$$
    \end{enumerate}
\end{enumerate} Combining \eqref{eq:proj_null_proof_statement}, \eqref{eq:remaind_null_proof_statement}, we establish the Theorem \ref{thm:unif_weak_conv}. Now, we now prove our claims. 
    
\textbf{\textit{Proof of \eqref{eq:proj_null_proof_statement}}:}
    Let us consider a simpler process of partial sums for $k \in \calI_{cp}$: $S_T^0(k) \defeq \sqrt{12} \sum_{i=m+1}^{k} \firstbigg{ \hat{W}^X_i - 1/2 }$, and the corresponding piece-wise constant stochastic process, $\{\wt{S}_T^0(r)\}_{r\in [\gamma, 1-\gamma]}$, where, $\wt{S}_T^0(r) = S_T^0(\floor{Tr})$. Note that $\{\hat{W}_i^X\}_{i=m+1}^{T-m}$ is iid under $\P_*$, with $\E_*(\hat{W}^X_{m+1})=1/2$ and $\var_*(\hat{W}^X_{m+1})=1/12$, by the  Donsker's  theorem, it follows that, $\secondbigg{\wt{S}_T^0(r)/\sqrt{T} }_{r \in [\gamma, 1-\gamma]} \ucd \{B(r)-B(\epsilon)\}_{r \in [\gamma, 1-\gamma]}$.
     Note that, $$\wt{P}_T^0(r) = \frac{1}{\sqrt{12}}\thirdbigg{ \frac{\wt{S}_T^0(1-\epsilon) - \wt{S}_T^0(r)}{T-\floor{T\epsilon}-\floor{Tr}} - \frac{\wt{S}_T^0(r)}{\floor{Tr}-\floor{T\epsilon}} },$$ which is a continuous function of $\{\wt{S}_T^0(r)\}_{r \in [\gamma, 1-\gamma]}$. As $T\to\infty$, it follows that, $(T-\floor{T\epsilon}-\floor{Tr})/T \to (1-\epsilon-r)$ and $(\floor{Tr}-\floor{T\epsilon})/T \to (r-\epsilon)$. Therefore, we apply the continuous mapping theorem with Slutsky's theorem to establish \eqref{eq:proj_null_proof_statement}.
    
\textbf{\textit{Proof of \eqref{eq:remaind_null_proof_statement}:}}
By Lemma \ref{lemma:cross_cov_remaind_terms_uniform}, we have $\E_* \hat{h}^{0}(\Z_i, \Z_j) \hat{h}^{0}(\Z_{i'}, \Z_{j'}) = 5/12$ for all $(i,j) = (i', j')$ and $0$ otherwise, where $i, j, i', j' \in \{m+1, \ldots, T-m\}$. Hence, \begin{equation*}
    \E_* \{R_T^{0}(\floor{Tr})\}^2 = \frac{1}{(k-m)^2(T-m-k)^2} \sum_{i=m+1}^{k} \sum_{j=k+1}^{T-m} \E_*
         \{\hat{h}^{0}(\Z_i, \Z_j)\}^2 \lesssim \frac{1}{T^2}.
\end{equation*} By Markov's inequality, we establish, $\sqrt{T} {R}_T^0(\floor{Tr}) = O_{\P_*}(1/\sqrt{T})$ for all $r \in [\gamma, 1-\gamma]$.
    To prove the asymptotic stochastic equicontinuity, we first show for any $r,r' \in [\gamma, 1-\gamma]$: $\E_* \thirdbigg{ \sqrt{T} \firstbigg{\wt{R}_T^0(r) - \wt{R}_T^0(r')} }^2 \lesssim T^{-1}.$ Then, we modify Lemma A.1 in \cite{kley2016quantile} (see Lemma \ref{lemma:kley_et_al_extn}) to establish stochastic equicontinuity of $\secondbigg{\sqrt{T}\wt{R}_{T}^0(r)}_{r \in [\gamma, 1-\gamma]}$. 
    Recall, $\wt{R}_T^0(r) \defeq R_T^0(\floor{Tr}),  \forall r \in [\gamma, 1-\gamma]$ and consider, $r, r' \in [\gamma, 1-\gamma]$ with $ r' < r$, and $k = \floor{Tr}, k' = \floor{Tr'}$. If $k=k'$, then, $\wt{R}_T^0(r) = \wt{R}_T^0(r')$. Therefore, we fix $r, r' \in [\gamma, 1-\gamma]$, such that for $k' < k$, \begin{align} \label{eq:diff_rtk_null}
    \begin{split}
        &\wt{R}_T^0(r) - \wt{R}_T^0(r')=R_T^0(k) - R_T^0(k') \\
        =& \firstbigg{\frac{1}{(k-m)(T-m-k)} - \frac{1}{(k'-m)(T-m-k')}} \sum_{i=m+1}^{k'}\sum_{j=k+1}^{T-m} \hat{h}^{0}(\Z_i, \Z_j) \\
        & + \frac{1}{(k-m)(T-m-k)} \sum_{i=k'+1}^{k}\sum_{j=k+1}^{T-m} \hat{h}^{0}(\Z_i, \Z_j)  \\
        & - \frac{1}{(k'-m)(T-m-k')} \sum_{i=m+1}^{k'}\sum_{j=k+1}^{T-m} \hat{h}^{0}(\Z_i, \Z_j).
    \end{split}
\end{align} By Lemma \ref{lemma:cross_cov_remaind_terms_uniform}, the cross-covariance between and within the three summations is $0$,  hence, \begin{flalign*}
            &\E_* \thirdbigg{ \sqrt{T} \firstbigg{\wt{R}_T^0(r) - \wt{R}_T^0(r')} }^2  \\
            =& \firstbigg{\frac{1}{(k-m)(T-m-k)} - \frac{1}{(k'-m)(T-m-k')}}^2 T \sum_{i=m+1}^{k'}\sum_{j=k+1}^{T-m} \E_*\{\hat{h}^{0}(\Z_i, \Z_j)\}^2 \\
            & + \frac{T}{(k-m)^2(T-m-k)^2} \sum_{i=k'+1}^{k}\sum_{j=k+1}^{T-m} \E_*\{\hat{h}^{0}(\Z_i, \Z_j)\}^2 \\
            & + \frac{T}{(k'-m)^2(T-m-k')^2} \sum_{i=m+1}^{k'}\sum_{j=k+1}^{T-m} \E_*\{\hat{h}^{0}(\Z_i, \Z_j)\}^2 \lesssim 1/T,
    \end{flalign*} where we use the fact that, $\absbigg{\hat{h}^{0}(\Z_i, \Z_j)} \leq 4$. We thus have $$
\E_* \thirdbigg{ \sqrt{T} \firstbigg{\wt{R}_T^0(r) - \wt{R}_T^0(r')} }^2\lesssim |r-r'|, \quad  \forall r,r'\in [\gamma, 1-\gamma]: |r-r'|\geq 1/T.
$$ Now, it is possible to modify Lemma A.1 in \cite{kley2016quantile} as in Lemma \ref{lemma:kley_et_al_extn}, to establish the stochastic equicontinuity, and therefore we complete the proof of \eqref{eq:remaind_null_proof_statement}.
\end{proof}

\subsection{Proof of Theorem \ref{thm:unif_weak_conv_alt}} \label{subsec:unif_weak_conv_alt_proof}
We first provide an outline of the theorem followed by detailed proofs of our claims. In some parts, we borrow ideas developed in the Proof of Theorem \ref{thm:unif_weak_conv} and refer to specific sections to avoid repetitions. Without loss of generality,  we consider the candidate change-point, $k \in \calI_{cp}: k < t_0$. Proof for the other case $k \geq t_0$ follows using a similar argument. 

\begin{proof}
Under $H_1$, we have  $\{\Z_t\}_{t=1}^{t_0} \simiid \cP_X$ and $\{\Z_t\}_{t=t_0+1}^{T} \simiid \cP_Y$. At a candidate change-point, $k <t_0$, the AUC $\hat{\Psi}(k)$ can be written as follows: \begin{equation} \label{eq:auc_decomp_local_alt}
    \hat{\Psi}(k) =\frac{\sum_{i=m+1}^{k} \thirdbigg{ \sum_{j=k+1}^{t_0}\indic\secondbigg{\hat{\theta}(\Z_i) < \hat{\theta}(\Z_j)} + \sum_{j=t_0+1}^{T-m} \indic\secondbigg{\hat{\theta}(\Z_i) < \hat{\theta}(\Z_j)} }}{(k-m)(T-k-m)}.
\end{equation} Terms in the first summation include random elements $\Z_i, \Z_j\sim \cP_X$. The second summation includes random elements $\Z_i\sim \cP_X, \Z_j\sim\cP_Y$ with different distributions. We consider two terms separately:
Given $\theta(\cdot)$, let $F_X(\cdot)$ and $F_Y(\cdot)$ be the CDF of $\theta(\Z)$ when $\Z \sim \cP_X$ and $\Z \sim \cP_Y$, respectively. Define $W_t^X \defeq F_X(\theta(\Z_t)), \ W_t^Y \defeq F_Y(\theta(\Z_t))$, and recall their conditional (on the training set) counterparts as $\hat{W}^X_j$ and $\hat{W}^Y_j$, respectively.  For independent $\Z_X \sim \cP_X, \Z_Y \sim \cP_Y$, we recall, $\mu \defeq \P(\theta(\Z_X) < \theta(\Z_Y))$, and  $\mu_* \defeq \P_*(\hat{\theta}(\Z_X) < \hat{\theta}(\Z_Y))$.
Therefore, if $\Z_i, \Z_j \simiid \cP_X$,  from \eqref{eq:null_remaind_term} in the proof of Theorem \ref{thm:unif_weak_conv}, we have: \begin{flalign} \label{eq:decomp_single1}
    \begin{split}
      &\indic\secondbigg{\hat{\theta}(\Z_i) < \hat{\theta}(\Z_j)}  - 1/2 = (\hat{W}^X_j - 1/2) + (1/2 - \hat{W}^X_i) + \hat{h}^0(\Z_i, \Z_j) \\
         = &\underbrace{(W^X_j - 1/2) + (1/2 - W^X_i)}_\text{Projection} + \underbrace{(\hat{W}^X_j - W^X_j) + (W^X_i - \hat{W}^X_i)}_\text{Approximation Error} + \underbrace{\hat{h}^0(\Z_i, \Z_j)}_\text{Remainder}.
    \end{split}
\end{flalign} Next, if $\Z_i\sim \cP_X, \Z_j\sim \cP_Y$, let $\hat{h}^{\dagger}(\Z_i, \Z_j) \defeq \indic\secondbigg{\hat{\theta}(\Z_i) < \hat{\theta}(\Z_j)}  - \hat{W}^X_j + \hat{W}^Y_i - 1 + \mu_*$, then 
\begin{flalign} \label{eq:decomp_single2}
    \begin{split}
\indic\secondbigg{\hat{\theta}(\Z_i) < \hat{\theta}(\Z_j)}  - 1/2 &= \underbrace{(\mu - 1/2)}_\text{Signal} + \underbrace{(W_j^X - \mu) + (1 - \mu - W_i^Y)}_\text{Projection} \\
        & \ + \underbrace{(\mu - \mu_*) + (\hat{W}_j^X - W_j^X) + (W_i^Y - \hat{W}_i^Y)}_\text{Approximation Error} + \underbrace{\hat{h}^{\dagger}(\Z_i, \Z_j)}_\text{Remainder}.
    \end{split}
\end{flalign} 
We substitute \eqref{eq:decomp_single1} and  \eqref{eq:decomp_single2} into \eqref{eq:auc_decomp_local_alt} to write:  \begin{flalign} \label{eq:ultimate_decomp_alt}
    \hat{\Psi}(k) - \frac{1}{2} = \underbrace{ w_T(k) (\mu - 1/2) }_\text{Signal} + \underbrace{ P_T(k) }_\text{Projection} + \underbrace{ R_T(k) }_\text{Remainder} + \underbrace{ w_T(k) (\mu - \mu_*) + A_T(k) }_\text{Approximation Error},
\end{flalign} where $w_T(k) = (T-m-t_0)/(T-m-k)$, \begin{flalign}
       \label{PT} P_T(k) &\defeq \frac{1}{T-k-m} \thirdbigg{ \sum_{j=k+1}^{t_0} ({W}_j^X - 1/2) + \sum_{j=t_0+1}^{T-m} (W_j^X - \mu) }, \\
            \notag & \ \ \ \ \ + \frac{1}{k-m} \sum_{i=m+1}^{k} \thirdbigg{ (1-w_T(k))(1/2 - {W}_i^X) + w_T(k)(1 - \mu - W_i^Y) }, \\
       \label{RT} R_T(k) &\defeq \frac{1}{(k-m)(T-m-k)} \sum_{i=m+1}^{k} \thirdbigg{ \sum_{j = k+1}^{t_0} \hat{h}^0(\Z_i, \Z_j) + \sum_{j=t_0+1}^{T-m}\hat{h}^{\dagger}(\Z_i, \Z_j) }, \\
     \label{AT}   A_T(k) &\defeq \frac{1}{T-k-m} \thirdbigg{ \sum_{j=k+1}^{t_0} (\hat{W}_j^X - W_j^X) + \sum_{j=t_0+1}^{T-m} (\hat{W}_j^X - W_j^X) } \\
      \notag  & \ \ \ \ \ + \frac{1}{k-m} \sum_{i=m+1}^{k} \thirdbigg{(1-w_T(k))(W_i^X - \hat{W}_i^X)+ w_T(k)(W_i^Y - \hat{W}_i^Y)  }.
\end{flalign} 
The rest of the proof hinges on the following facts for $\delta_T \to 0$ and $ T\to \infty$, 
\begin{flalign}
\label{eq:proj_alt_proof_statement}
    &\secondbigg{\sqrt{T}P_T(\floor{Tr})}_{r \in [\gamma, 1-\gamma]} \ucd \secondbigg{G_0(r)}_{r \in [\gamma, 1-\gamma]}, \\
\label{eq:remaind_alt_proof_statement}
    &\sup_{r \in [\gamma,1-\gamma]}\secondbigg{\sqrt{T} R_T(\floor{Tr})} \cp 0,\\
\label{eq:error_alt_proof_statement}
       & \sup_{r \in [\gamma,1-\gamma]} \sqrt{T} A_T(\floor{Tr}) \cp 0.
\end{flalign}   Finally, we apply Lemma \ref{lemma:reln_mu_delta} to observe that, $w_T(k) (\mu - 1/2) = w_T(k) \delta_T/4 \to (1-\epsilon-\tau)\delta_T/[4(1-\epsilon-r)]$. Provided, $T \to \infty, \ \sqrt{T}\delta_T \to C \in (0,\infty)$, we have: \begin{equation*}
    \secondbigg{\sqrt{T}\firstbigg{\hat{\Psi}(\floor{Tr}) - \frac{1}{2}}}_{r \in [\gamma,1-\gamma]} \ucd \secondbigg{\Delta_G(r;\tau, \epsilon) + G_0(r)}_{r \in [\gamma,1-\gamma]},
 \end{equation*} where, $\Delta_G(r;\tau) \defeq C(1-\epsilon-\tau)/[4(1-\epsilon-r)]$, for $k \defeq \floor{Tr} < \floor{T\tau} \defeq t_0$. By symmetry of our argument, we have $\Delta_G(r;\tau) \defeq C(\tau-\epsilon)/[4(r-\epsilon)]$, for $k \geq  t_0$, and we conclude the proof of Theorem \ref{thm:unif_weak_conv_alt}.
Now, we establish our claims, i.e., \eqref{eq:proj_alt_proof_statement}, \eqref{eq:remaind_alt_proof_statement}, \eqref{eq:error_alt_proof_statement}.

\textbf{\textit{Proof of \eqref{eq:proj_alt_proof_statement}:}}
For $r \in [\gamma, 1-\gamma]$, define $\wt{P}_T(r) \defeq P_T(\floor{Tr})$,  where in \eqref{PT}, \begin{flalign*}
    P_T(k) &=\frac{1}{T-k-m} \thirdbigg{ \sum_{j=k+1}^{t_0} (W_j^X - 1/2) + \sum_{j=t_0+1}^{T-m} (W_j^X - \mu) }, \\
        & \ \ \ \ \ + \frac{1}{k-m} \sum_{i=m+1}^{k} \thirdbigg{ (1-w_T(k))(1/2 - W_i^X) + w_T(k)(1 - \mu - W_i^Y) }.
\end{flalign*} 
Note that the above three terms are independent with respect to the conditional probability measure $\P_*$. Therefore, it suffices to show the uniform weak convergence separately. Then, we can use the continuous mapping theorem to establish the uniform weak convergence of $\{\wt{P}_T(r)\}_{r \in [\gamma,1-\gamma]}$. In what follows, we provide all the arguments with respect to $\P_*$.
Note that, $\{W_j^X\}_{j=k+1}^{t_0} \simiid \text{Uniform}(0,1)$ with $\E_* (W^X_j) = 1/2, \ \var_* (W^X_j) = 1/12$. Therefore, we apply Donsker's theorem to establish: $$\secondbigg{ \frac{1 }{\sqrt{T}} \sum_{j=\floor{Tr}+1}^{t_0} (W_j^X - 1/2) }_{r \in [\gamma,1-\gamma]} \ucd  \frac{1}{\sqrt{12}} \secondbigg{B(\tau) - B(r)}_{r \in [\gamma,1-\gamma]}.$$ 
Let $\Z_Y \sim \cP_Y$, denote $ \ W^X_Y \defeq F_{X}(\theta(\Z_Y))$, $\sigma_{XY} \defeq \var (W^X_Y)$. In Lemma \ref{lemma:sigmas_asymp}, we show that, $\sigma_{XY} = 1/12 + O(\delta_T)$. Since $\delta_T \to 0$, we have $\sigma_{XY} \to 1/12$. Also, $\mu = \E W^X_Y$. Therefore, we apply the classical central limit theorem to establish: $$\frac{1}{\sqrt{T}} \sum_{j=t_0+1}^{T-m} (W_j^X - \mu) \cd \frac{1}{\sqrt{12}} [B(1-\epsilon) - B(\tau)], \quad \text{as $\delta_T \to 0$.}$$ In a similar spirit, we define $\Z_X \sim \cP_X$,  $ \ W^Y_X \defeq F_{Y}(\theta(\Z_X))$, $\sigma_{YX} \defeq \var (W^Y_X)$. By symmetry of argument in Lemma \ref{lemma:sigmas_asymp}, it is possible to show $\sigma_{YX} = 1/12 + O(\delta_T)$. Note that for $k<t_0$, $\{W^Y_i\}_{i=m+1}^{k}$ is a sequence of iid random variables with mean $1-\mu$ and variance $\sigma_{YX}$. Using Lemma \ref{lemma:cross_cov_asymp}, we have, $\E (W^X_i W^Y_i) \to 1/3$, as $\delta_T \to 0$. Therefore, it follows that, $\E \thirdbigg{ (1-w_T(k))(1/2 - W_i^X) + w_T(k)(1 - \mu - W_i^Y) } = 0$ and, 
$\var [(1-w_T(k))(1/2 - W_i^X) + w_T(k)(1 - \mu - W_i^Y)] = 1/12 + O(\delta_T).$ Hence, as $\delta_T \to 0$, \begin{flalign*}
    &\secondbigg{ \frac{1}{\sqrt{T}} \sum_{i=m+1}^{\floor{Tr}} \thirdbigg{ (1-w_T(\floor{Tr}))(1/2 - W_i^X) + w_T(\floor{Tr})(1 - \mu - W_i^Y) } }_{r \in [\gamma,1-\gamma]}\\ \ucd& \frac{1}{\sqrt{12}} \{B(r) - B(\epsilon)\}_{r \in [\gamma,1-\gamma]},
\end{flalign*} due to Donsker's Theorem. Now, we apply the continuous mapping theorem and Slutsky's theorem to have the following as $\delta_T\to0$: $$\secondbigg{\sqrt{T}\wt{P}_T(r)}_{r \in [\gamma,1-\gamma]} \ucd \secondbigg{G_0(r)}_{r \in [\gamma,1-\gamma]},$$ where, we also use the fact that, $(T-k-m)/T \to (1-r-\epsilon)$ and $(k-m)/T \to (r-\epsilon)$, as $T\to \infty$. Hence, we complete the proof of \eqref{eq:proj_alt_proof_statement}.

\textbf{\textit{Proof of \eqref{eq:remaind_alt_proof_statement}}:}
We follow the same steps as in the proof of \eqref{eq:remaind_null_proof_statement}. First, we establish the finite-dimensional weak convergence to 0 followed by the proof of stochastic equicontinuity.  For $r \in [\gamma, 1-\gamma]$, define $\wt{R}_T(r) \defeq R_T(\floor{Tr})$.  We recall $R_T(k)$ in \eqref{RT}, and   apply Lemma \ref{lemma:cross_cov_remaind_terms_uniform}, \ref{lemma:cross_cov_remaind_terms}, \ref{lemma:cross_cov_remaind_terms_mix} to have $$\E_* R_T^2(k) = \frac{ \sum_{i=m+1}^{k}\left\{ \sum_{j = k+1}^{t_0} \E_* [\hat{h}^{0}(\Z_i, \Z_j) ]^2+ \sum_{j=t_0+1}^{T-m} \E_* [\hat{h}^{\dagger}(\Z_i, \Z_j)]^2 \right\}}{(k-m)^2(T-m-k)^2} =O_{\P_*}(1/T^2).$$ Therefore, $\sqrt{T} \wt{R}_T(r) = o_{\P_*}(1)$, for   $r \in [\gamma,1-\gamma]$. 
For stochastic equicontinuity, we fix $r, r' \in [\gamma,1-\gamma]: r' < r$, with $k = \floor{Tr}, k' = \floor{Tr'}$, and show the following:\\ $\E \thirdbigg{ \sqrt{T} \firstbigg{\wt{R}_T(r) - \wt{R}_T(r')} }^2 \lesssim T^{-1}.$ Similar to \eqref{eq:diff_rtk_null}, we have \begin{flalign*}
        &\wt{R}_T(r) - \wt{R}_T(r') = R_T(k) - R_T(k') \\
        =& \firstbigg{\frac{1}{(k-m)(T-m-k)} - \frac{1}{(k'-m)(T-m-k')}} \sum_{i=m+1}^{k'}\sum_{j=k+1}^{t_0} \hat{h}^{0}(\Z_i, \Z_j) \\
        & + \frac{1}{(k-m)(T-m-k)} \sum_{i=k'+1}^{k}\sum_{j=k+1}^{t_0} \hat{h}^{0}(\Z_i, \Z_j) \\
        &- \frac{1}{(k'-m)(T-m-k')} \sum_{i=m+1}^{k'}\sum_{j=k+1}^{t_0} \hat{h}^{0}(\Z_i, \Z_j) \\
        &  + \firstbigg{\frac{1}{(k-m)(T-m-k)} - \frac{1}{(k'-m)(T-m-k')}} \sum_{i=m+1}^{k'}\sum_{j=t_0+1}^{T-m} \hat{h}^{\dagger}(\Z_i, \Z_j) \\
        & + \frac{1}{(k-m)(T-m-k)} \sum_{i=k'+1}^{k}\sum_{j=t_0+1}^{T-m} \hat{h}^{\dagger}(\Z_i, \Z_j).
\end{flalign*} Hence, we have written $R_T(k) - R_T(k')$ into terms with zero cross-covariance. Now, we apply Lemma \ref{lemma:cross_cov_remaind_terms_uniform}, \ref{lemma:cross_cov_remaind_terms}, \ref{lemma:cross_cov_remaind_terms_mix} in the following: \begin{align} \label{eq:rtk_diff_var}
\begin{split}
        &\E_* \thirdbigg{\sqrt{T}(R_T(k) - R_T(k'))}^2 \\
        = &\firstbigg{\frac{1}{(k-m)(T-m-k)} - \frac{1}{(k'-m)(T-m-k')}}^2 \sum_{i=m+1}^{k'}\sum_{j=k+1}^{t_0} \E_* [\hat{h}^{0}(\Z_i, \Z_j)]^2 \\
        &   + \frac{1}{(k-m)^2(T-m-k)^2} \sum_{i=k'+1}^{k}\sum_{j=k+1}^{t_0} \E_*[\hat{h}^{0}(\Z_i, \Z_j)]^2 \\
        &   + \frac{1}{(k'-m)^2(T-m-k')^2} \sum_{i=m+1}^{k'}\sum_{j=k+1}^{t_0} \E_* [\hat{h}^{0}(\Z_i, \Z_j)]^2 \\
        &   + \firstbigg{\frac{1}{(k-m)(T-m-k)} - \frac{1}{(k'-m)(T-m-k')}}^2 \sum_{i=m+1}^{k'}\sum_{j=t_0+1}^{T-m} \E_* [\hat{h}^{\dagger}(\Z_i, \Z_j)]^2 \\
        & + \frac{1}{(k-m)^2(T-m-k)^2} \sum_{i=k'+1}^{k}\sum_{j=t_0+1}^{T-m} \E_* [\hat{h}^{\dagger}(\Z_i, \Z_j)]^2 \\
        &  \lesssim 1/T,
        \end{split}
\end{align} where, we use the fact that, $\absbigg{\hat{h}^{0}(\Z_i, \Z_j)}, \absbigg{\hat{h}^{\dagger}(\Z_i, \Z_j)} \leq 4$. By modifying Lemma A.1 in \cite{kley2016quantile} as in Lemma \ref{lemma:kley_et_al_extn}, we establish the stochastic equicontinuity and \eqref{eq:remaind_alt_proof_statement}. 

\textbf{\textit{Proof of \eqref{eq:error_alt_proof_statement}}:}
Recall, \begin{align} \label{eq:recall_atk}
    \begin{split}
        A_T(k) &\defeq \frac{1}{T-k-m} \sum_{j=k+1}^{t_0} (\hat{W}_j^X - W_j^X) + \frac{1}{T-k-m} \sum_{j=t_0+1}^{T-m} (\hat{W}_j^X - W_j^X)  \\
        & \ \ \ \ \ + \frac{1}{k-m} \sum_{i=m+1}^{k} \thirdbigg{ w_T(k)(W_i^Y - \hat{W}_i^Y) + (1-w_T(k))(W_i^X - \hat{W}_i^X) }.
    \end{split}
\end{align} We establish the uniform weak convergence to zero for the three terms separately. For the first term, note that, $\hat{W}^X_j, W^X_j \in [0, 1]$, which implies  for some $\zeta > 0$,  $$\E_* (\hat{W}_j^X - W_j^X)^2 \leq \E_* |\hat{W}_j^X - W_j^X| \leq \E_* \absbigg{\indic\{\hat{\theta}(\Z_X) < \hat{\theta}(\Z_j)\} - \indic\{\theta(\Z_X) < \theta(\Z_j)\}} = O_{\P_*}(T^{-\zeta}),$$ where the second inequality holds due to the triangle inequality and the last equality holds by Assumption \ref{assump:local_alt}(1). Now $\{\hat{W}^X_j - W^X_j\}_{j=m+1}^{t_0}$ is iid with respect to $\P_*$ and $\E_*\hat{W}^X_j = 1/2 = \E W^X_j, \ \forall \ j \in \{m+1, \ldots, t_0\}$. For any fixed $r \in [\gamma,1-\gamma]$, we have for some $\zeta>0$, $$\var_* \left\{ \frac{1}{\sqrt{T}} \sum_{j=\floor{Tr}+1}^{t_0} (\hat{W}_j^X - W_j^X)\right\} = \frac{1}{T} \sum_{j=\floor{Tr}+1}^{t_0} \E_*(\hat{W}_j^X - W_j^X)^2 = O_{\P_*}(T^{-\zeta}),$$ where the first equality holds due to independence. Therefore, the finite-dimensional convergence holds for the first term in \eqref{eq:recall_atk}.
To establish stochastic equicontinuity using Lemma \ref{lemma:kley_et_al_extn}, it suffices to show for any $r, r' \in [\gamma,1-\gamma]: k' =\floor{Tr'} < \floor{Tr} =k$, we have  \begin{flalign*}
    &\var_*\left\{\frac{1}{\sqrt{T}}\left[ \sum_{j=\floor{Tr}+1}^{t_0} (\hat{W}_j^X - W_j^X) - \sum_{j=\floor{Tr'}+1}^{t_0} (\hat{W}_j^X - W_j^X) \right]\right\} \\
    &=\frac{1}{T} \sum_{j=\floor{Tr'+1}}^{\floor{Tr}} \E_*(\hat{W}_j^X - W_j^X)^2 = O_{\P_*}(T^{-\zeta}),
\end{flalign*} for some $\zeta>0$, where we use independence and Assumption \ref{assump:local_alt}(1). Therefore, we have established that the first term in \eqref{eq:recall_atk} converges uniformly to zero. 
$\E_* \hat{W}^X_j = \mu_*, \ \E W^X_j = \mu $ for $j>t_0$. By Assumption \ref{assump:local_alt}(2), we have, $\mu_* - \mu = o_{\P_*}(1/\sqrt{T})$. Hence, for $j\in \{t_0+1, \ldots, T-m\}$, we follow similar steps to show the finite-dimension weak convergence since $\var_* \{ \frac{1}{\sqrt{T}} \sum_{j=t_0+1}^{T-m} (\hat{W}_j^X - W_j^X) \}^2 = o_{\P_*}(1)$ due to Assumption \ref{assump:local_alt}. 
The proof of the third term in \eqref{eq:recall_atk} is similar. Hence, combining all the above results, we complete the proof of \eqref{eq:error_alt_proof_statement}, and hence we conclude the proof of Theorem \ref{thm:unif_weak_conv_alt}.
\end{proof}

\subsection{Proof of Theorem \ref{thm:fixed_alt}} \label{subsec:fixed_alt_proof}
The proof of Theorem \ref{thm:fixed_alt} borrows some major ideas presented in the proof of Theorem \ref{thm:unif_weak_conv_alt}. However, as stated in Assumption \ref{assump:fixed_alt}, it holds under a much-relaxed condition.

\begin{proof}
Note that $\sup_{k \in \calI_{cp}} \secondbigg{ \sqrt{T} \firstbigg{ \hat{\Psi}(k) - \frac{1}{2} } } \geq \sqrt{T} \firstbigg{ \hat{\Psi}(t_0) - \frac{1}{2} }.$ Hence, it is enough to show that, under fixed alternative, the right-hand side diverges to infinity as $T\to \infty$. 
Following similar steps as in \eqref{eq:ultimate_decomp_alt} in the Proof of Theorem \ref{thm:unif_weak_conv_alt}, we write: \begin{align} \label{eq:ultimate_decomp_alt_fixed}
    \begin{split}
        \hat{\Psi}(t_0) - \frac{1}{2} = \underbrace{(\mu - 1/2)}_\text{Signal} + \underbrace{P_T(t_0)}_\text{Projection} + \underbrace{R_T(t_0)}_\text{Remainder} + \underbrace{(\mu - \mu_{*}) + A_T(t_0)}_\text{Approximation Error}, \quad \text{where},
    \end{split}
\end{align} \begin{flalign*}
    P_T(t_0) =& \frac{1}{T-m-t_0} \sum_{j=t_0+1}^{T-m} \firstbigg{W^X_j - \mu} - \frac{1}{t_0-m} \sum_{i=m+1}^{t_0}\secondbigg{ W^Y_i - (1-\mu) }, \\R_T(t_0) = &\frac{1}{(t_0-m)(T-m-t_0)} \sum_{i=m+1}^{t_0} \sum_{j=t_0+1}^{T-m} \hat{h}^{\dagger}(\Z_i, \Z_j), \\A_T(t_0) =& \frac{1}{T-m-t_0} \sum_{j=t_0+1}^{T-m} \firstbigg{ \hat{W}^X_j - W^X_j } - \frac{1}{t_0-m} \sum_{i=m+1}^{t_0} \firstbigg{ \hat{W}^Y_i - W^Y_i }.
\end{flalign*}
Additionally,  let us define: \begin{flalign*}
    A_T^{'}(t_0) &\defeq A_T(t_0) - 2(\mu_{*} - \mu) \\=& \frac{1}{T-m-t_0} \sum_{j=t_0+1}^{T-m} \left\{ \firstbigg{ \hat{W}^X_j - W^X_j } - (\mu_{*} - \mu) \right\} \\
    &- \frac{1}{t_0-m} \sum_{i=m+1}^{t_0} \left\{ \firstbigg{ \hat{W}^Y_i - W^Y_i } - (\mu - \mu_{*})\right \},
\end{flalign*} so that, $\E_{*} A_T^{'}(t_0) = 0$. Therefore, we can rewrite \eqref{eq:ultimate_decomp_alt_fixed} as: \begin{align} \label{eq:ultimate_decomp_alt_fixed1}
        \hat{\Psi}(t_0) - \frac{1}{2} = \underbrace{(\mu - 1/2)}_\text{Signal} + \underbrace{P_T(t_0)}_\text{Projection} + \underbrace{R_T(t_0)}_\text{Remainder} + \underbrace{(\mu_{*} - \mu) + A^{'}_T(t_0)}_\text{Approximation Error}.
\end{align} We claim: (1) $\sqrt{T} P_T(t_0) = O_{\P_*}(1), \ (2) \sqrt{T} R_T(t_0) = o_{\P_*}(1) \ \text{and,} \  (3) \sqrt{T} A^{'}_T(t_0) = O_{\P_*}(1)$. Since $t_0$ is fixed, we borrow ideas presented in proving the finite-dimensional convergence results of the respective terms from the Proof of Theorem \ref{thm:unif_weak_conv_alt} in Appendix \ref{subsec:unif_weak_conv_alt_proof}. 

(1) Note  by construction,  $\E_* P_T(t_0)=0$. Using independence of $\{\Z_t\}_{t=m+1}^{T-m}$, we have: $$\E_* \thirdbigg{ \sqrt{T}P_T(t_0) }^2 = \frac{T\E_* \thirdbigg{ W^X_{t_0} - \mu }^2}{T-m-t_0}  + \frac{T\E_* \thirdbigg{ W^Y_{m+1} - (1-\mu) }^2}{t_0-m}  \lesssim \frac{1-2\epsilon}{(\tau-\epsilon)(1-\tau-\epsilon)} < \eta^{-2},$$ for large $T$ since $T/(T-m-t_0) \to 1/(1 - \tau - \epsilon) \leq 1/\eta, \ T/(t_0-m) \to 1/(\tau-\epsilon) \leq 1/\eta$ and expectation terms are bounded. Therefore, $\sqrt{T} P_T(t_0) = O_{\P_*}(1)$, due to Markov's inequality.

(2) \eqref{eq:remaind_alt_proof_statement} implies $\sqrt{T} R_T(t_0) = o_{\P_*}(1)$. 

(3) Now, \begin{flalign} \label{eq:approx_diff_t0}
    A'_T(t_0) 
    &= \frac{\sum_{j=t_0+1}^{T-m} \thirdbigg{ \firstbigg{ \hat{W}^X_j - W^X_j } - (\mu_{*} - \mu) }}{T-m-t_0}  - \frac{\sum_{i=m+1}^{t_0} \thirdbigg{ \firstbigg{ \hat{W}^Y_i - W^Y_i } - (\mu - \mu_{*}) }}{t_0-m}.
\end{flalign} Hence $\E_* A'_T(t_0) = 0$. Utilizing independence of $\{\Z_t\}_{t=m+1}^{T-m}$, similar to the proof of (1), we have $\E_* \thirdbigg{ \sqrt{T}A'_T(t_0) }^2  \lesssim \eta^{-2},$ for large $T$. Hence, $\sqrt{T} A'_T(t_0) = O_{\P_*}(1)$ due to Markov's inequality. Combining, we have $\sqrt{T} \firstbigg{ \hat{\Psi}(t_0) - 1/2 } = \sqrt{T}\thirdbigg{ (\mu - 1/2) + (\mu_{*} - \mu) } + O_{\P_*}(1) = \sqrt{T} \firstbigg{ \hat{\Psi}(t_0) - 1/2 } = \sqrt{T}\thirdbigg{ (\mu - 1/2) + (\mu_{*} - \mu) } + O_{\P_*}(1)$ due to Lemma \ref{lemma:reln_mu_delta}. Furthermore, $\sqrt{T} \firstbigg{ \hat{\Psi}(t_0) - 1/2 } \geq \sqrt{T}[\delta_T/4 - |\mu_{*} - \mu|] + O_{\P_*}(1) \geq \sqrt{T} c_T + O_{\P_*}(1)$ due to Assumption \ref{assump:fixed_alt}. Now, $O_{\P_*}(1) = O_{\P}(1)$ and hence $\sqrt{T} \firstbigg{ \hat{\Psi}(t_0) - 1/2 } \geq \sqrt{T} c_T + O_{\P}(1)$. Since $c_T \gg T^{-1/2}$ under  Assumption \ref{assump:fixed_alt}, the lower bound diverges to $\infty$ in probability, and thus we conclude.
\end{proof}
\subsection{ Proof of Theorem \ref{thm:weak_consistency+} }
Recall, $\calI_{cp} = \{\floor{rT}: r \in [\epsilon + \eta, 1 - \epsilon - \eta]\}$. Without loss of generality, assume $l = T(\epsilon+\eta) \in \naturals$. Then, $\calI_{cp} = \{l+1, \ldots, T-l\}$.
\begin{lemma}(\textbf{Hoeffding's Inequality}) \label{lemma:hoeff_ineq}
    Let $Z_1, \ldots, Z_T$ be independent random variables such that $Z_i \in [a, b], \ \forall i \in [T]$. Then, it follows that, \begin{equation*}
        \P\thirdbigg{ \sum_{i=1}^{T} (Z_i - \E(Z_i)) \geq (b-a) \sqrt{T \log(T)} } \leq \frac{1}{T^2}.
    \end{equation*}
\end{lemma}
\begin{definition}(\textbf{Bounded Difference Property}) \label{defn:bdd_diff_prop}
    A function $f:\reals^n \mapsto \reals$ satisfies the Bounded Difference Property if there exists positive constants $L_1, \ldots , L_n$  such that for all $(x_1, \ldots, x_n)$ in the domain of $f$ and for all $k \in \{1, \ldots, n\}$, \begin{equation*}
        \sup_{x, y} \absbigg{ f(x_1, \ldots, x_{k-1}, x, x_{k+1}, \ldots, x_n) - f(x_1, \ldots, x_{k-1}, y, x_{k+1}, \ldots, x_n) } \leq L_k.
    \end{equation*}
\end{definition}
\begin{lemma}(\textbf{McDiarmid's Inequality}) \label{lemma:bdd_diff_ineq}
    Let $X_1, \ldots, X_n$ be independent random variables, $f:\reals^n \mapsto \reals$ a function that satisfies the Bounded Difference Property as stated in Definition \ref{defn:bdd_diff_prop}, with constants $L_1, \ldots, L_n$, and $Z = f(X_1, \ldots, X_n)$. Then, \begin{equation*}
        \P\firstbigg{Z - \E(Z) \geq \sqrt{\log(T) \sum_{k=1}^{n} L_k^2 }} \leq \frac{1}{T^2}.
    \end{equation*}
\end{lemma}
We omit the proof of Lemma \ref{lemma:hoeff_ineq} and \ref{lemma:bdd_diff_ineq}. Now, we are ready to prove Theorem \ref{thm:weak_consistency+}. 
\begin{proof}
From \eqref{eq:ultimate_decomp_alt} we recall the following decomposition for any $k \in \calI_{cp}: k < t_0$, \begin{flalign*}
    \hat{\Psi}(k) - \frac{1}{2} = \underbrace{ w_T(k) (\mu - 1/2) }_\text{Signal} + \underbrace{ P_T(k) }_\text{Projection} + \underbrace{ R_T(k) }_\text{Remainder} + \underbrace{ w_T(k) (\mu - \mu_*) + A_T(k) }_\text{Approximation Error}.
\end{flalign*} 
The above decomposition also holds for all $k \in \calI_{cp}: k \geq t_0$. For simplicity, we focus on the case  $k \in \calI_{cp}: k < t_0$, where we note that, \begin{equation} \label{eq:expr_wtk}
    w_T(k) \defeq \begin{cases}
        (T-m-t_0)/(T-m-k) \quad &\text{if} \quad k < t_0, \\
        (t_0-m)/(k-m) \quad &\text{if} \quad k \geq t_0.
    \end{cases}
\end{equation} Let us recall $A_T^{'}(k) \defeq A_T(k) - 2\omega_T(k)(\mu_{*} - \mu)$ so that, $\E_{*} A_T^{'}(k) = 0$. Therefore, we have the following decomposition, for any $k \in \calI_{cp}: k < t_0$, \begin{align} \label{eq:auc_diff_decomp}
    \begin{split}
        \hat{\Psi}(t_0) - \hat{\Psi}(k) &= \underbrace{ (\mu - 1/2) (1 - w_T(k)) }_\text{Signal} + \underbrace{ P_T(t_0) - P_T(k) }_\text{Projection} + \underbrace{ R_T(t_0) - R_T(k) }_\text{Remainder} \\
        & \ \ + \underbrace{(\mu_{*} - \mu)(1 - w_T(k))}_\text{Approximation Error of Signal} + \underbrace{  A^{'}_T(t_0) - A^{'}_T(k) }_\text{Approximation Error}.
    \end{split}
\end{align} We first outline the overall structure of the proof. We show that the terms `Projection', `Remainder', and `Approximation Error'  concentrate toward 0 with high probability, faster than the `Signal' and `Approximation Error of Signal'. Then, we use the standard M-estimation argument to argue that the estimated location of the change-point is consistent as long as the term `Approximation Error of Signal'  remains smaller than `Signal'. In what follows, we state the concentration inequality of the three terms above. Then, we use the stated claims to complete the proof. In the end, we prove our claims.

\begin{itemize}
    \item \textbf{High-probability Bound for the Projection Term:} \begin{equation} \label{eq:conc_proj}
        \P\underbrace{ \thirdbigg{ \bigcap_{k \in \calI_{cp}} \secondbigg{ P_T(t_0) - P_T(k) \geqsim - \frac{\sqrt{|t_0 - k| \log(T)}}{T} } }}_\text{$:=\mc{A}_1$}  \geq 1-\frac{1}{T}.
    \end{equation}
    \item \textbf{High-probability Bound for the Approximation Error Term:} \begin{equation} \label{eq:conc_approx}
        \P_{*}\underbrace{ \thirdbigg{ \bigcap_{k \in \calI_{cp}} \secondbigg{ A^{'}_T(t_0) - A^{'}_T(k) \geqsim - \frac{\sqrt{|t_0 - k| \log(T)}}{T} } }}_\text{$:=\mc{A}_2$} \geq 1-\frac{1}{T}.
    \end{equation}
    \item \textbf{High-probability Bound for the Remainder Term:} \begin{equation} \label{eq:conc_remaind}
        \P_{*}\underbrace{\thirdbigg{  \bigcap_{k \in \calI_{cp}} \secondbigg{  R_T(t_0) - R_T(k) \geqsim - \frac{\sqrt{|t_0-k| \log(T)}}{T} } } }_\text{$:=\mc{A}_3$} \geq 1-\frac{1}{T}.
    \end{equation}
\end{itemize} Since $(T-m-k), (k-m) < T, \ \forall k \in \calI_{cp}$, we have, $1-w_T(k) > |t_0-k|/T, \ \forall k, t_0 \in \calI_{cp}$. In addition, we have $(\mu - 1/2) = \delta_T/4$ due to Lemma \ref{lemma:reln_mu_delta}. Therefore, we can that, $\forall k \in \calI_{cp}$: \begin{align*} 
    \begin{split}
        \hat{\Psi}(t_0) - \hat{\Psi}(k) &\geq \frac{|t_0 - k|}{T} \firstbigg{ \frac{\delta_T}{4} - |\mu - \mu_*| } + \underbrace{ P_T(t_0) - P_T(k) }_\text{Projection} + \underbrace{ R_T(t_0) - R_T(k) }_\text{Remainder}  \\
        &+ \underbrace{ A^{'}_T(t_0) - A^{'}_T(k) }_\text{Approximation Error}. \quad \text{Now},
    \end{split}
\end{align*} $$\bigcap_{i=1}^{3} \mc{A}_i \implies\underbrace{  \bigcap_{k \in \calI_{cp}} \secondbigg{ \hat{\Psi}(t_0) - \hat{\Psi}(k) \geqsim \frac{|t_0 - k|}{T} \firstbigg{ \frac{\delta_T}{4} - |\mu - \mu_*| } - \frac{\sqrt{|t_0 - k| \log(T)}}{T} } }_\text{$:=\mc{B}_1$},$$  \begin{align} \label{eq:prob_b1}
\P_*(\mc{B}_1) \geq \P_*\firstbigg{ \bigcap_{i=1}^{3} \mc{A}_i } \geq 1 - 3/T.
\end{align} We recall that $c_T\gg\sqrt{\log T/T}$, and  define two events: \begin{flalign*}
    \mc{B}_2 \defeq \secondbigg{  \frac{\delta_T}{4} - |\mu_* - \mu|  > c_T  } \quad \mc{B}_3\defeq \bigcap_{k \in \calI_{cp}}\left\{\hat{\Psi}(t_0) - \hat{\Psi}(k) \geqsim \frac{|t_0 - k|}{T} c_T - \frac{\sqrt{|t_0 - k| \log(T)}}{T}\right \}.
\end{flalign*} Note that, $\mc{B}_1 \cap \mc{B}_2 \subset \mc{B}_3$. By Assumption \ref{assump:fixed_alt}, we have $\P(\mc{B}_2) \geq 1 - o(1)$. By \eqref{eq:prob_b1} we have $\P(\mc{B}_1) = \P_*(\mc{B}_1) \geq 1 - 3/T$. Therefore, $\P(\mc{B}_3) \geq \P(\mc{B}_1 \cap \mc{B}_2) \geq \P(\mc{B}_1) + \P(\mc{B}_2) - 1 \geq 1 - o(1)$. Note that if $\mathcal{B}_3$ holds, then we have that 
$$
\left\{k\in \calI_{cp}:\frac{|t_0 - k|}{T} c_T - \frac{\sqrt{|t_0 - k| \log(T)}}{T} \geq 0 \right\}\subset \left\{k\in \calI_{cp}: \hat{\Psi}(t_0) > \hat{\Psi}(k) \right\},
$$
$$
\text{i.e., under} \ \mc{B}_3, \quad \left\{k\in \calI_{cp}:  \hat{\Psi}(k) \geq \hat{\Psi}(t_0) \right\}\subset \left\{k\in \calI_{cp}:\frac{|t_0 - k|}{T} c_T - \frac{\sqrt{|t_0 - k| \log(T)}}{T}< 0 \right\},
$$
Recall, $\hat{R}_T = \argmax_{k \in \calI_{cp}} \hat{\Psi}(k)$ so that $\hat{\Psi}(\hat{R}_T)\geq \hat{\Psi}(t_0)$, therefore $ \absbigg{\hat{R}_T - t_0} < \log(T)/c_T^2 ,$ hence the theorem holds since $\P(\mc{B}_3) \geq 1 - o(1).$ 
Now, we prove our claims. 

\textbf{\textit{Proof of \eqref{eq:conc_proj}}:}
The projection term in \eqref{eq:auc_diff_decomp} can be written $ \forall k \in \calI_{cp}: k < t_0$ as: \begin{align}
    \begin{split}
        &P_T(t_0) - P_T(k) = \thirdbigg{ \frac{1}{T-m-t_0} - \frac{1}{T-m-k} } \sum_{i=t_0+1}^{T-m} \firstbigg{W_i^X - \mu} \\
        &   + \sum_{i=m+1}^{k} \underbrace{ \thirdbigg{ \frac{1}{t_0-m}\firstbigg{1-\mu-W_i^Y} - \frac{1}{k-m}\secondbigg{ (1-w_T(k))(1/2 - W_i^X) + w_T(k)(1 - \mu - W_i^Y) } } }_\text{$A(i)$} \\
        & + \sum_{i=k+1}^{t_0} \underbrace{ \thirdbigg{ \frac{1}{t_0-m} (1 - \mu - W_i^Y) + \frac{1}{T-k-m}(1/2 - W_i^X) } }_\text{$B(i)$}.
    \end{split}
\end{align} The rest of the proof involves carefully applying Lemma \ref{lemma:hoeff_ineq} and a union bound on each of the above three terms. Let us consider the first term above. 

Since, $\{W_i^X\}_{i=t_0+1}^{T-m} \in [0, 1]$, we apply Lemma \ref{lemma:hoeff_ineq} on $\{-(W_i^X - \mu)\}_{i=t_0+1}^{T-m}$ to have, \begin{equation} \label{eq:proj_conc_first}
    \P\thirdbigg{ \thirdbigg{ \frac{1}{T-m-t_0} - \frac{1}{T-m-k} } \sum_{i=t_0+1}^{T-m} \firstbigg{W_i^X - \mu} \leq - \frac{ (t_0 - k) \sqrt{\log(T)} }{(T-m-k) \sqrt{(T-m-t_0)}} } \leq \frac{1}{T^2}.
\end{equation} Next, for the second term, since $\{W_i^X\}_{i=m+1}^{k}, \ \{W_i^Y\}_{i=m+1}^{k} \in [0, 1]$, we have \begin{equation*}
    \frac{-\mu}{t_0-m} - \frac{1-w_T(k)}{2(k-m)} - \frac{w_T(k) (1-\mu)}{k-m} \leq A(i) \leq \frac{1 - \mu}{t_0 - m} + \frac{1 - w_T(k)}{2(k-m)} + \frac{w_T(k) \mu}{k-m}.
\end{equation*} In a similar spirit, we apply Lemma \ref{lemma:hoeff_ineq} on $\{-A(i)\}_{m+1}^{k}$ to have the following, \begin{equation} \label{eq:proj_conc_second}
    \P\thirdbigg{ \sum_{i=m+1}^{k} A(i) \leq -\firstbigg{ \frac{1}{t_0-m} + \frac{1}{k-m} } \sqrt{ (k-m) \log(T) } } \leq \frac{1}{T^2}.
\end{equation} Similarly, we note that $\{W_i^X\}_{i=k+1}^{t_0} \in [0, 1]$, we have the following \begin{equation*}
    \frac{-\mu}{t_0 - m} + \frac{-1}{2(T-k-m)} \leq B(i) \leq \frac{1-\mu}{t_0 - m} + \frac{1}{2(T-k-m)}.
\end{equation*}
\begin{equation} \label{eq:proj_conc_third}
    \text{Hence,} \quad \P\thirdbigg{ \sum_{i=k+1}^{t_0} B(i) \leq -\firstbigg{ \frac{1}{t_0-m} + \frac{1}{T-k-m} } \sqrt{ (t_0-k) \log(T) } } \leq \frac{1}{T^2}.
\end{equation} Combining \eqref{eq:proj_conc_first}, \eqref{eq:proj_conc_second} and \eqref{eq:proj_conc_third}, we have, \begin{equation}
    \P\thirdbigg{ P_T(t_0) - P_T(k) \leq - p_T(k; t_0, m) } \leq \frac{3}{T^2}, \quad \text{where},
\end{equation} \begin{align*}
    p_T(k; t_0, m) &= \frac{ (t_0 - k) \sqrt{\log(T)} }{(T-m-k) \sqrt{(T-m-t_0)}} + \firstbigg{ \frac{1}{t_0-m} + \frac{1}{k-m} } \sqrt{ (k-m) \log(T) } \\
        & \ \ + \firstbigg{ \frac{1}{t_0-m} + \frac{1}{T-k-m} } \sqrt{ (t_0-k) \log(T) }.
\end{align*} Now, we recall that $k \in \calI_{cp}: k < t_0$ and $\calI_{cp} :=\{l+1, \ldots, T-l\}$. In what follows, we obtain a uniform upper bound of $p_T(k; t_0, m)$ that holds for all $k \in \calI_{cp}: k < t_0$: \begin{align} \label{eq:ptk_ubd}
    \begin{split}
        p_T(k; t_0, m) &= \frac{\sqrt{(t_0 - k) \log(T)}}{T} \left[ \frac{T\sqrt{(t_0-k)}}{(T-m-k)\sqrt{T-m-t_0}} + \firstbigg{ \frac{T}{t_0-m} + \frac{T}{k-m} } \sqrt{\frac{k-m}{t_0-k}} \right. \\
        & \left. \ \ + \firstbigg{ \frac{T}{t_0-m} + \frac{T}{T-k-m} } \right] \\
        & \leq \frac{\sqrt{(t_0 - k) \log(T)}}{T} \left[ \frac{T\sqrt{(t_0-l)}}{(T-m-t_0)\sqrt{T-m-t_0}} + \firstbigg{ \frac{T}{t_0-m} + \frac{T}{l-m} } \sqrt{\frac{t_0-m}{t_0-l}} \right. \\
        & \left. \ \ + \firstbigg{ \frac{T}{t_0-l} + \frac{T}{T-t_0-m} } \right] \\
        & \leq C_1 \frac{\sqrt{(t_0 - k) \log(T)}}{T},
    \end{split}
\end{align} for some constant $C_1 > 0$. The last inequality holds since, $(l-m), (t_0 - l), (t_0-m)$ are of same order as $T$. More importantly, $C_1$ only depends on $\epsilon, \eta$ and $\tau$. Therefore, \begin{equation*}
    \P\thirdbigg{ P_T(t_0) - P_T(k) \leq - C_1\frac{\sqrt{(t_0 - k) \log(T)}}{T} } \leq P\thirdbigg{ P_T(t_0) - P_T(k) \leq - p_T(k; t_0, m) } \leqsim \frac{1}{T^2},
\end{equation*} and we apply the union bound, \begin{equation} \label{eq:boole_1}
    \P\thirdbigg{ \bigcup_{k \in \calI_{cp}: k < t_0} \secondbigg{ P_T(t_0) - P_T(k) \leqsim - \frac{\sqrt{(t_0 - k) \log(T)}}{T} } } \leqsim \frac{t_0-m}{T^2} < \frac{t_0}{T^2}.
\end{equation} Following similar steps, it is possible to show the following $\forall k\in \calI_{cp}: k \geq t_0$: \begin{equation*}
    \P\thirdbigg{ P_T(t_0) - P_T(k) \leqsim - \frac{\sqrt{(k - t_0) \log(T)}}{T} } \leqsim \frac{T-t_0}{T^2}.
\end{equation*} We apply the union bound again and combine with \eqref{eq:boole_1} to obtain the following, \begin{equation} \label{eq:high_prob_proj_diff}
    \P\thirdbigg{ \bigcup_{k\in \calI_{cp}} \secondbigg{ P_T(t_0) - P_T(k) \leqsim - \frac{\sqrt{|t_0 - k| \log(T)}}{T} } } \leqsim \frac{t_0}{T^2} + \frac{T-t_0}{T^2} = \frac{1}{T}.
\end{equation} 

\textbf{\textit{Proof of \eqref{eq:conc_approx}}:}
Proof of \eqref{eq:conc_approx} follows a similar argument as the proof of \eqref{eq:conc_proj}; hence, we explain the similarity and omit a detailed proof to avoid repetition. 
For all $k \in \calI_{cp}: k < t_0$, we can write the approximation error term in \eqref{eq:auc_diff_decomp} as, \begin{align*}
    \begin{split}
        A^{'}_T(k) &= \frac{1}{T-k-m} \thirdbigg{ \sum_{j=k+1}^{t_0} (\hat{W}_j^X - W_j^X) + \sum_{j=t_0+1}^{T-m} \secondbigg{ (\hat{W}_j^X - W_j^X) - (\mu_* - \mu) } } \\
        & \ \ \ \ \ + \frac{1}{k-m} \sum_{i=m+1}^{k} \thirdbigg{(1-w_T(k))(W_i^X - \hat{W}_i^X)+ w_T(k) \secondbigg{ (W_i^Y - \hat{W}_i^Y) - (\mu_* - \mu) } },
    \end{split}
\end{align*} and we recall \eqref{eq:approx_diff_t0}.  Then, \begin{flalign*}
    \begin{split}
        & A^{'}_T(t_0) - A^{'}_T(k) \\
        & = \thirdbigg{ \frac{1}{T-m-t_0} - \frac{1}{T-m-k} } \sum_{i=t_0+1}^{T-m} \secondbigg{ (\hat{W}_j^X - W_j^X) - (\mu_* - \mu) } \\
        & \ \ \ + \sum_{i=m+1}^{k}  \frac{1}{t_0-m} \secondbigg{ \firstbigg{ W^Y_i - \hat{W}^Y_i } - (\mu_{*} - \mu) } \\
        & \ \ \ - \sum_{i=m+1}^{k} \frac{1}{k-m}\secondbigg{ (1-w_T(k))(W_i^X - \hat{W}_i^X)+ w_T(k) \secondbigg{ (W_i^Y - \hat{W}_i^Y) - (\mu_* - \mu) } }  \\
        & \ \ \ + \sum_{i=k+1}^{t_0} \thirdbigg{ \frac{1}{t_0-m} \secondbigg{ \firstbigg{ W^Y_i - \hat{W}^Y_i } - (\mu_{*} - \mu) } + \frac{1}{T-k-m} (\hat{W}_j^X - W_j^X) }.
    \end{split}
\end{flalign*} Since, $\E_* \thirdbigg{ A^{'}_T(t_0) - A^{'}_T(k) } = 0$, we use a crude bound of $[-4, 4]$ for each term inside the summation, with respect to $\P_*$. Following similar steps as before, we apply Lemma \ref{lemma:hoeff_ineq} with the union bound repeatedly followed by repeating the same after extending the above decomposition for all $k \in \calI_{cp}: k \geq t_0$, we have the following: \begin{equation} \label{eq:high_prob_approx_diff}
    \P_*\thirdbigg{ \bigcup_{k\in \calI_{cp}} \secondbigg{ A^{'}_T(t_0) - A^{'}_T(k) \leqsim - \frac{\sqrt{|t_0 - k| \log(T)}}{T} } } \leqsim \frac{1}{T}.
\end{equation} 

\textbf{\textit{Proof of \eqref{eq:conc_remaind}}:}
The main challenge in proving \eqref{eq:conc_remaind} is that it cannot be written as a sum of independent random variables. Instead, it is an incomplete U-statistic of independent random variables with respect to $\P_*$. The rest of the proof involves a careful decomposition of $R_T(t_0) - R_T(k)$ into three independent incomplete U-statistics, followed by application of Lemma \ref{lemma:bdd_diff_ineq} on each incomplete U-statistic, after satisfying the Bounded Difference Property as outlined in Definition \ref{defn:bdd_diff_prop}. Note that, we can write the following for all $k \in \calI_{cp}: k < t_0$: \begin{equation} \label{eq:diff_rtk_consistency}
    \begin{split}
        &R_T(t_0) - R_T(k) \\
        & = \firstbigg{ \frac{1}{(t_0 - m)(T-m-t_0)} - \frac{1}{(k-m)(T-m-k)} } \sum_{i=m+1}^{k} \sum_{j=t_0+1}^{T-m} \hat{h}^{\dagger}(\Z_i, \Z_j) \\
        & \ \ + \frac{1}{(t_0-m)(T-m-t_0)} \sum_{i=k+1}^{t_0} \sum_{j=t_0+1}^{T-m} \hat{h}^{\dagger}(\Z_i, \Z_j) \\
        & \ \ - \frac{1}{(k-m)(T-m-k)} \sum_{i=m+1}^{k} \sum_{i=k+1}^{t_0} \hat{h}^{0}(\Z_i, \Z_j) \\
        & = R_1(k) + R_2(k) + R_3(k), \quad \text{where,}
    \end{split}
\end{equation} \begin{equation*}
    \begin{split}
        R_1(k) &= - \frac{(t_0-k)(T-k-t_0)}{(t_0 - m)(T-m-t_0) (k-m)(T-m-k)}  \sum_{i=m+1}^{k} \sum_{j=t_0+1}^{T-m} \hat{h}^{\dagger}(\Z_i, \Z_j), \\
        R_2(k) &= -\frac{1}{(t_0-m)(T-m-t_0)} \sum_{i=k+1}^{t_0} \sum_{j=t_0+1}^{T-m} \hat{h}^{\dagger}(\Z_i, \Z_j), \\
        R_3(k) &= -\frac{1}{(k-m)(T-m-k)} \sum_{i=m+1}^{k} \sum_{i=k+1}^{t_0} \hat{h}^{0}(\Z_i, \Z_j).
    \end{split}
\end{equation*} We now focus on the three terms separately. Let us consider the first term, i.e., $R_1(k)$. Denote $\hat{V}_t := \hat{\theta}(\Z_t)$. We can write $R_1$ as a function of $\secondbigg{\hat{V}_t}_{t=m+1}^{k}$ and $\secondbigg{\hat{V}_t}_{t=t_0+1}^{T-m}$, which are independent with respect to $\P_*$. Let \begin{equation*}
    \sum_{i=m+1}^{k} \sum_{j=t_0+1}^{T-m} \hat{h}^{\dagger}(\Z_i, \Z_j) =: f\firstbigg{\hat{V}_{m+1}, \ldots, \hat{V}_{k}, \hat{V}_{t_0+1}, \ldots, \hat{V}_{T-m}},
\end{equation*} and denote $T' = T-2m+k-t_0$. There are $T'$ arguments in $f(\cdot)$. We now show that $f(\cdot)$ satisfies Definition \ref{defn:bdd_diff_prop}.  It follows that, \begin{align*}
    \begin{split}
        &\absbigg{f(x_1, \ldots, x_{l-1}, x, x_{l+1}, \ldots, x_{T'}) - f(x_{m+1}, \ldots, x_{l-1}, y, x_{l+1}, \ldots, x_{T'})} \\
        &\leq 8 \begin{cases}
            T-m-t_0, \quad \text{if} \quad 1 \leq l \leq k-m, \\
            k-m, \quad \text{if} \quad k-m+1 \leq l \leq T',
        \end{cases}
    \end{split}
\end{align*} since, $\absbigg{\hat{h}^{\dagger}(\Z_i, \Z_j)} \leq 4$ for any $i,j \in \{m+1, \ldots, T-m\}: i \neq j$. Therefore, we apply can Lemma \ref{lemma:bdd_diff_ineq}  on $-R_1$ to obtain the following concentration bound: 

\begin{align*}
    \begin{split}
       &\P_*\biggl( R_1(k) \leqsim - \underbrace{\scriptstyle\frac{(t_0-k)(T-k-t_0)\sqrt{\log(T) \bigl( (k-m)(T-m-t_0)^2 + (T-m-t_0)(k-m)^2 \bigr) }}{(t_0 - m)(T-m-t_0) (k-m)(T-m-k)} }_\text{:= $r_T(k;t_0, m)$}  \biggr) \\
        &\leq \frac{1}{T^2} \\
        &\iff \P_*\biggl( R_1(k) \leq -C_2\frac{(t_0-k)\sqrt{\log(T)}}{T\sqrt{T}} \biggr) \leq \frac{1}{T^2}.
    \end{split}
\end{align*}

Following similar steps as in \eqref{eq:ptk_ubd}, it is possible to show that, for some constant $C_2>0$, $r_T(k; t_0, m) \leq C_2 (t_0-k)\sqrt{\log(T)}/(T\sqrt{T})$ holds uniformly for all $k \in \calI_{cp}: k < t_0$. Therefore, \begin{equation*}
    \P_*\firstbigg{ R_1(k)\leqsim - \frac{(t_0-k)\sqrt{\log(T)}}{T\sqrt{T}} } \leq \P_*\firstbigg{ R_1(k) \leqsim - r_T(k; t_0, m) } \leq \frac{1}{T^2}.
\end{equation*} Now, we apply union bound for all $k \in \calI_{cp}: k < t_0$ to have the following: \begin{equation*}
    \P_*\firstbigg{ \bigcup_{k \in \calI_{cp}: k < t_0} \secondbigg{ R_1(k) \leqsim - C_2\frac{(t_0-k)\sqrt{\log(T)}}{T\sqrt{T}} } } \leq \frac{t_0}{T^2}.
\end{equation*} Following similar steps, it is possible to show, \begin{equation*}
    \P_*\firstbigg{ \bigcup_{k \in \calI_{cp}: k \geq t_0} \secondbigg{ R_1(k) \leq  - C_3\frac{(k-t_0)\sqrt{\log(T)}}{T\sqrt{T}} } } \leq \frac{T-t_0}{T^2},
\end{equation*} for some constant $C_3>0$. Therefore, we have \begin{equation} \label{eq:conc_r1}
    \P_*\firstbigg{ \bigcup_{k \in \calI_{cp}} \secondbigg{ R_1(k) \leqsim -\frac{\absbigg{k-t_0}\sqrt{\log(T)}}{T\sqrt{T}} } } \leq \frac{1}{T}
\end{equation} We now consider $R_2(k)$ and omit the proof for $R_3(k)$ as it can be shown using similar steps.

Let us recall, \begin{equation*}
    R_2(k) := \frac{1}{(t_0-m)(T-m-t_0)} \underbrace{ \sum_{i=k+1}^{t_0} \sum_{j=t_0+1}^{T-m} \hat{h}^{\dagger}(\Z_i, \Z_j) }_\text{$:= f(\hat{V}_{k+1}, \ldots, \hat{V}_{T-m})$}.
\end{equation*} Denote $T' = T-m-k$ since $f(\cdot)$ has $T'$ many arguments. We observe that, \begin{align*}
    \begin{split}
        &\absbigg{f(x_1, \ldots, x_{l-1}, x, x_{l+1}, \ldots, x_{T'}) - f(x_{m+1}, \ldots, x_{l-1}, y, x_{l+1}, \ldots, x_{T'})} \\
        &\leq 8 \begin{cases}
            T-m-t_0, \quad \text{if} \quad 1 \leq l \leq t_0-k, \\
            t_0-k, \quad \text{if} \quad t_0-k+1 \leq l \leq T'.
        \end{cases}
    \end{split}
\end{align*} Therefore, we apply Lemma \ref{lemma:bdd_diff_ineq} on $-R_1$ to obtain the following concentration bound: \begin{align*}
    \begin{split}
        &\P_*\firstbigg{ R_2(k) \leqsim -\frac{ \sqrt{\log(T) \firstbigg{ (t_0-k)(T-m-t_0)^2 + (T-m-t_0)(t_0-k)^2 }}}{(t_0 - m)(T-m-t_0)}  } \leq \frac{1}{T^2} \\
        &\iff \P_*\firstbigg{ R_2(k) \leq -C_4\frac{\sqrt{(t_0-k)}\sqrt{\log(T)}}{T} } \leq \frac{1}{T^2},
    \end{split}
\end{align*} for some constant $C_4>0$ that does not depend on $k$. Applying union bound and following similar steps, it is possible to show: \begin{equation} \label{eq:conc_r2}
    \P_*\firstbigg{ \bigcup_{k \in \calI_{cp}} \secondbigg{ R_2(k) \leqsim -\frac{\sqrt{\absbigg{k-t_0}}\sqrt{\log(T)}}{T} } } \leq \frac{1}{T}
\end{equation} For $R_3(k)$, we can follow the same steps to arrive at the following concentration bound: \begin{equation} \label{eq:conc_r3}
    \P_*\firstbigg{ \bigcup_{k \in \calI_{cp}} \secondbigg{ R_3(k) \leqsim -\frac{\sqrt{\absbigg{k-t_0}}\sqrt{\log(T)}}{T} } } \leq \frac{1}{T}
\end{equation} Combining \eqref{eq:conc_r1}, \eqref{eq:conc_r2} and \eqref{eq:conc_r3} we prove our claim since, \begin{align*}
    \begin{split}
        &\bigcap_{k \in \calI_{cp}}\bigcap_{i=1}^{3}\secondbigg{ R_i(k) \geqsim -\frac{\sqrt{\absbigg{k-t_0}}\sqrt{\log(T)}}{T} } \\
        &\implies  \bigcap_{k \in \calI_{cp}}\secondbigg{ R_T(t_0) - R_T(k) \geqsim -\frac{\sqrt{\absbigg{k-t_0}}\sqrt{\log(T)}}{T} }.
    \end{split}
\end{align*}
 
\end{proof}

\section{Technical Lemmas} \label{sec:lemmas}
Define, $\hat{h}^0(\Z_1, \Z_2) \defeq \indic\{\hat{\theta}(\Z_1) < \hat{\theta}(\Z_{2})\} - \hat{W}^X_2 + \hat{W}^X_1 - 1/2$ and, $\hat{h}^{\dagger}(\Z_1, \Z_2) \defeq \indic\{\hat{\theta}(\Z_1) < \hat{\theta}(\Z_{2})\} - \hat{W}^X_2 + \hat{W}^Y_1 - 1 + \mu_*$. The proof of Lemma \ref{lemma:cross_cov_remaind_terms_uniform}, \ref{lemma:cross_cov_remaind_terms} and \ref{lemma:cross_cov_remaind_terms_mix} can be shown using similar steps. To avoid repetitions, we only include the proof of Lemma \ref{lemma:cross_cov_remaind_terms_mix}.

\begin{lemma} \label{lemma:cross_cov_remaind_terms_uniform}
    Suppose, $\Z_i, \Z_{i'}, \Z_j, \Z_{j'} \simiid \cP_X$. Then, $\E_* \hat{h}^0(\Z_i, \Z_j) \hat{h}^0(\Z_{i'}, \Z_{j'}) = 5/12$ if $(i,j) = (i' , j')$ and $0$ otherwise. 
\end{lemma}

\begin{lemma} \label{lemma:cross_cov_remaind_terms}
    Suppose, $\Z_i, \Z_{i'} \simiid \cP_X$ and $\Z_j, \Z_{j'} \simiid \cP_Y$ are independent random elements. Then, $\E_{*} \hat{h}^{\dagger}(\Z_i, \Z_j) \hat{h}^{\dagger}(\Z_{i'}, \Z_{j'}) =  O(1)$ if $(i, j) = (i', j')$ and $0$ otherwise.
\end{lemma}

\begin{lemma} \label{lemma:cross_cov_remaind_terms_mix} (Proof in  Appendix \ref{subsec:cross_cov_remaind_terms_mix_proof})
    Suppose, $\Z_i, \Z_j \simiid \cP_X$ and $\Z_{j'} \simiid \cP_Y$ are independent random elements. Then, $\E_{*} \hat{h}^{0}(\Z_{i}, \Z_{j}) \hat{h}^{\dagger}(\Z_{i}, \Z_{j'}) = 0.$
\end{lemma}

\begin{lemma} \label{lemma:kley_et_al_extn} (Proof in Appendix \ref{subsec:kley_et_al_extn_proof})
    Let $\gamma\in (0, 1/2)$. Define the index set $\mathcal{T}_T=\{k/T: k\in [\gamma T,1-\gamma T]\cap \mathbb{Z}\}$. Let, $\{G_T(r): r \in \mc{T}_T\}$ be a separable stochastic process with $\E \thirdbigg{ G_T(r) - G_T(r') }^2 < C/T$ for any $r, r' \in \mc{T}_T$ and some constant $C>0$. Then, $\{G_T(r): r \in \mc{T}_T\}$ is stochastic equicontinuous, i.e., for any $x > 0$, it follows that, \begin{equation*}
        \lim_{\delta\to 0}\limsup_{T\to\infty}  \mathbb{P}\left(\sup_{|r-r'|\leq \delta, \gamma\leq r,r'\leq 1-\gamma}| G_T(r) - G_T(r') |>x \right)=0.
    \end{equation*}
\end{lemma}

\begin{lemma} \label{lemma:reln_mu_delta} (Proof in Appendix \ref{subsec:reln_mu_delta_proof})
    Recall that we define $\delta \defeq \E \absbigg{ L(\Z_X) - L(\Z'_X) }$, and $\mu = \P\firstbigg{\theta(\Z_X < \theta(\Z_Y))}$ for independent $\Z_X, \Z'_X \simiid \cP_X$, $\Z_Y \sim \cP_Y$. Then,  $ \delta=4\mu - 2$.
\end{lemma}

\begin{lemma} \label{lemma:sigmas_asymp} (Proof in Appendix \ref{subsec:sigmas_asymp_proof})
    For independent $\Z_X, \Z'_X \simiid \cP_X$ and $\Z_Y, \Z'_Y \sim \cP_Y$,  let  $\E F_{X}^{2}(\theta(\Z_Y)) = \P(\theta(\Z_X)<\theta(\Z_Y), \theta(\Z_X') < \theta(\Z_Y)) \defeq \mu_{XY}$, $\E F_{Y}^{2}(\theta(\Z_X))) = \P(\theta(\Z_Y)<\theta(\Z_X), \theta(\Z_Y') < \theta(\Z_X)) \defeq \mu_{YX}$ and, $\var F_{X}( \theta(\Z_Y)) = \mu_{XY} - \mu^2 \defeq \sigma_{XY}$ and \\$\var F_{Y}( \theta(\Z_X)) = \mu_{YX} - (1-\mu)^2 \defeq \sigma_{YX}$. Then, it follows that, $\sigma_{XY} = 1/12 + O(\delta) = \sigma_{YX}$ and $\mu_{XY} = 1/3 + O(\delta) = \mu_{YX},$ where, $\delta= \E \absbigg{ L(\Z_X) - L(\Z'_X) }$, as defined in \eqref{eq:defn_delta}.
\end{lemma}

\begin{lemma} \label{lemma:cross_cov_asymp} (Proof in Appendix \ref{subsec:cross_cov_asymp_proof})
    Let $\Z_X \sim \cP_X$, then, \\$\frac{1}{3} (1 - \frac{\delta}{2}) \leq \E F_X(\theta(\Z_X)) F_Y(\theta(\Z_X)) \leq \frac{1}{3}.$ Define $W \defeq F_Y(\theta(\Z_X)) - (1-\mu) + F_X(\theta(\Z_X)) - 1/2$, then, $\var(W) \to 1/3$ if $\delta \to 0$.
\end{lemma}

\section{Proof of Technical Lemmas} \label{sec:lemma_proofs}
For the reader's convenience, we recall some notations defined previously. For any $\Z_l$, we let $\hat{W}^X_l = F_{X,*}( \hat{\theta}(\Z_l))$, and, $\hat{W}^Y_l = F_{Y,*}( \hat{\theta}(\Z_l))$. For some $i, j \in \{m+1, \ldots, T-m\}$, suppose, $\Z_i, \Z_j \simiid \cP_X$, then $\E_* [\indic\{\hat{\theta}(\Z_i) < \hat{\theta}(\Z_j)\} |\Z_j] = \hat{W}^X_j$, $\E_* [\indic\{\hat{\theta}(\Z_i) < \hat{\theta}(\Z_j)\} |\Z_i] = 1 - \hat{W}^X_i$, and, $\E_* \hat{W}^X_i = 1/2$. Moreover, suppose $\Z_i \sim \cP_X, \ \Z_j \sim \cP_Y$. Then, it follows that, $\E_* [\indic\{\hat{\theta}(\Z_i) < \hat{\theta}(\Z_j)\}|\Z_j] = \hat{W}^X_j$, $\E_* [\indic\{\hat{\theta}(\Z_i) < \hat{\theta}(\Z_j)\}|\Z_i] = 1 - \hat{W}^Y_i$, and, $\E_* \hat{W}^X_j = \mu_*= 1 - \E_* \hat{W}^Y_i$.
\subsection{Proof of Lemma \ref{lemma:cross_cov_remaind_terms_mix}} \label{subsec:cross_cov_remaind_terms_mix_proof}
\begin{proof}
 Note that, $\Z_i, \Z_j \simiid \cP_X$ and $\Z_{j'} \sim \cP_Y$ are independent random elements. Now we can write, \begin{flalign*}
        \begin{split}
            &\hat{h}^{0}(\Z_i, \Z_j) \hat{h}^{\dagger}(\Z_{i}, \Z_{j'}) \\=& \firstbigg{ \indic\{\hat{\theta}(\Z_i) < \hat{\theta}(\Z_j)\} - \hat{W}^X_{j} + \hat{W}^X_i - 1/2 } \firstbigg{ \indic\{\hat{\theta}(\Z_i) < \hat{\theta}(\Z_{j'})\} - \hat{W}^X_{j'} - \hat{W}^Y_i - 1 + \mu_* } \\
            =& \firstbigg{ \indic\{\hat{\theta}(\Z_i) < \hat{\theta}(\Z_j)\} - \hat{W}^X_j } \firstbigg{ \indic\{\hat{\theta}(\Z_i) < \hat{\theta}(\Z_{j'})\} - \hat{W}^X_{j'} } - \firstbigg{\hat{W}^X_i - 1/2} \firstbigg{ \hat{W}^Y_i - (1-\mu_{*}) } \\
            &- \firstbigg{ \indic\{\hat{\theta}(\Z_i) < \hat{\theta}(\Z_j)\} - \hat{W}^X_j } \firstbigg{ \hat{W}^Y_i - (1-\mu_*) } \\
            & + \firstbigg{ \indic\{\hat{\theta}(\Z_i) < \hat{\theta}(\Z_{j'})\} - \hat{W}^X_{j'} } \firstbigg{ \hat{W}^X_i - 1/2 } \\
            \defeq& \mathrm{I} - \mathrm{II} -\mathrm{III} +\mathrm{IV} 
            \end{split}
    \end{flalign*} We now simplify each of the four terms using the tower property of expectation. \begin{flalign*}
        \E_{*} I &= \E_{*} \thirdbigg{ \firstbigg{ \indic\{\hat{\theta}(\Z_i) < \hat{\theta}(\Z_j)\} - \hat{W}^X_j } \E_{*} \thirdbigg{ \indic\{\hat{\theta}(\Z_i) < \hat{\theta}(\Z_{j'})\} - \hat{W}^X_{j'} |\Z_i, \Z_j} } \\
            &= \E_{*} \thirdbigg{ \firstbigg{ \indic\{\hat{\theta}(\Z_i) < \hat{\theta}(\Z_j)\} - \hat{W}^X_j } \E_{*} \thirdbigg{ \indic\{\hat{\theta}(\Z_i) < \hat{\theta}(\Z_{j'})\} - \hat{W}^X_{j'} |\Z_i} } \\
            &= \E_{*} \thirdbigg{ \firstbigg{ \indic\{\hat{\theta}(\Z_i) < \hat{\theta}(\Z_j)\} - \hat{W}^X_j } \firstbigg{ 1 - \hat{W}^Y_i - \mu_* } } \\
            &= \E_{*} \thirdbigg{ \firstbigg{ 1 - \hat{W}^Y_i - \mu_* } \E_{*}\thirdbigg{ \firstbigg{ \indic\{\hat{\theta}(\Z_i) < \hat{\theta}(\Z_j)\} - \hat{W}^X_j }|\Z_i } } \\
            &= \E_{*} \thirdbigg{ \firstbigg{ 1 - \hat{W}^Y_i - \mu_* } \firstbigg{ 1 - \hat{W}^X_i - 1/2 } } \\
            &= \E_{*} \mathrm{II}
    \end{flalign*} Moreover, we simplify the third term, \begin{flalign*}
        \begin{split}
            \E_{*} \mathrm{III} &= \E_{*} \thirdbigg{ \firstbigg{ \hat{W}^Y_i - (1-\mu_*) } \E_{*} \thirdbigg{ \firstbigg{ \indic\{\hat{\theta}(\Z_i) < \hat{\theta}(\Z_j)\} - \hat{W}^X_j }|\Z_i } } \\
            &= \E_{*} \thirdbigg{ \firstbigg{ \hat{W}^Y_i - (1-\mu_*) } \E_{*}\firstbigg{ 1 - \hat{W}^X_i - 1/2 } } \\
            &= -\E_{*} \mathrm{II}.
        \end{split}
    \end{flalign*} Following similar steps, we simplify the fourth term, \begin{flalign*}
        \begin{split}
            \E_{*} \mathrm{IV} &= \E_{*}\thirdbigg{ \firstbigg{ \hat{W}^X_i - 1/2 } \E_{*}\thirdbigg{ \firstbigg{ \indic\{\hat{\theta}(\Z_i) < \hat{\theta}(\Z_{j'})\} - \hat{W}^X_{j'} }| \Z_i } } \\
            &= \E_{*}\thirdbigg{ \firstbigg{ \firstbigg{ \hat{W}^X_i - 1/2 } \E_{*}\thirdbigg{ \firstbigg{ 1 - \hat{W}^Y_i - \mu_* } } } } \\
            &= - \E_{*} \mathrm{II}.
        \end{split}
    \end{flalign*} Combining all the simplifications, we have, $\E_{*} \hat{h}^{0}(\Z_i, \Z_j) \hat{h}^{\dagger}(\Z_{i}, \Z_{j'}) = 0$. 
\end{proof}
\subsection{Proof of Lemma \ref{lemma:kley_et_al_extn}} \label{subsec:kley_et_al_extn_proof}
\begin{proof}
    For all $r, r' \in \mc{T}_T$ with $\absbigg{r-r'} \geq 1/T$, we note that, $\E \thirdbigg{G_T(r) - G_T(r')}^2 \lesssim \absbigg{r-r'}.$ Moreover, we have, \begin{equation*}
        \sup_{|r-r'|\leq \delta, \gamma \leq r,r'\leq 1-\gamma} \absbigg{G_T(r) - G_T(r')} \leq \sup_{r,r'\in \mathcal{T}_T:|r-r'|\leq\delta+T^{-1}}\absbigg{G_T(r) - G_T(r')}.
    \end{equation*} Note that the packing number of $\mathcal{T}_T$ with respect to the metric $d(r,r')=|r-r'|$ satisfies that $D_{\mathcal{T}_T}(\tilde{\epsilon},d) \leq D_{[\gamma,1-\gamma]}(\tilde{\epsilon},d) \leq 2\tilde{\epsilon}^{-1}.$ Hence, by Lemma A.1 in \cite{kley2016quantile}  with $\Psi(x)=x^2$, $d(r,r')=|r-r'|$, $\bar{\eta}=T^{-1}/2$, we have that for any $\tilde{\eta}\geq\bar{\eta}$, there exists a random variable $S_T(\tilde{\eta},\delta)$ such that 
\begin{equation}\label{eq:kley_eq}
    \sup_{|r-r'|\leq \delta+T^{-1}} \left|G_T(r) - G_T(r')\right|\leq S_T(\tilde{\eta},\delta) + 2 \sup_{r,r'\in\mathcal{T}_T:|r-r'|\leq \bar{\eta} } \left|G_T(r) - G_T(r')\right|, \quad \text{with}
\end{equation} 
$$\E S^2_T(\tilde{\eta},\delta) \lesssim \int_{\bar{\eta}}^{\tilde{\eta}}D^{1/2}_{\mathcal{T}_T}(\tilde{\epsilon},d)\mathrm{d}\tilde{\epsilon}+(\delta+T^{-1}+2\bar{\eta})D_{\mathcal{T}_T}(\tilde{\epsilon},d) \lesssim \tilde{\eta}^{1/2}+\tilde{\epsilon}^{-1}(\delta+2T^{-1}).$$
Note that $\inf_{r,r'\in\mathcal{T}_T,r\neq r'}|r-r'|\geq T^{-1}$, it follows that ${r,r'\in\mathcal{T}_T:|r-r'|\leq \bar{\eta}} $ implies $r=r'$. Therefore, the second term in \eqref{eq:kley_eq} vanishes, and
$$
\sup_{|r-r'|\leq \delta+T^{-1}} \left| G_T(r) - G_T(r') \right|\leq S_T(\tilde{\eta},\delta).
$$
Then, applying the Markov inequality, we have 
\begin{flalign*}
      &\lim_{\delta\to 0}\limsup_{T\to\infty}  \mathbb{P}\left(\sup_{|r-r'|\leq \delta, \gamma\leq r,r'\leq 1-\gamma} \absbigg{G_T(r) - G_T(r')}>x \right)\\&\leq  \lim_{\delta\to 0}\limsup_{T\to\infty}  \mathbb{P}\left(|S_T(\tilde{\eta},\delta) |>x \right)\\
      &\leq \lim_{\delta\to 0}\limsup_{T\to\infty} 2x^{-2} [\tilde{\eta}+\tilde{\epsilon}^{-2}(\delta+2T^{-1})^2]\leq 2x^{-2}\tilde{\eta}.
\end{flalign*}
Since $\tilde{\eta}$ can be arbitrarily chosen, the stochastic equicontinuity is established.
\end{proof}
\subsection{Proof of Lemma \ref{lemma:reln_mu_delta}} \label{subsec:reln_mu_delta_proof}
\begin{proof}
    Let us recall, $L(\cdot) = \frac{d\cP_Y}{d\cP_X}(\cdot)$ and consider, $\Z_X, \Z_{X}^{'}, \Z_{X}^{''} \simiid \cP_X$, $\Z_Y \sim \cP_Y$. Then, \begin{flalign*}
    \begin{split}
        \mu &= \P \firstbigg{ \theta(\Z_X) < \theta(\Z_Y) } = \P \firstbigg{ L(\Z_X) < L(\Z_Y) } \\
        &= \int_{x,x'\in E} \indic\secondbigg{ L(x) < L(x') } d\cP_X(x) d\cP_Y(x') \\
        &= \int_{x,x'\in E}  \indic\secondbigg{ L(x) < L(x') } \frac{d\cP_Y}{d\cP_X}(x') d\cP_X(x) d\cP_X(x') \\
        &= \int_{x,x'\in E}  \indic\secondbigg{ L(x) < L(x')} L(x') d\cP_X(x) d\cP_X(x') \\
        &= \E L(\Z_X^{'}) \indic\secondbigg{ L(\Z_X) < L(\Z_X^{'}) }.
    \end{split}
\end{flalign*} Since $\Z_X$ and $\Z_X^{'}$ are iid and $\E L(\Z_X) = \int_{x\in E}(d\cP_Y/d\cP_X)(x) d\cP_X(x)=1$, then, \begin{flalign*}
    \begin{split}
        \mu &= \E L(\Z_X^{'}) \indic\secondbigg{ L(\Z_X) < L(\Z_X^{'}) } \\&= \frac{1}{2} \firstbigg{ \E L(\Z_X^{'}) \indic\secondbigg{ L(\Z_X) < L(\Z_X^{'}) } + \E L(\Z_X) \indic\secondbigg{ L(\Z_X^{'}) < L(\Z_X) } } \\
    &= \frac{1}{2} \firstbigg{ 1 - \E L(\Z_X^{'}) \indic\secondbigg{ L(\Z_X) > L(\Z_X^{'}) } + \E L(\Z_X) \indic\secondbigg{ L(\Z_X^{'}) < L(\Z_X) } } \\
    &= \frac{1}{2} \firstbigg{ 1 + \E \firstbigg{L(\Z_X) - L(\Z_X^{'})} \indic\secondbigg{ L(\Z_X^{'}) < L(\Z_X) } } \\
    &= \frac{1}{2} + \frac{1}{4} \E \absbigg{ L(\Z_X) - L(\Z_X^{'}) }= \frac{1}{2} + \frac{1}{4} \delta.
    \end{split}
\end{flalign*}
\end{proof}
\subsection{Proof of Lemma \ref{lemma:sigmas_asymp}} \label{subsec:sigmas_asymp_proof}
\begin{proof}
    Note, for  $\Z_X''$ being an independent copy of $\Z_X$, \begin{flalign*}
        \mu_{XY} &= \P \firstbigg{ \theta(\Z_X) < \theta(\Z_Y), \theta(\Z_X^{'}) < \theta(\Z_Y) } \\
        &= \P \firstbigg{ L(\Z_X) < L(\Z_Y), L(\Z_X^{'}) < L(\Z_Y) } \\
        &= \int_{x,x',y\in E}\indic \secondbigg{ L(x) < L(y), L(x') < L(y) } d\cP_X(x) d\cP_X(x') d\cP_Y(y) \\
        &= \int_{x,x',y\in E} \indic \secondbigg{ L(x) < L(y), L(x{'}) < L(y) } \frac{d\cP_Y}{d\cP_X}(y) d\cP_X(x) d\cP_X(x') d\cP_X(y) \\
        &= \E L(\Z_X^{''}) \secondbigg{ L(\Z_X) < L(\Z_X^{''}), L(\Z_X^{'}) < L(\Z_X^{''}) }.
\end{flalign*} It is hard to find an exact expression of $\mu_{XY}$ in terms of $\delta$. Instead, we provide a bound and show, $\mu_{XY} = 1/3 + O(\delta)$, and  $\sigma_{XY} = 1/12 + O(\delta)$. The same conclusion holds for $\sigma_{YX}$ by the symmetry of our argument. First, we obtain an upper bound for $\mu_{XY}$. To simplify the notations, let $L = L(\Z_X), \ L' = L(\Z_X^{'})$ and $L'' = L(\Z_X^{''})$. We use Lemma \ref{lemma:reln_mu_delta} and note the following after eliminating ties between $L, L', L''$: $\frac{1}{2} + \frac{\delta}{4} = \E L'' \indic\secondbigg{L < L''} = \E L'' \indic\secondbigg{L<L'', L'<L''} + \E L'' \indic\secondbigg{L<L'', L' > L''}.$ Therefore, we have, \begin{equation} \label{eq:simple_trick}
  \E L'' \indic\secondbigg{L<L'', L'<L''} = \firstbigg{ \frac{1}{2} + \frac{\delta}{4} } - \E L'' \indic\secondbigg{L < L'' < L'}.
\end{equation} Moreover, we note that: \begin{align} \label{eq:another_trick}
    \begin{split} 
        \E L' \indic\secondbigg{L<L', L''<L'} &= \E L' \indic\secondbigg{L<L', L''<L', L < L''} + \E L' \indic\secondbigg{L<L', L''<L', L > L''} \\
        &= \E L' \indic\secondbigg{L<L''<L'} + \E L' \indic\secondbigg{L'' < L < L'}.
    \end{split}
\end{align} Since $L, L', L''$ are non-negative, i.i.d. random variables,  we can write: \begin{equation} \label{eq:simple_inequality}
    0 \leq \E (L' - L'') \indic\secondbigg{L<L''<L'} \leq \E (L' - L'') \indic\secondbigg{L''<L'}
\end{equation} The first inequality holds since $L'-L'' > 0$ inside the support and the second holds since $\indic\secondbigg{L<L''<L'}\leq \indic\secondbigg{L''<L'}.$
Now, we derive a similar bound for $\mu_{XY}$: \begin{flalign*}
    \begin{split}
        \mu_{XY} &= \E L'' \indic\{ L < L'', L' < L'' \} \\
        &\stackrel{\text{using iid}}{=} \frac{1}{3} \thirdbigg{ \E L'' \indic \{ L < L'', L' < L'' \} + \E L' \indic \{ L < L', L'' < L' \} + \E L \indic \{ L'' < L, L' < L \} } \\
        &\stackrel{\eqref{eq:simple_trick}}{=} \frac{1}{3} \left[ \firstbigg{\frac{1}{2} + \frac{\delta}{4}} - \E L'' \indic\{L<L''<L'\} + \E L'\indic\{L<L', L'' < L'\} \right. \\
        & \left. \ \ \ \ \ \ \ + \firstbigg{\frac{1}{2} + \frac{\delta}{4}} - \E L \indic\secondbigg{L'' < L < L'} \right] \\
        &\stackrel{\eqref{eq:another_trick}}{=} \frac{1}{3} \left[ 1 + \frac{\delta}{2} - \E L'' \indic\{L<L''<L'\} + \E L'\indic\{L<L''< L'\} \right. \\
        & \left. \ \ \ \ \ \ \ + \E L'\indic\{L''<L< L'\} - \E L \indic\secondbigg{L'' < L < L'} \right] \\
        &= \frac{1}{3} \thirdbigg{ 1 + \frac{\delta}{2} + \E (L'-L'') \indic\{L<L''<L'\} + \E (L'-L) \indic\{L''<L<L'\} } \\
        &\stackrel{\text{using iid}}{=} \frac{1}{3} + \frac{\delta}{6} + \frac{2}{3} \E (L'-L'') \indic\{L<L''<L'\} \\
        &\stackrel{\eqref{eq:simple_inequality}}{\leq} \frac{1}{3} + \frac{\delta}{6} + \frac{2}{3} \E (L'-L'') \indic\{L''<L'\} \\
        &\stackrel{\text{using iid}}{=} \frac{1}{3} + \frac{\delta}{6} + \frac{2}{6} \E \absbigg{L'-L''} \\
        &= \frac{1}{3} + \frac{\delta}{2}.
    \end{split}
\end{flalign*} The second last inequality holds due to $\delta \defeq \E \absbigg{L' - L''}$. We now apply \eqref{eq:simple_inequality} to have, $\mu_{XY} = \frac{1}{3} + \frac{\delta}{6} + \frac{2}{3} \E (L'-L'') \indic\{L<L''<L'\} \geq \frac{1}{3} + \frac{\delta}{6}.$ Therefore, we have, $1/3 + \delta/6 \leq \mu_{XY} \leq 1/3 + \delta/2$. Hence, it follows that, $\mu_{XY} = 1/3 + O(\delta)$. Recall, $\sigma_{XY} = \mu_{XY} - \mu^2$. From Lemma \ref{lemma:reln_mu_delta}, we have, $\mu - 1/2 = \delta/4$. Therefore, $\sigma_{XY} = 1/12 + O(\delta)$. Following similar steps, it is possible to show the same for $\sigma_{YX}$.
\end{proof}
\subsection{Proof of Lemma \ref{lemma:cross_cov_asymp}} \label{subsec:cross_cov_asymp_proof}
\begin{proof}
    Suppose, $\Z_X, \Z'_X, \Z''_X \simiid \cP_X, \ \Z_Y \sim \cP_Y$. Let us recall, $L(\Z) = \frac{d\cP_Y}{d\cP_X}(\Z)$. We observe that, \begin{flalign*}
        \begin{split}
           &\E F_X(\theta(\Z_X)) F_Y(\theta(\Z_X))) \\
            =& \int_{x,x',y\in E} \indic\secondbigg{ \theta(x')< \theta(x) } \indic\secondbigg{ \theta(y)< \theta(x) } d\cP_X(x') d\cP_X(x) d\cP_Y(y) \\
            = &\int_{x,x',y\in E}\indic\secondbigg{ L(x') < L(x) } \indic\secondbigg{ L(y) < L(x) } L(y) d\cP_X(x') d\cP_X(x) d\cP_X(y) \\
            =& \E L(\Z''_X) \indic\secondbigg{ L(\Z'_X) < L(\Z_X), L(\Z''_X) < L(\Z_X) }.
        \end{split}
    \end{flalign*} To simplify the notations, let us define, $L = L(\Z_X)$, $L' = L(\Z'_X)$ and $L'' = L(\Z''_X)$, which are non-negative and iid random variables. From Lemma \ref{lemma:reln_mu_delta} recall, $\mu = \E L' \indic\secondbigg{L''<L'} = 1/2 + \delta/4$ and $\E L = 1$. Therefore, $\E L' \indic\secondbigg{L'<L''} = \E L' - \E L' \indic\secondbigg{L'>L''} = 1/2 - \delta/4$ and, $\E L' \indic\secondbigg{L'<L''}    = \E L' \indic\secondbigg{L'<L'', L < L''} + \E L' \indic\secondbigg{L'<L''<L}.$ Hence,\begin{equation} \label{eq:simple_trick1}
        \E L' \indic\secondbigg{L'<L'', L < L''} = 1/2 - \delta/4 - \E L' \indic\secondbigg{L'<L''<L},
    \end{equation} \begin{align} \label{eq:simple_trick2}
        \begin{split}
            \E L \indic\secondbigg{L'<L'', L<L''} = \E L \indic\secondbigg{L<L'<L''} + \E L \indic\secondbigg{L'<L<L''},
        \end{split} 
    \end{align}   \begin{align} \label{eq:simple_inequality1}
        \begin{split}
            \E (L-L') \indic\secondbigg{L'<L<L''}  \leq  \E (L' - L) \indic\secondbigg{L < L'} 
            &\stackrel{\text{using iid}}{=} \frac{1}{2} \E\absbigg{L' - L}.
        \end{split}
    \end{align} We now use \eqref{eq:simple_trick1}, \eqref{eq:simple_trick2} and \eqref{eq:simple_inequality1} to prove the lemma. Note, \begin{flalign*}
        \begin{split}
            &\E L' \indic\secondbigg{L<L'', L'<L''} \\
            &= \frac{1}{3} \thirdbigg{ \E L' \indic\secondbigg{L<L'', L'<L''} + \E L \indic\secondbigg{L'<L'', L<L''} + \E L'' \indic\secondbigg{L<L', L''<L'} } \\
            &\stackrel{\eqref{eq:simple_trick1}}{=} \frac{1}{3} \left[ \firstbigg{1/2 - \delta/4} - \E L' \indic\secondbigg{L'<L''<L} + \E L \indic\secondbigg{L'<L'', L<L''} \right. \\
            & \ \ \ \ \ \ \left. + \firstbigg{1/2 - \delta/4} - \E L'' \indic\secondbigg{L''<L'<L} \right] \\
            &\stackrel{\eqref{eq:simple_trick2}}{=} \frac{1}{3} \left[ \firstbigg{1 - \delta/2} - \E L' \indic\secondbigg{L'<L''<L} + \E L \indic\secondbigg{L<L'<L''} \right. \\
            & \ \ \ \ \ \ \left. + \E L \indic\secondbigg{L'<L<L''} - \E L'' \indic\secondbigg{L''<L'<L} \right] \\
            &\stackrel{\text{using iid}}{=} \frac{1}{3} \thirdbigg{ \firstbigg{1 - \delta/2} + \E L \indic\secondbigg{L'<L<L''} - \E L'' \indic\secondbigg{L''<L'<L} } \\
            &\stackrel{\text{using iid}}{=} \frac{1}{3} \thirdbigg{ \firstbigg{1 - \delta/2} + \E L \indic\secondbigg{L'<L<L''} - \E L' \indic\secondbigg{L'<L<L''} } \\
            &= \frac{1}{3} \thirdbigg{ \firstbigg{1 - \delta/2} + \E (L-L') \indic\secondbigg{L'<L<L''} }.
        \end{split}
    \end{flalign*} Hence, we observe $\E L' \indic\secondbigg{L<L'', L'<L''} \geq \frac{1}{3} \firstbigg{1 - \delta/2}$. Also, we use \eqref{eq:simple_inequality1} to obtain $\E L' \indic\secondbigg{L<L'', L'<L''} \leq \frac{1}{3} \thirdbigg{1 - \frac{\delta}{2} + \frac{1}{2} \E\absbigg{L'-L}} = \frac{1}{3},$ since, $\E\absbigg{L'-L} = \delta$ from Lemma \ref{lemma:reln_mu_delta}. Combining the last two inequality, we have: $\frac{1}{3} - \frac{\delta}{6} \leq \E F_X(\theta(\Z_X)) F_Y(\theta(\Z_X))  \leq \frac{1}{3}.$ Next, by definition, we have $\E W= \P(\theta(\Z_Y)<\theta(\Z_X))-(1-\mu)+\P(\theta(\Z_X')<\theta(\Z_X))-1/2=0$. Hence, $\var W = \E W^2 = \sigma_{YX} + \frac{1}{12} + 2 \firstbigg{ \E F_X(\theta(\Z_X)) F_Y(\theta(\Z_X)) - \frac{1-\mu}{2} }\stackrel{\delta \to 0}{\to} \frac{1}{12} + \frac{1}{12} + 2 \firstbigg{ \frac{1}{3} - \frac{1}{4} } = \frac{1}{3},$ where, we apply Lemma \ref{lemma:sigmas_asymp} and Lemma \ref{lemma:reln_mu_delta}.
\end{proof}
\section{Additional Simulation Results}\label{sec:add_simu}
\subsection{Performance Comparison with Additional Methods} \label{subsec:simu_additional_methods}
\begin{figure}[!htp]
    \centering
    \includegraphics[width=\textwidth]{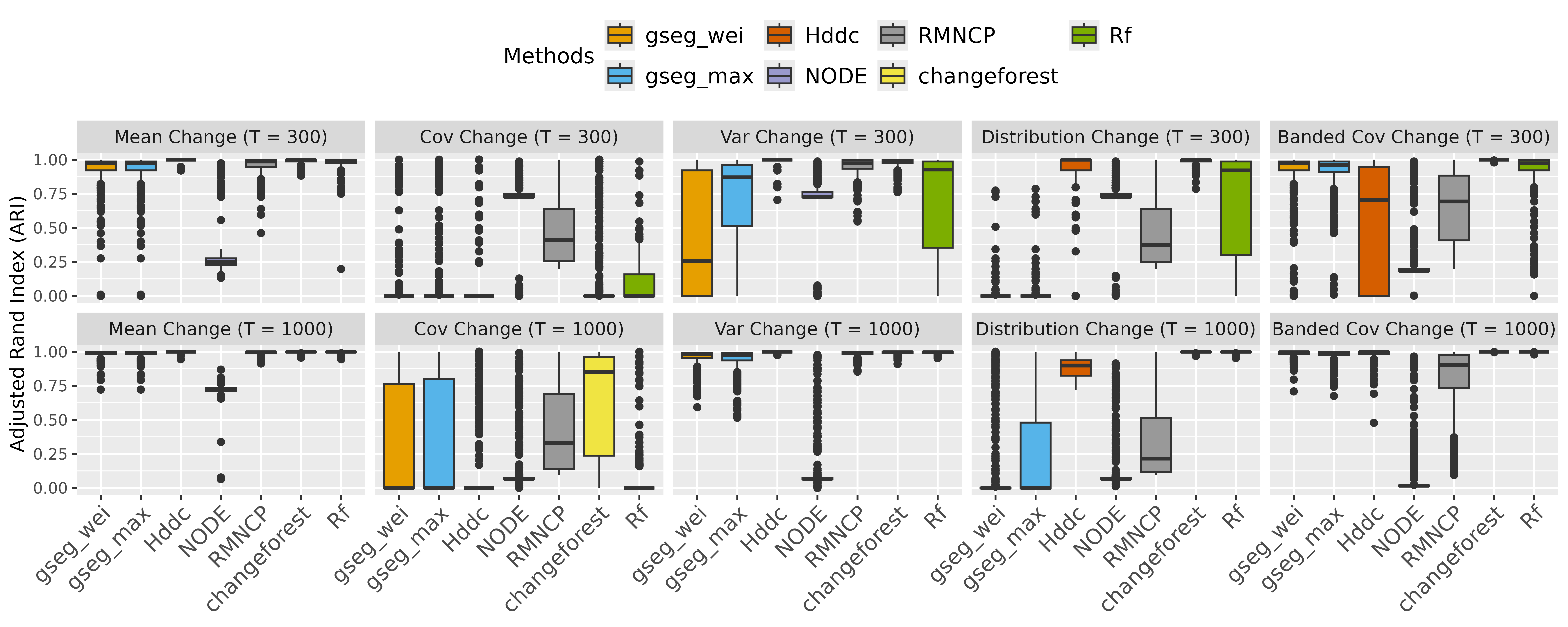}
     \caption{Boxplots of ARI values across 500 repetitions. Data are generated from the same setup outlined in Section \ref{subsec:sim_Euclidean} with $T \in \{300, 1000\}, \ p=30$. The first row depicts ARI boxplots for $T=300$, whereas the second corresponds to $T=1000$. The columns represent different data-generating processes under consideration. Each of the 7 grouped boxplots contains seven different boxplots arranged horizontally, which respectively correspond to \texttt{gseg\_wei}, \texttt{gseg\_max},  \texttt{Hddc}, \texttt{NODE} \citep{wang_et_al_2023_ieee}, \texttt{RMNCP} \citep{padilla_et_al_2021_ieee}, \texttt{changeforest} and \texttt{Rf}.}
    \label{fig:boxplot_sig_small}  
\end{figure}

In this section, we compare our proposed method \texttt{Rf} classifier with the other methods, with a particular focus on \texttt{RMNCP} \citep{padilla_et_al_2021_ieee}. Here we conduct a separate simulation study with $T\in \{300,1000\}$ and $p=30$ as considered in \cite{padilla_et_al_2021_ieee} using DGPs same as ours as outlined in \ref{subsec:sim_Euclidean}.  Figure \ref{fig:boxplot_sig_small} compares the performance when there is at most one change-point. Since $p=30$, the data generated from `Sparse Mean Change', `Sparse Diag Cov Change' and `Sparse Distribution Change' do not contain any change-points, and thus, we remove it from this experiment. We omit `Dense' from the names of DGPs and rename `Dense Diag Cov Change' as `Var Change.'
From Figure \ref{fig:boxplot_sig_small}, we observe good performance from all methods in `Mean Change' and equally bad performance in `Cov Change.' Both \texttt{Rf} and \texttt{changeforest} perform equally well in `Banded Cov Change,' while the former struggles to perform well in `Var Change' and `Distribution Change' due to the low training sample size. \texttt{RMNCP} perform well in `Mean Change' and `Var Change.' However, it struggles in other data-generating regimes. 
\subsection{Comparison of \texttt{changeAUC} and CUSUM Process using VGG Embedding} \label{subsec:vgg_cusum}
In this section, we include all the details on using the CUSUM process around classification probabilities and outline key technical details. 
This is motivated by a reviewer's comment, who pointed out that it is possible to use the conditional probabilities $\hat{\theta}(\cdot)$ inside a one-dimensional CUSUM process. This will serve as a baseline to understand the usefulness of AUC-based rank-sum comparisons in detecting a single change-point. Let us recall $\calI_{cp} = \{\floor{Tr}: r \in [\epsilon+\eta, 1-\epsilon-\eta]\}$, denote $\hat{W}_t := \hat{\theta}(\Z_t)$ and we define the CUSUM statistic constructed using $\{\hat{W}_t\}_{t \in \calI_{cp}}$ as the following for all $k \in \calI_{cp}$: \begin{equation*}
    \hat{\Phi}(k) := \frac{1}{\sqrt{\hat{V}}} \sqrt{\frac{T-2m}{(k-m)(T-k-m)}} \thirdbigg{ \sum_{i=m+1}^{k} \hat{W}_i - \firstbigg{ \frac{k-m}{T-2m} } \sum_{i=m+1}^{T-m} \hat{W}_i },
\end{equation*} where we recall $m = \floor{T\epsilon}$ and define $\hat{V} := \min_{m+1 \leq k \leq T-m} \hat{V}(k)$ with \begin{align*}
    \begin{split}
        \hat{V}(k) &= \frac{1}{T-2m} \secondbigg{ \sum_{i=m+1}^{k}\firstbigg{ \hat{W}_i - \Bar{\hat{W}}_{k, 1} }^2 + \sum_{i=k+1}^{T-m}\firstbigg{ \hat{W}_i - \Bar{\hat{W}}_{k, 2} }^2 }, \quad \text{for} \ k \in \calI_{cp}, \\
        \Bar{\hat{W}}_{k, 1} &= \frac{1}{k-m} \sum_{i=m+1}^{k} \hat{W}_i, \quad \Bar{\hat{W}}_{k, 2} = \frac{1}{T-m-k} \sum_{i=k+1}^{T-m} \hat{W}_i, \quad \text{for} \ k \in \calI_{cp}.
    \end{split}
\end{align*} Under suitable conditions, we conjecture that \begin{align*}
    \begin{split}
        \secondbigg{\hat{\Phi}(\floor{Tr})}_{r \in [\gamma, 1-\gamma]} \ucd \secondbigg{ H_0(r) }_{r \in [\gamma, 1-\gamma]}, \quad L^{\infty}([0, 1]), \quad \text{where} \ \gamma=\epsilon+\eta, \\
        H_0(r) := \sqrt{\frac{1-2\epsilon}{(r-\epsilon)(1-r-\epsilon)}} \thirdbigg{ \secondbigg{ B(r) - B(\epsilon) } + \firstbigg{ \frac{r-\epsilon}{1-2\epsilon} } \secondbigg{ B(1-\epsilon) - B(\epsilon) } },
    \end{split}
\end{align*} and $B(\cdot)$ is the standard Brownian Motion. 
\begin{table}
\centering
\begin{tabular}{rrrrrr}
  \hline
$\alpha$ & 20\% & 10\% & 5\% & 1\% & 0.5\% \\ 
  \hline
$Q(1-\alpha)$ & 2.170 & 2.529 & 2.828 & 3.413 & 3.626 \\ 
   \hline
\end{tabular}
\caption{Simulated quantiles of $\sup_{r\in [\gamma, 1-\gamma]} H_0(r)$ for $\epsilon=0.15, \eta=0.05$.}
\label{tab:theo_quantiles_cusum}
\end{table}
In Table \ref{tab:theo_quantiles_cusum}, we provide the quantiles of $\sup_{r \in [\gamma, 1-\gamma]} H_0(r)$ with $10^5$ knots of the standard Brownian motion using $10^5$ replications. $Q(0.95)$ is used in Table \ref{tab:cifar10_v2} to obtain the empirical rejection rate when the CUSUM statistic is used with classification probabilities obtained from \texttt{vgg16} and \texttt{vgg19}. 

\vskip 0.2in
\bibliography{ref}

\end{document}